\documentclass[useAMS,usegraphicx,usenatbib]{mn2e}
\usepackage[T1]{fontenc}
\usepackage{times,amssymb,calligra}
\DeclareMathAlphabet{\mathcalligra}{T1}{calligra}{m}{n}
\DeclareFontShape{T1}{calligra}{m}{n}{<->s*[2.2]callig15}{}
\newcommand\bb[1]{\mbox{\boldmath{$#1$}}}
\newcommand\del{\bb{\nabla}}
\newcommand\bcdot{\bb{\cdot}}

\title[Nonisothermal magnetic star formation]{The nonisothermal stage of magnetic star formation. II. Results}
\author[M. W. Kunz \& T. Ch. Mouschovias]{Matthew W. Kunz\thanks{Current Address: Rudolf Peierls Centre for Theoretical Physics, University of Oxford, 1 Keble Road, Oxford OX1 3NP, UK; E-mail: kunz@thphys.ox.ac.uk} and Telemachos Ch. Mouschovias \\ Departments of Physics and Astronomy, University of Illinois at Urbana-Champaign, 1002 W. Green Street, Urbana, IL 61801}

\date{Accepted ?. Received ?}
\pagerange{\pageref{firstpage}--\pageref{lastpage}} \pubyear{2010}
\def\LaTeX{L\kern-.36em\raise.3ex\hbox{a}\kern-.15em
    T\kern-.1667em\lower.7ex\hbox{E}\kern-.125emX}

\begin{document}

\label{firstpage} \maketitle

\begin{abstract}

In a previous paper we formulated the problem of the formation and evolution of fragments (or cores) in magnetically-supported, self-gravitating molecular clouds in axisymmetric geometry, accounting for the effects of ambipolar diffusion and Ohmic dissipation, grain chemistry and dynamics, and radiative transfer. Here we present results of star formation simulations that accurately track the evolution of a protostellar fragment over eleven orders of magnitude in density (from $300~{\rm cm}^{-3}$ to $\approx 10^{14}~{\rm cm}^{-3}$), i.e., from the early ambipolar-diffusion--initiated fragmentation phase, through the magnetically supercritical, dynamical-contraction phase and the subsequent magnetic decoupling stage, to the formation of a protostellar core in near hydrostatic equilibrium. As found by \citet{fm93}, gravitationally-driven ambipolar diffusion leads to the formation and subsequent dynamic contraction of a magnetically supercritical core. Moreover, we find that ambipolar diffusion, not Ohmic dissipation, is responsible for decoupling all the species except the electrons from the magnetic field, by a density $\approx 3\times 10^{12}~{\rm cm}^{-3}$. Magnetic decoupling precedes the formation of a central stellar object and ultimately gives rise to a concentration of magnetic flux (a `magnetic wall') outside the hydrostatic core --- as also found by \citet{tm05a,tm05b} through a different approach. At approximately the same density at which Ohmic dissipation becomes more important than ambipolar diffusion ($\gtrsim 7\times 10^{12}~{\rm cm}^{-3}$), the grains carry most of the electric charge {\em as well as} the electric current. The prestellar core remains disclike down to radii $\sim 10~{\rm AU}$, inside which thermal pressure becomes important. The magnetic flux problem of star formation is resolved for at least strongly magnetic newborn stars by this stage of the evolution, i.e., by a central density $\approx 10^{14}~{\rm cm}^{-3}$. The hydrostatic core has radius $\approx 2~{\rm AU}$, density $\approx 10^{14}~{\rm cm}^{-3}$, temperature $\approx 300~{\rm K}$, magnetic field strength $\approx 0.2~{\rm G}$, magnetic flux $\approx 5\times 10^{18}~{\rm Wb}$, luminosity $\sim 10^{-3}~{\rm L}_\odot$, and mass $\sim 10^{-2}~{\rm M}_\odot$. 
\end{abstract}


\section{Introduction}\label{section:introduction}

In a previous paper \citep[hereafter Paper I]{km09a} we formulated the problem of the formation and subsequent evolution of fragments (or cores) in magnetically-supported, self-gravitating molecular clouds in two spatial dimensions. The formulation allowed for the detailed study of both ambipolar diffusion and Ohmic dissipation, grain chemistry and dynamics, and radiation. Here we present results from the first star formation simulation to rigorously and simultaneously take into account both nonideal magnetohydrodynamic (MHD) effects and radiative transfer.

The thermodynamic evolution of protostellar cores in the absence of magnetic fields has been studied numerically in spherical geometry \citep{bodenheimer68,larson69,larson72b,az75,yk77,yorke79,wn80a,wn80b,sst80a,sst80b,mmi98,lbca05}, in axisymmetric geometry \citep{larson72a,bb76,tscharnuter75,tscharnuter78,boss84,byrt90,ybl93,ybl95}, and in three-dimensional geometry \citep{boss86,boss88,boss93,bm95,bate98,wb06}, using widely varying initial conditions, approximations, and numerical techniques. Despite these differences, a consensus has emerged regarding the qualitative evolution of hydrodynamic, nonisothermal prestellar objects that is in broad agreement with the pioneering work of \citet{larson69}. Such objects have been shown to remain isothermal until central densities in the approximate range $2\times 10^{8}~{\rm cm}^{-3}$ -- $2\times 10^{10}~{\rm cm}^{-3}$ (depending on the simulation) are reached. The subsequent rise in temperature ultimately leads to the formation of a hydrostatic core at a temperature $\sim 100~{\rm K}$ and central density $\approx 2\times 10^{12}~{\rm cm}^{-3}$. The initial mass and radius of the hydrostatic core are typically found to be $\approx 0.005~{\rm M}_\odot$ and $\approx 4~{\rm AU}$, respectively. A thermal shock forms at the core boundary as mass is accreted onto the hydrostatic core, while small-amplitude oscillations occur about equilibrium. Finally, a temperature of $1000~{\rm K}$ is reached at a central density $\approx 2\times 10^{15}~{\rm cm}^{-3}$. Soon thereafter, molecular hydrogen dissociates (at $2000~{\rm K}$), the ratio of specific heats $\gamma$ dips below the critical value $4/3$, and material at the centre of the core becomes unstable and collapses dynamically. Despite the ensuing highly dynamic evolution, the temperature rises only slowly since most of the gravitational energy goes into molecular dissociation. Once molecular hydrogen is almost all dissociated, $\gamma$ rises above $4/3$ and the thermal pressure increases rapidly, decelerating and ultimately halting the collapse at the centre. A second (stellar) core forms, accompanied by another small rebound and subsequent radial pulsations. The mass and size of this core are $\approx 0.0015~{\rm M}_\odot$ and $\approx 1.3~{\rm R}_\odot$, respectively. Its central density $\approx 0.02~{\rm g~cm}^{-3}$ and temperature $\approx 20,000~{\rm K}$.

The addition of magnetic fields to the picture has brought a myriad of interesting effects \citep[e.g., see reviews by][]{mouschovias87,mouschovias96,cb00}. Aside from the support that magnetic fields provide against self-gravity, the inclusion of magnetic fields has led to a qualitatively different picture of molecular cloud fragmentation and evolution than earlier hydrodynamic calculations had suggested. In an initially magnetically subcritical cloud, gravitational infall occurs only as rapidly as allowed by ambipolar diffusion, the relative drift of neutral and charged particles. Ambipolar diffusion allows thermally supercritical fragments to separate out, in which the neutral particles fall in toward local gravity centres, impeded by collisions with the charged particles, which are essentially `held in place' by magnetic forces. Magnetic braking is very effective during this phase of contraction and is responsible for reducing the angular momenta of cloud cores to their observed low values and for resolving the angular momentum problem during the early, isothermal stage of contraction. The central mass-to-flux ratio of each forming fragment eventually increases to the critical value for collapse (typically when the central density reaches values in the range $\approx 10^4~{\rm cm}^{-3}$ -- $10^5~{\rm cm}^{-3}$). The now thermally and magnetically supercritical fragments begin to collapse more rapidly than their surroundings and they are simply referred to as {\it supercritical fragments or cores}. Each core evolves dynamically (though slower than free fall) under near flux freezing, until the resurrection of ambipolar diffusion causes magnetic decoupling\footnote{Complete magnetic decoupling refers to conditions such that the magnetic field has no significant effect on the dynamics of the neutral matter {\em and} the motion of the neutral matter no longer affects the magnetic field --- see footnote 3 of \citet{dm01}.} to set in at a density $\approx 10^{10}~{\rm cm}^{-3}$. Magnetic decoupling occurs over a few orders of magnitude in central density enhancement and precedes the formation of a central stellar object. The isothermal evolution of a magnetic molecular cloud has been followed in detail up to central densities of $2\times 10^{12}~{\rm cm}^{-3}$ by \citet{dm01}, who find that the magnetic field strength asymptotically approaches $\sim 0.1~{\rm G}$ in the innermost $\approx 20~{\rm AU}$ of the cloud.

The numerical simulations of \citet{boss97,boss99,boss02,boss05,boss07,boss09} followed the thermodynamic evolution of a magnetic protostellar core with a reasonable degree of accuracy, via the Eddington approximation, while treating magnetic field effects and ambipolar diffusion crudely through various approximations and parametrisations based on previous isothermal MHD calculations. \citet{tm07a,tm07b,tm07c} took the opposite approach by incorporating a detailed treatment of nonideal MHD and chemistry into their numerical simulations, while approximating the thermodynamic evolution via a piecewise adiabatic equation of state. More recently, \citet{chact09} followed the evolution of a $1~{\rm M}_\odot$ dense core in the presence of magnetic fields, including flux-limited diffusion radiative transfer but neglecting both nonideal-MHD effects and chemistry.

The present paper follows the detailed nonideal MHD and grain physics and chemistry of \citet{tm07a,tm07b,tm07c}, but also adds radiative transfer, which is essential for making accurate predictions that can be compared with observations pertaining to the opaque phase of star formation. In Section \ref{section:model} we summarise the formulation of the problem and method of solution (detailed in Paper I), and we provide the initial and boundary conditions of the numerical simulation. Section \ref{section:results} presents the results. Contact is made with both observations and prior theoretical work where appropriate. Section \ref{section:summary} summarises the results and predictions, and discusses their limitations.

\vspace{-4ex}
\section{Physical Model}\label{section:model}

A detailed discussion of the physical processes included in the calculations was presented in Paper I. In summary, we consider a two-dimensional, nonrotating, nonisothermal model molecular cloud, whose axis of symmetry is aligned with the $z$-axis of a cylindrical-polar coordinate system $(r,\phi,z)$. The cloud is initially threaded by a uniform magnetic field oriented along the symmetry axis. The abundances of all species (except the neutrals) are determined from an extensive chemical-equilibrium network that includes UV radiation, cosmic rays, radioactive decays, thermal ionisation, dissociative and radiative recombination, atomic and molecular ion charge transfer, electron and ion attachment onto grains, and charge transfer by grain--grain collisions. While the code described in Paper I can handle an MRN grain size distribution, we present here only the simpler case of a uniform grain size distribution. Results of numerical calculations that employ an MRN distribution will be presented as part of a parameter study in a future publication.

The model cloud is evolved by using the six-fluid RMHD equations (Equations 72a-g in Paper I) from a nonequilibrium, uniform `reference' state, characterised by a number density of neutrals $n_{\rm n,ref} = 300~{\rm cm}^{-3}$ (corresponding to a neutral mass density $\rho_{\rm n,ref} \simeq 1.17\times 10^{-21}~{\rm g~cm}^{-3}$), temperature $T_{\rm ref} = 10~{\rm K}$ (corresponding to an isothermal sound speed $C_{\rm ref} = 0.188~{\rm km~s}^{-1}$), and magnetic field strength $B_{\rm ref} = 15~\mu{\rm G}$ (corresponding to an Alfv\'{e}n speed $v_{\rm A}= 1.24~{\rm km~s}^{-1}$). The part of the cloud whose evolution is followed numerically has radius $R$ and half-thickness $Z$, both equal to $0.75~{\rm pc}$, implying a total mass of $45.5~{\rm M}_\odot$. The central mass-to-flux ratio of the reference state in units of the critical central value is then
\begin{eqnarray}\label{equation:muo}
\mu_{\rm ref} &\equiv & \frac{(dM/d\Phi_{\rm B})_{\rm c}}{(dM/d\Phi_{\rm B})_{\rm c,cr}} = \frac{2Z\rho_{\rm n,ref}/B_{\rm ref}}{(3/2)(63G)^{-1/2}} \nonumber\\*
&= & 0.49\left(\frac{n_{\rm n,ref}}{300~{\rm cm}^{-3}}\right)\left(\frac{Z}{0.75~{\rm pc}}\right)\left(\frac{15~\mu{\rm G}}{B_{\rm ref}}\right) ,
\end{eqnarray}
where $(dM/d\Phi_{\rm B})_{\rm c,cr} = (3/2)(63G)^{-1/2}$ is the {\em central} critical value for collapse originally determined by \citet{ms76}.

If the magnetic field were frozen in the matter, the `reference' state would relax along field lines and oscillate about an equilibrium state (denoted by a subscript `0') in which gravity is balanced by thermal-pressure forces along field lines and mainly by magnetic forces perpendicular to field lines. An important dimensionless parameter in this equilibrium state is the ratio of the magnetic and thermal pressures,
\begin{eqnarray}\label{equation:alpha}
\alpha_0 &\equiv &\frac{B^2_0}{8\pi\rho_{\rm n,0}C^2_0} \nonumber\\*
&= &6.48\left(\frac{B_0}{15~\mu{\rm G}}\right)^2\left(\frac{10^3~{\rm cm}^{-3}}{n_{\rm n,0}}\right)\left(\frac{10~{\rm K}}{T_0}\right) .
\end{eqnarray}
In equilibrium \citep[see][]{mouschovias91b}, these two dimensionless free parameters satisfy the relation: 
\begin{equation}\label{equation:mu0alpha0}
\alpha_0 \mu_0^2 \approx 0.71 .
\end{equation}
The physical meaning of this equation is that, since the same force (gravity) is balanced by the thermal-pressure force along field lines and by the magnetic force perpendicular to the field lines, the thermal-pressure and magnetic forces must also be comparable in magnitude. A point related to this equation cannot be overemphasised: for subcritical central flux tubes ($\mu_0 < 1$), the {\em local} magnetic pressure exceeds the thermal and gravitational pressures. This does {\em not} imply, however, that such flux tubes will expand. It simply means that the magnetic field in the core is comparable to that in the envelope, and is confined by the massive cloud envelope, not the gravity of the core. Virial-theorem based arguments miss this essential point, leading to the incorrect conclusion that magnetically subcritical fragments (or clouds) cannot exist. After all, \citet{mouschovias76a,mouschovias76b} calculated rigorously the equilibria of such objects, as did \citet{tin88a,tin88b,tin89}.

Since $\mu_{\rm ref} = \mu_0$ (i.e., contraction along field lines does not alter a fragment's mass-to-flux ratio), the central density in the equilibrium state will be a factor $\approx 1.4\alpha_{\rm ref}\mu^2_{\rm ref}\approx 7$ greater than that of the reference state. As in \citet{fm93}, we have found little quantitative difference (for typical parameters) between a run in which ambipolar diffusion operates from the outset and one in which the cloud is allowed to reach equilibrium under flux-freezing before ambipolar diffusion is switched on. The physical reason for this is that the relaxation time along field lines is short compared to the ambipolar diffusion timescale, particularly for the low densities characteristic of the relaxation phase ($n_{\rm n}\lesssim 10^3~{\rm cm}^{-3}$ --- see below), during which ionisation due to UV radiation is important.

Once the heat generated by released gravitational energy during core contraction is unable to escape freely (at a central number density $n_{\rm opq} \simeq 10^7~{\rm cm}^{-3}$), we use radiative transfer to determine the thermal evolution of the core and its effect on the dynamics. For this, we employ the grey flux-limited diffusion (FLD) approximation. Realistic temperature-dependent grain opacities are used that account for a variety of grain compositions (see Paper I, fig.~2).

Due to the assumed symmetry of the problem, we need only follow the evolution of the region bounded by $0\le r \le R$ and $0\le z \le Z$. Boundary conditions on the density and velocity are chosen so that no mass, magnetic flux, or thermal energy crosses any boundary, as employed and tested by \citet{fm92,fm93}. The magnetic field lines are constrained to cross the midplane ($z=0$) and upper boundary ($z=Z$) normally, and to have no radial component at the axis ($r=0$) and outer boundary ($r=R$). In order to prevent unrealistic densities near the upper boundary of the computational domain ($z=Z$), a floor is imposed on the neutral number density at $n_{\rm n,floor} = 1~{\rm cm}^{-3}$. In addition, a floor of $10~{\rm K}$ is imposed on the gas and radiation temperatures.

The number of computational cells is fixed ($80\times 80$), and their positions are spaced logarithmically, so that the spacing between the $i$th and $i{\rm th} + 1$ cells is a number greater than the spacing between the $i{\rm th}-1$ and $i$th cells. This ratio is kept spatially uniform (with the same value in both the $r$ and $z$ directions) but is allowed to vary in time. Its value is such that the innermost cell is constrained to always have a width $\lambda_{\rm T,cr}/5$ -- $\lambda_{\rm T,cr}/7$, where $\lambda_{\rm T,cr} \equiv 1.4 C \tau_{\rm ff}$ is the critical thermal lengthscale \citep{mouschovias91a} and $\tau_{\rm ff}\equiv(3\pi/32G\rho_{\rm n,c})^{1/2}$ is the central spherical free-fall time. (The quantity $\rho_{\rm n,c}$ is the neutral mass density at the cloud centre, and the local isothermal sound speed $C$ in this calculation is a function of time.) The critical thermal lengthscale is the smallest scale on which there can be spatial structure in the density without thermal-pressure forces smoothing it out. At the end of the simulation, the minimum cell size is $\approx 0.2~{\rm AU}\times 0.2~{\rm AU}$.

\section{Results}\label{section:results}

\subsection{Overall Evolution}\label{section:overall}

\begin{figure}
\center
\includegraphics[width=2.8in]{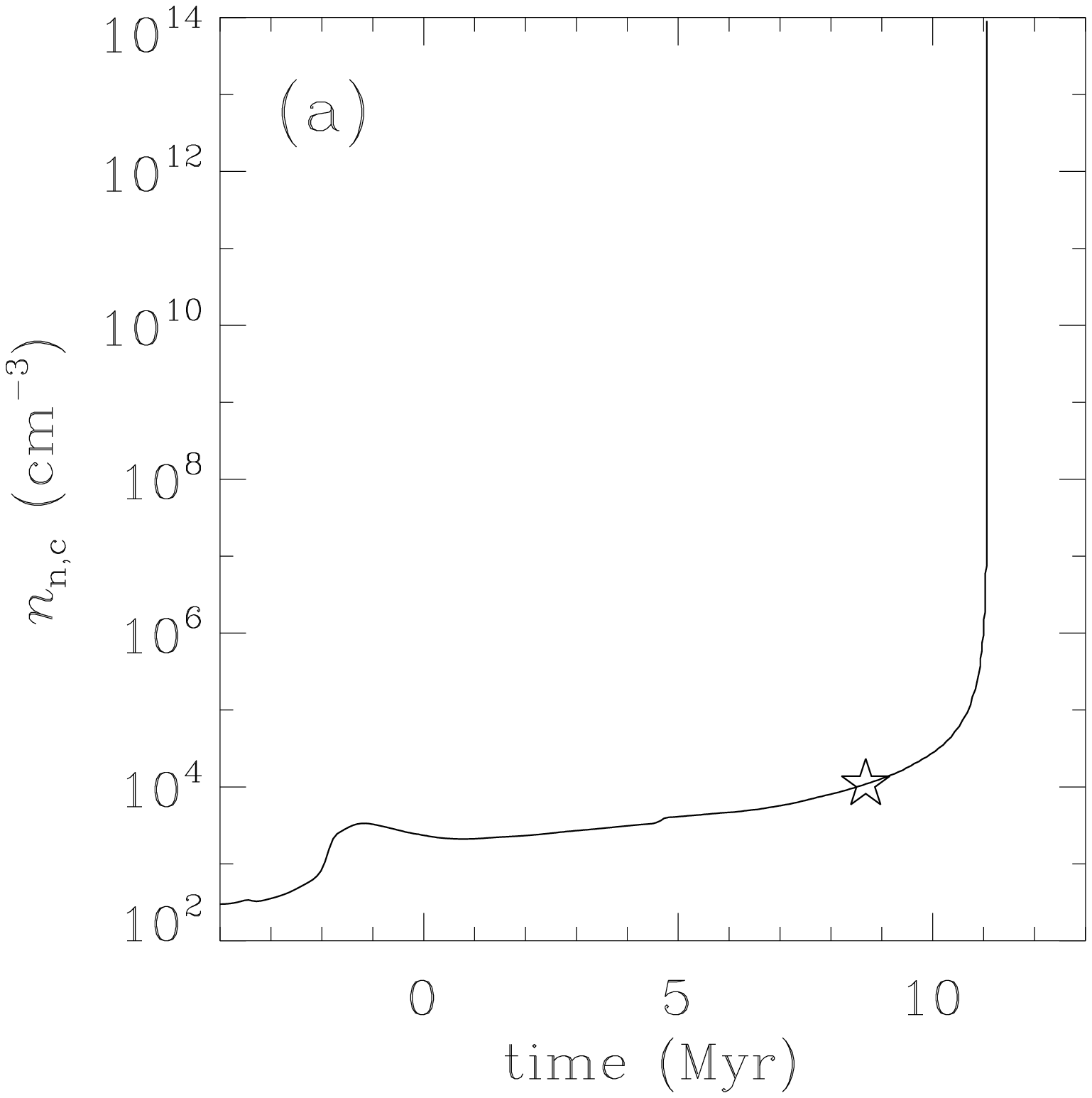}
\newline
\includegraphics[width=2.8in]{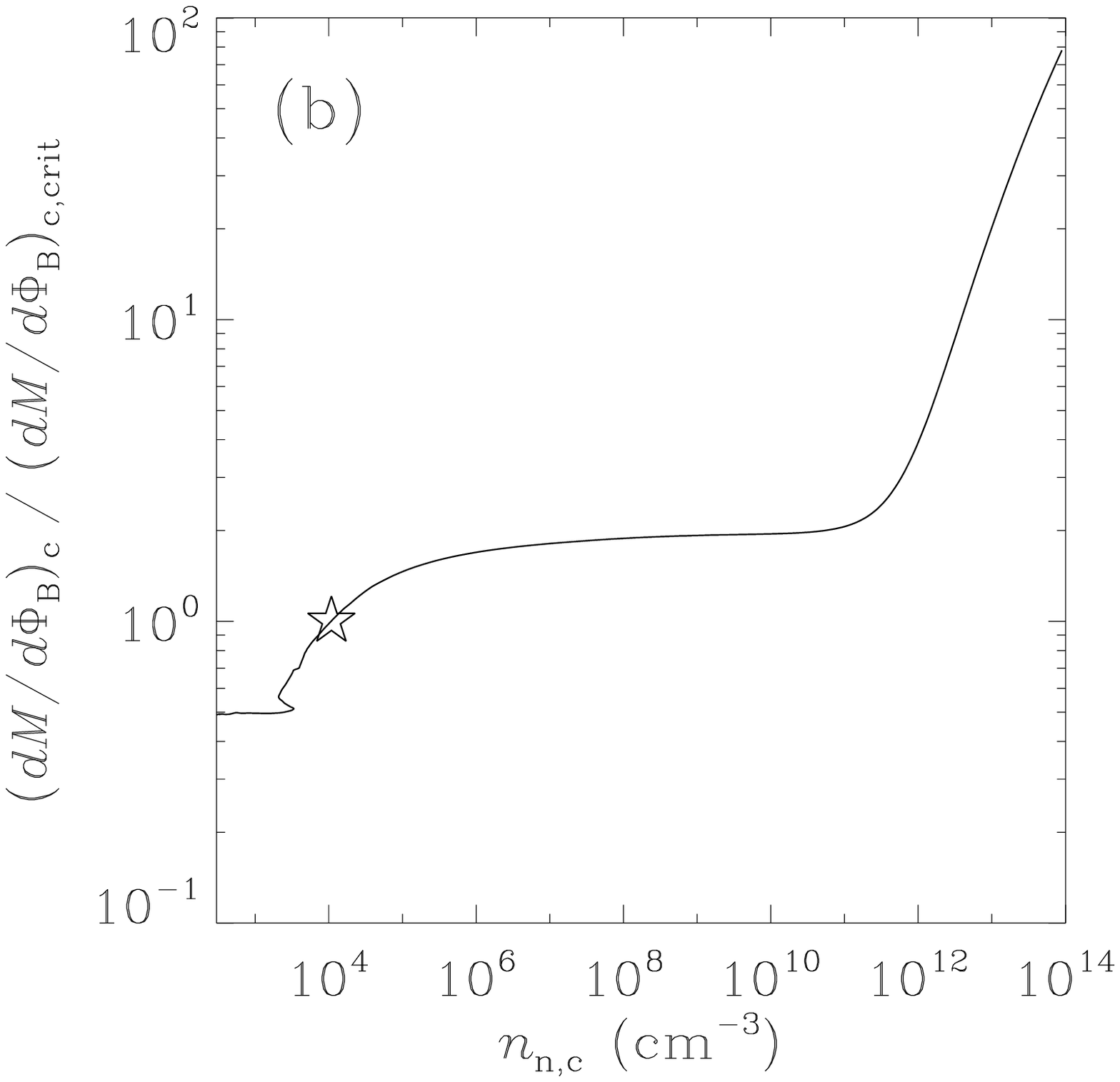}
\newline
\caption{(a) Central number density of neutrals, $n_{\rm n,c}$, as a function of time. (b) Central mass-to-flux ratio (normalised to the central critical value for collapse) as a function of $n_{\rm n,c}$. The `star' marks the time at which a supercritical core forms.}\label{figure:overall:density}
\end{figure}

In Fig.~\ref{figure:overall:density}, we show (a) the time evolution of the central number density of neutrals, $n_{\rm n,c}$, and (b) the central mass-to-flux ratio, $(dm/d\Phi_{\rm B})_{\rm c}$, (normalised to the central critical value for collapse) as a function of $n_{\rm n,c}$. (Note that our normalisation is greater than that of \citet{tm07a,tm07b,tm07c}, which referred to a uniform, thin disc, by a factor $\approx 2$.) There are three distinct phases of evolution: relaxation, quasistatic, and dynamic. First, the cloud relaxes along magnetic field lines from its uniform reference state to a quasi-equilibrium state whose central density $n_{\rm n,c}\simeq 2280~{\rm cm}^{-3}$. Ambipolar diffusion is negligible during this phase, since it is characterised by relatively low densities ($\lesssim 10^3~{\rm cm}^{-3}$) and therefore relatively large degrees of ionisation ($\sim 10^{-6}$ -- $10^{-5}$) due to UV radiation. Hence, the mass-to-flux ratio remains roughly constant. Then ambipolar diffusion sets in and the fragment contracts quasistatically (i.e., with negligible acceleration) under its own self-gravity until a supercritical core forms at $n_{\rm n,c} \simeq 1.1\times 10^4~{\rm cm}^{-3}$ and time $\approx 9~{\rm Myr}$ (marked in the figure by a `star'). The subsequent evolution is dynamic, although significantly slower than free fall. The central mass-to-flux ratio asymptotes to roughly twice its critical value until ambipolar diffusion is reawakened at a central density $n_{\rm n,c}\approx 10^{11}~{\rm cm}^{-3}$. The mass-to-flux ratio then increases dramatically, reaching $\simeq 80$ times the central critical value for collapse by the end of the calculation (at a density $\simeq 10^{14}~{\rm cm}^{-3}$). During the entire supercritical phase of evolution, the central density increases by ten orders of magnitude in $2.38~{\rm Myr}$.

Note that soon after the core becomes magnetically supercritical (at a central density $1.1\times 10^4~{\rm cm}^{-3}$), it will also become observable as a well-defined ammonia core. Therefore, Zeeman observations of ammonia cores are predicted to preferentially measure slightly supercritical mass-to-flux ratios. This is in excellent agreement with observations (e.g., by \citealt{crutcher99}). Also, the detectable (using ammonia as a tracer) lifetime of the core formed in this simulation is $\approx 1~{\rm Myr}$, which is also in excellent agreement with the $\sim 1~{\rm Myr}$ lifetimes of starless cores advocated by \citet{lm99} and \citet{jma99}. (See \citealt{tm04} for further discussion of this point and its importance in interpreting starless core statistics.)

\begin{figure*}
\center
\includegraphics[height=2.65in]{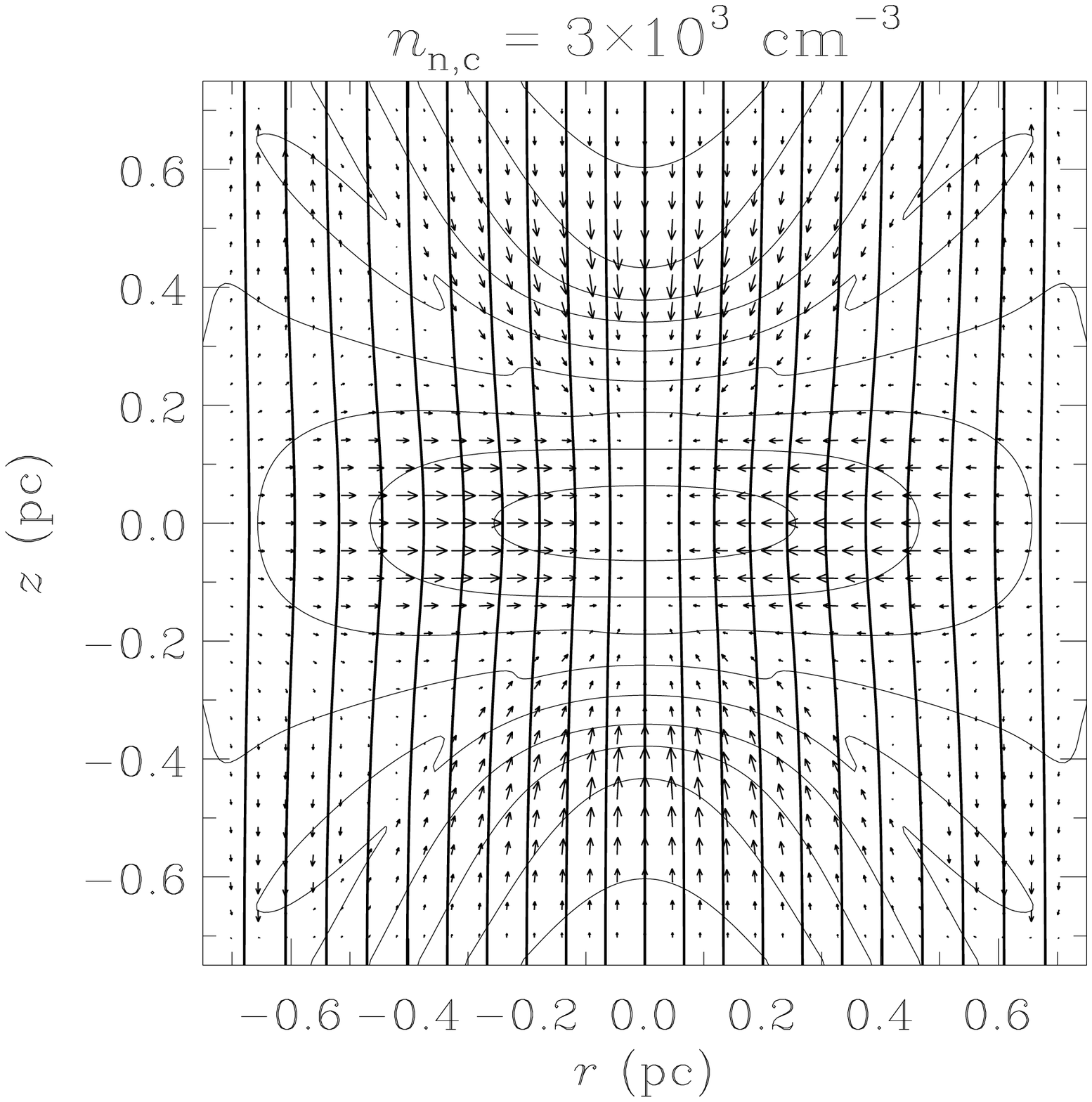}
\includegraphics[height=2.65in]{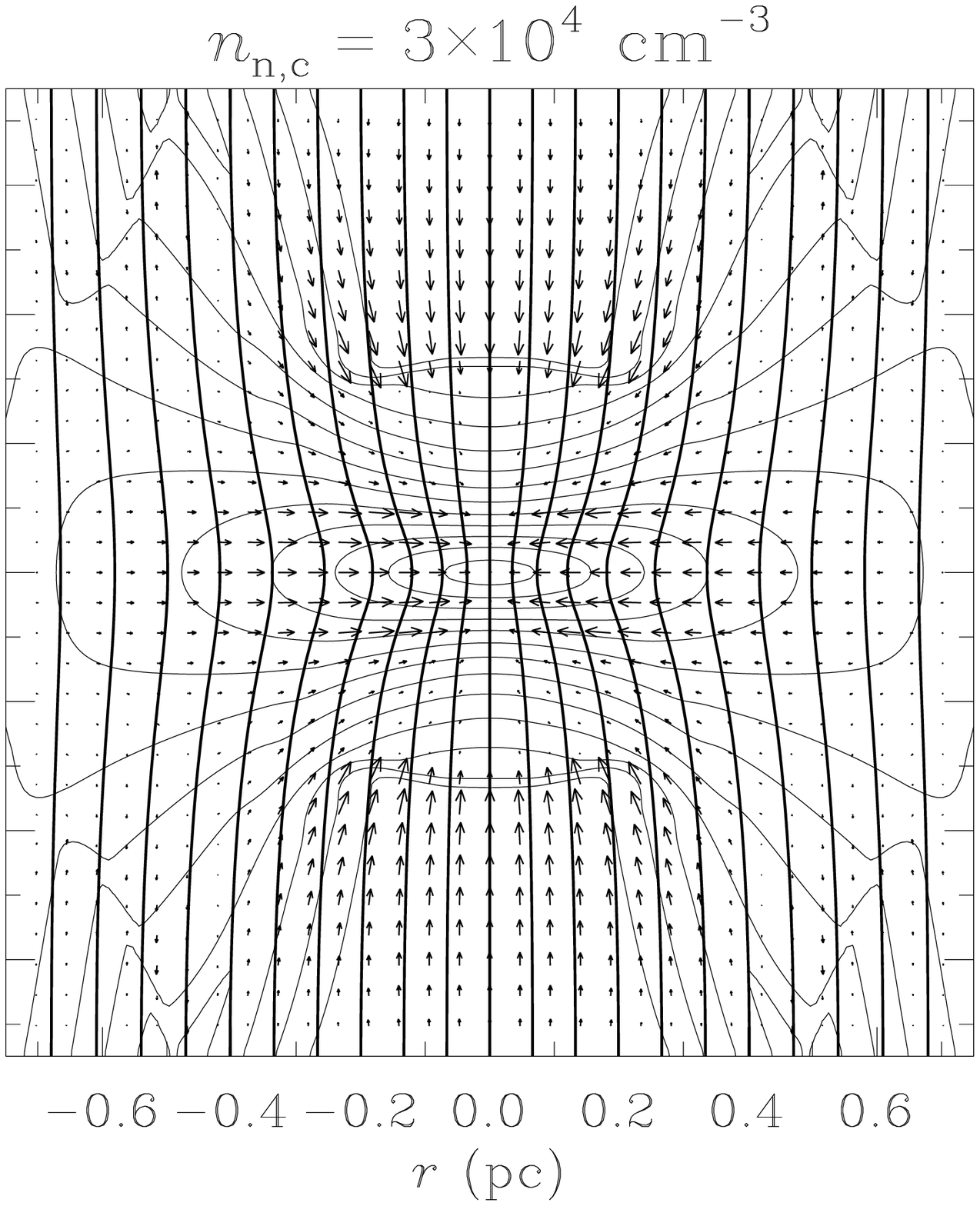}
\includegraphics[height=2.65in]{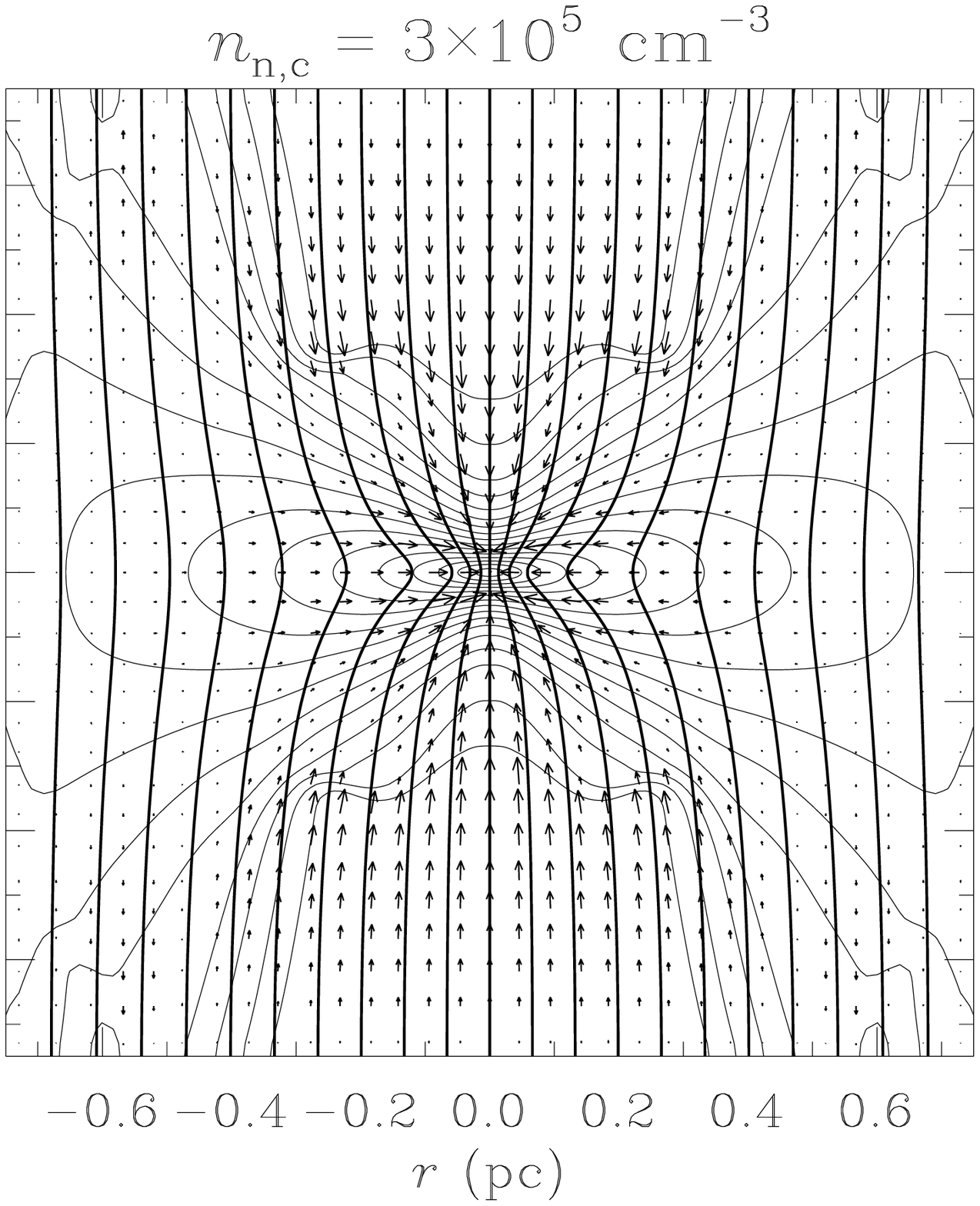}
\newline\center
\qquad\quad
\includegraphics[height=2.65in]{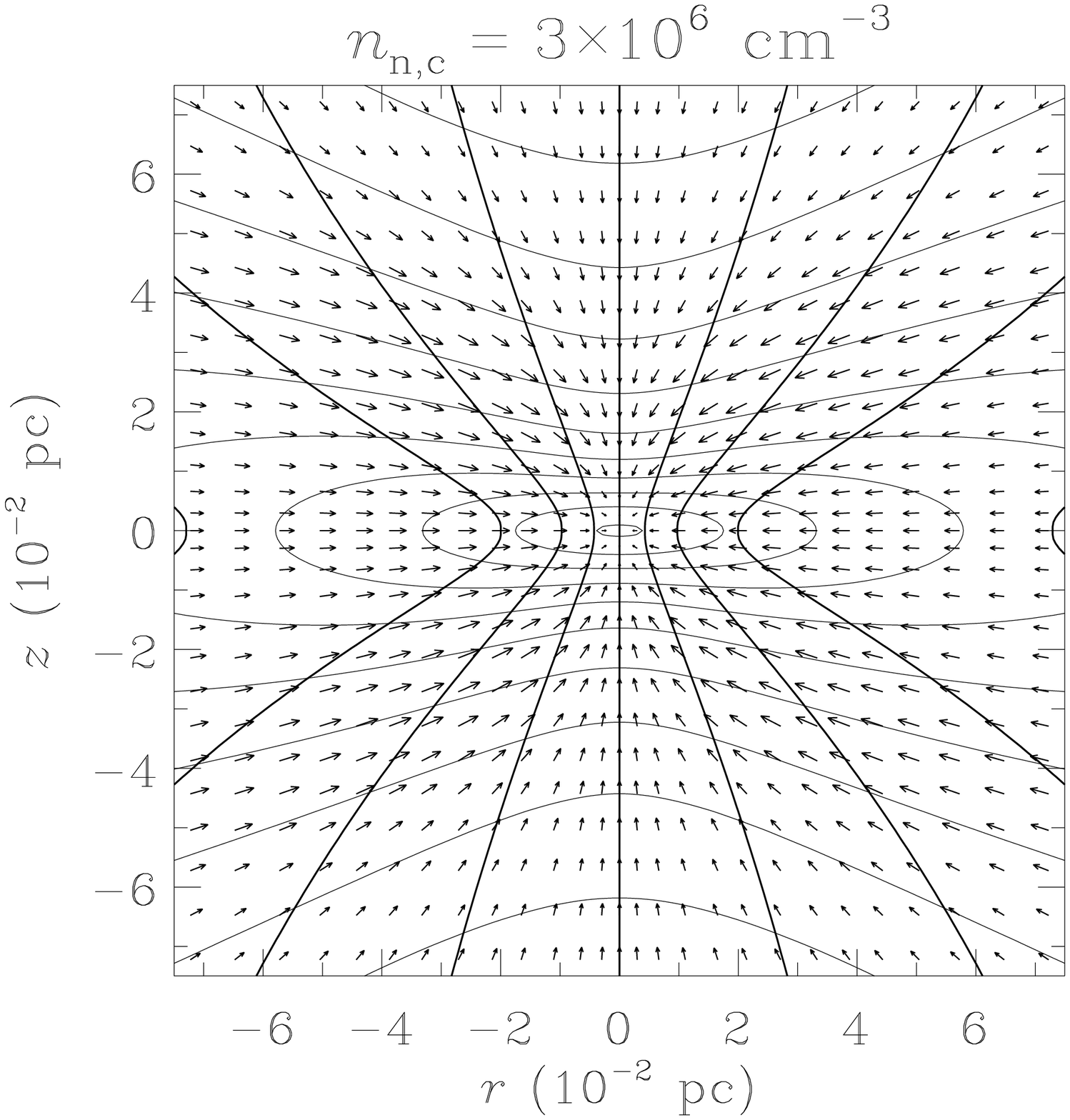}
\qquad\qquad
\includegraphics[height=2.65in]{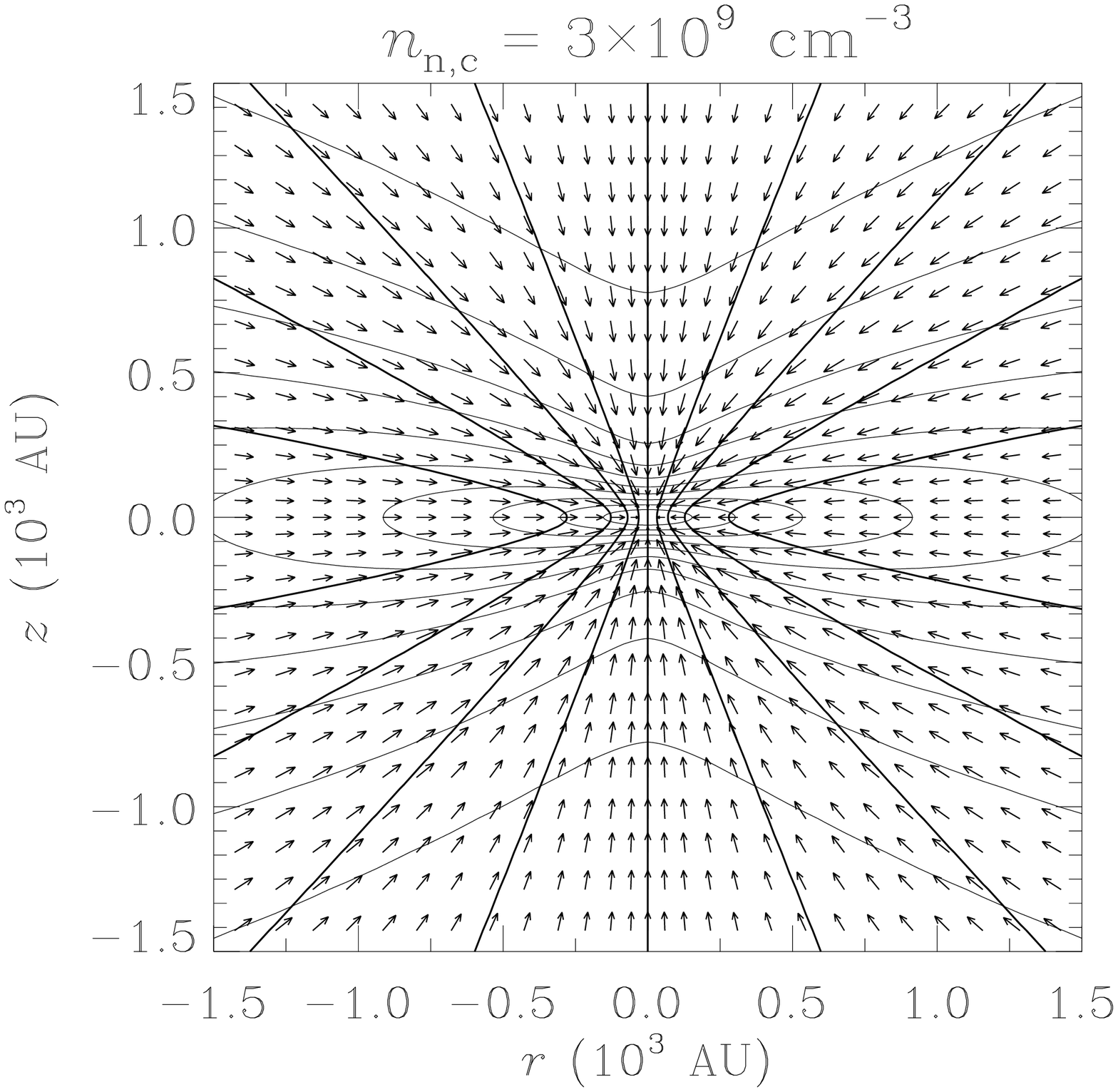}
\newline\center
\includegraphics[height=2.65in]{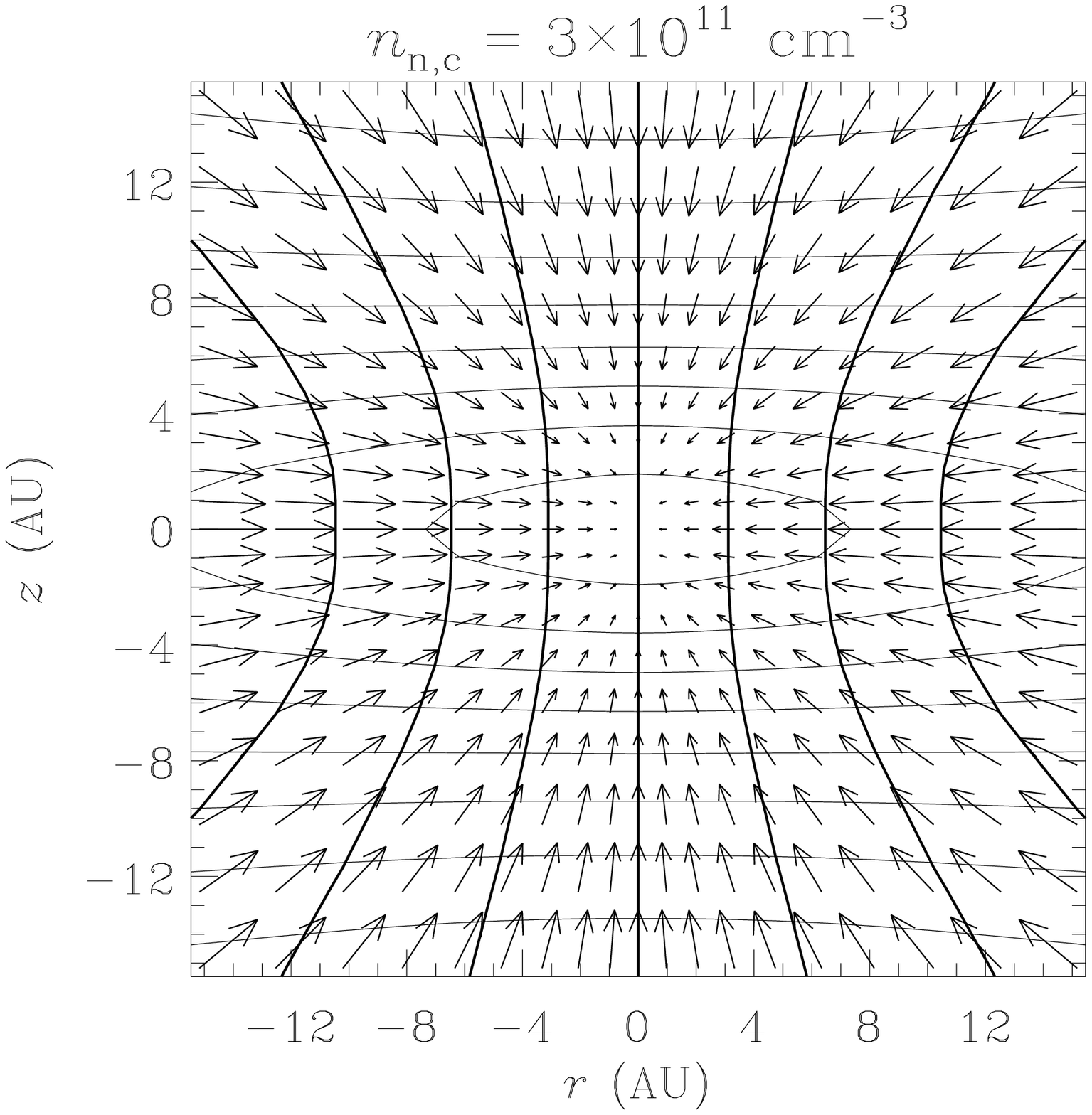}
\includegraphics[height=2.65in]{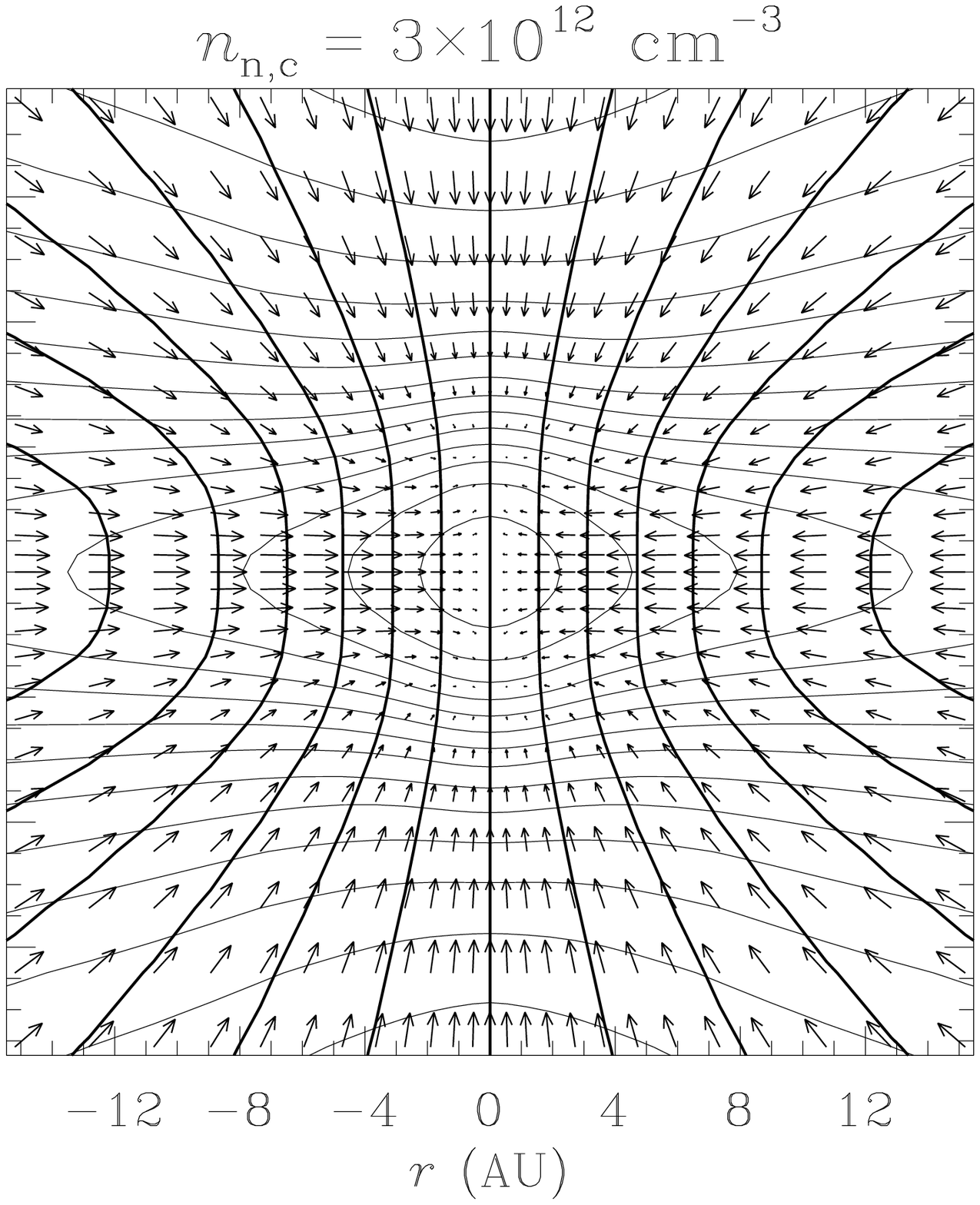}
\includegraphics[height=2.65in]{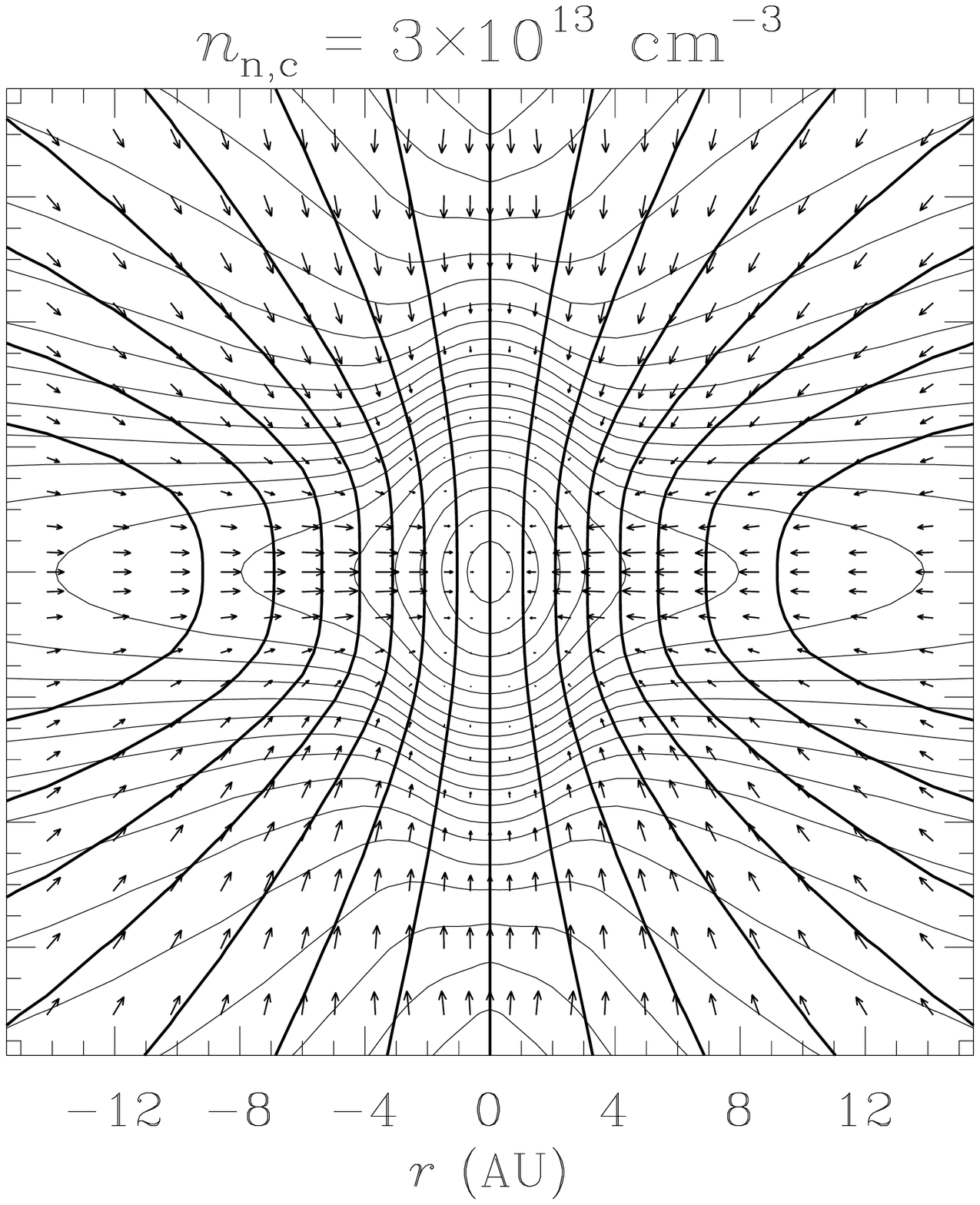}
\newline
\caption{Overall spatial and temporal evolution of the model molecular cloud. The density (thin solid lines), magnetic field (thick solid lines), and velocity field (arrows) are given in each of the eight frames, which show portions of the cloud at the times when the central density $n_{\rm n,c} = 3\times (10^3,~10^4,~10^5,~10^6,~10^9,~10^{11},~10^{12},~10^{13})~{\rm cm}^{-3}$ (left to right, top to bottom). The velocity vectors are normalised to the maximum velocity in each frame (see text). Every third isodensity contour denotes a change in density by a factor of $10$.}
\label{figure:overall:contour}
\end{figure*}

\begin{figure}
\center
\includegraphics[width=2.8in]{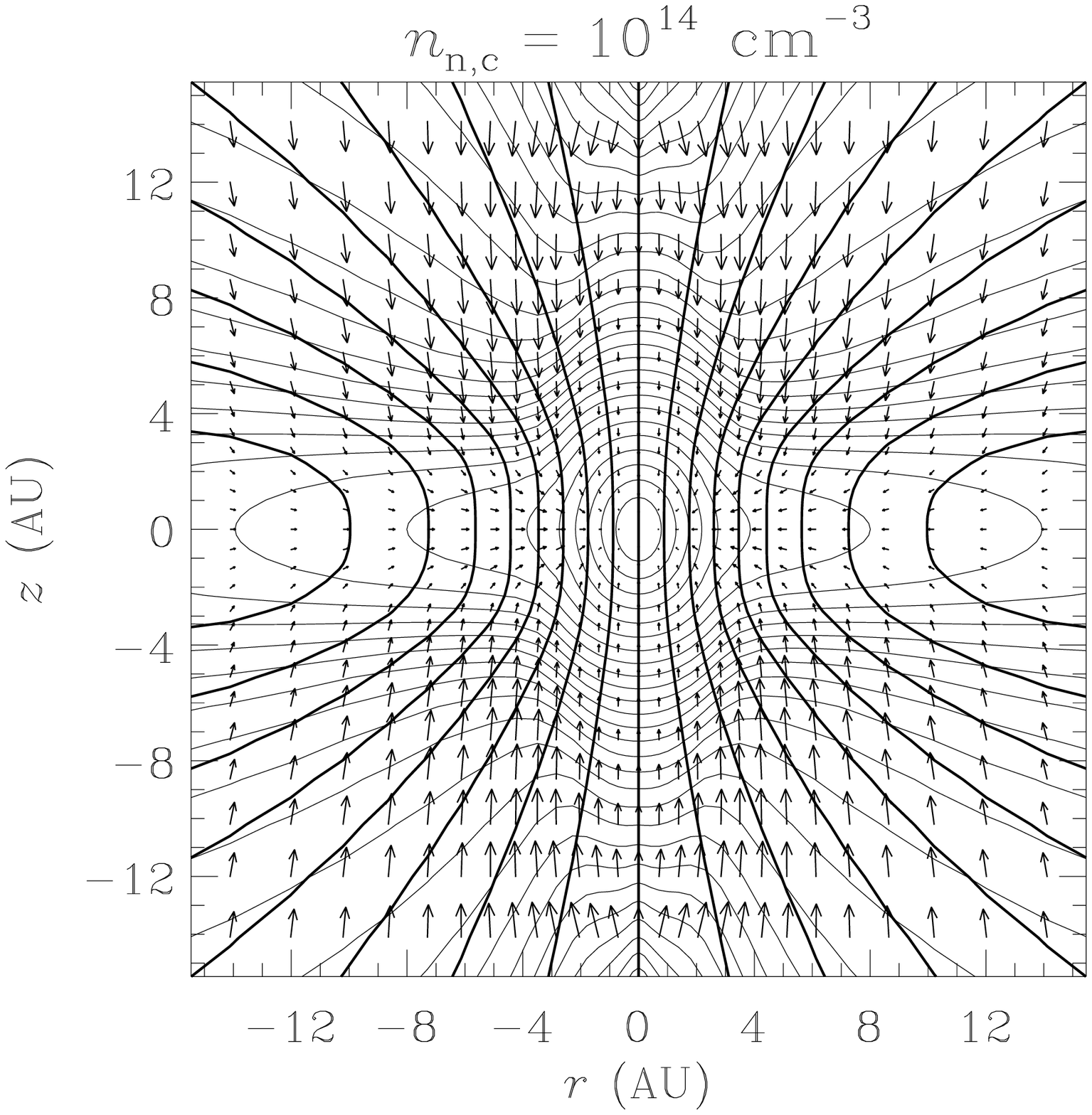}
\newline
\includegraphics[width=2.8in]{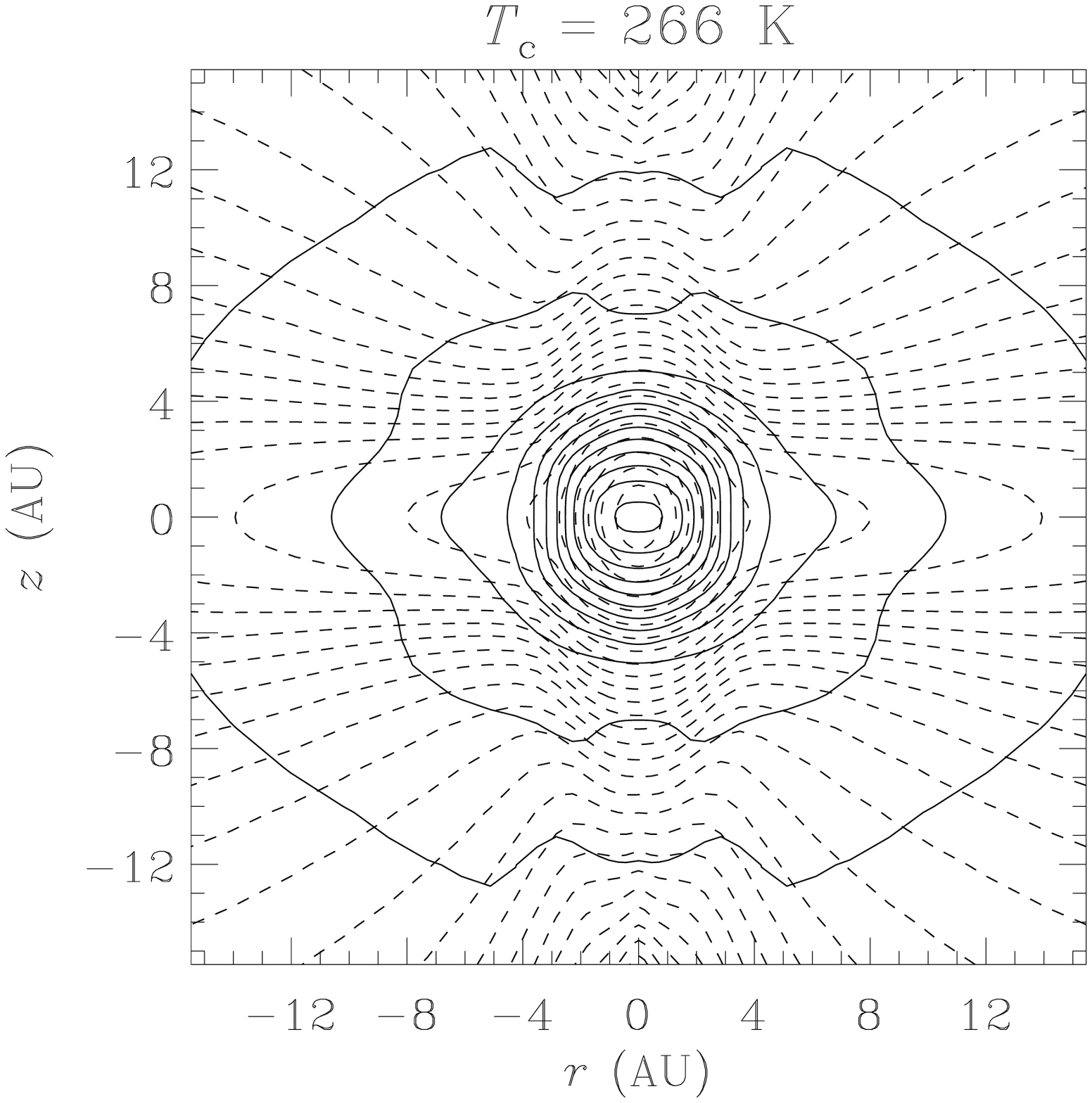}
\newline
\caption{Overall structure of the innermost 0.01\% of the model molecular cloud at the end of simulation ($n_{\rm n,c}=10^{14}~{\rm cm}^{-3}$). (a) Density (thin solid lines), magnetic field (thick solid lines), and velocity field (arrows). The velocity vectors are normalised to the maximum velocity in the frame ($1.44~{\rm km~s}^{-1}$); every third isodensity contour indicates a change in density by a factor of $10$. (b) Density (dashed lines) and temperature (solid lines). Every tenth isotherm denotes a change in temperature by a factor of $10$. The maximum temperature is $266~{\rm K}$.}
\label{figure:overall:final}
\end{figure}

The overall spatial and temporal evolution of the cloud is shown in Figures \ref{figure:overall:contour} and \ref{figure:overall:final}. In Fig.~\ref{figure:overall:contour}, the density (thin solid lines), magnetic field (thick solid lines), and velocity field (arrows) are displayed in each of the eight frames, which show portions of the cloud at different times, when the central density $n_{\rm n,c} = 3\times (10^3,~10^4,~10^5,~10^6,~10^9,~10^{11},~10^{12},~10^{13})~{\rm cm}^{-3}$ (left to right, top to bottom). Fig.~\ref{figure:overall:final} shows the innermost 0.01\% of the cloud at the end of the simulation (when $n_{\rm n,c}\simeq 10^{14}~{\rm cm}^{-3}$). In Figures \ref{figure:overall:contour} and \ref{figure:overall:final}a, the velocity vectors are normalised to the maximum velocity in each frame (given below) and every third isodensity contour denotes a change in density by a factor of $10$. In Fig.~\ref{figure:overall:final}b, the density (dashed lines) and temperature (solid lines) are displayed, with every tenth isotherm denoting a change in temperature by a factor of $10$.

In the first row of Fig. \ref{figure:overall:contour}, we show the entire computational domain ($0\le r \le R$, $0\le z \le Z$) at the three instances at which the central density is $n_{\rm n,c} = 3\times (10^3,~10^4,~10^5)~{\rm cm}^{-3}$. In the first frame, the cloud has settled into a quasi-equilibrium state and ambipolar diffusion has already set in. The magnetic field lines remain relatively undeformed, while the neutrals contract through them under their self-gravity -- their maximum radial velocity at this moment is $0.012~{\rm km~s}^{-1}$). For $r>0.5~{\rm pc}$ and $z>0.2~{\rm pc}$, there is a small outward motion as the cloud rebounds from overshooting its equilibrium state. In the second frame, the central magnetic flux tubes of the cloud have just become magnetically supercritical and begun to bend into an hourglass morphology. Such a field morphology has been observed via dust polarimetry by, e.g., \citet{schleuning98}, \citet{hddsv99}, \citet{gcr99}, \citet{svhdndd00}, \citet{lcgr02}, \citet{mw02}, \citet{hdhdvppb04}, \citet{cc06}, \citet{grm06}, \citet{vaillancourt08}, \citet{thkglr09}, and \citet{kirby09}. The maximum radial velocity has increased to $0.033~{\rm km~s}^{-1}$. The outermost envelope of the cloud remains magnetically supported, whence the radial velocities there are very small. A shock has formed at $z\simeq 0.3~{\rm pc}$ (maximum vertical velocity $0.39~{\rm km~s}^{-1}\approx 2 C$) because of the rapid contraction along field lines. In the third frame, the magnetically supercritical core is well into its dynamic (though slower than free-fall) stage of evolution and the maximum radial velocity has reached $\approx 0.1~{\rm km~s}^{-1}$.

By this point, the qualitative features of the outer envelope of the cloud, which is magnetically supported, have been decided and hardly change at subsequent times. The result is a highly nonhomologous collapse. We therefore focus on progressively smaller radii and track the local evolution of the magnetically supercritical core. In the first frame of the second row, we show the innermost 10\% of the computational region when $n_{\rm n,c} = 3\times 10^6~{\rm cm}^{-3}$. The maximum radial (vertical) velocity at this moment is $0.18~{\rm km~s}^{-1}$ ($0.47~{\rm km~s}^{-1}$). It is clear that the core remains disclike down to the smallest scales, even in the absence of rotation, a consequence of the presence of magnetic forces. In the second frame of this row, we show the innermost 1\% of the computational region when $n_{\rm n,c} = 3\times 10^9~{\rm cm}^{-3}$. The maximum radial (vertical) velocity at this moment is $0.38~{\rm km~s}^{-1}$ ($0.80~{\rm km~s}^{-1}$). The similarities between these two figures suggest that the supercritical phase of core contraction is nearly self-similar, a fact that has been exploited by, e.g., \citet{basu97}, \citet{cck98}, and \citet{kk02}, who derived semi-analytic self-similar solutions for this stage of contraction that reproduce the main qualitative features found in detailed simulations.

The evolution departs from being nearly self-similar at higher densities, once the gas begins to decouple from the magnetic field and a transition from disclike to spherical geometry takes place. For the third row of frames ($n_{\rm n,c}=3\times 10^{11},~10^{12},~10^{13}~{\rm cm}^{-3}$), we zoom in on the innermost 0.01\% of the computational region ($\simeq 15.5~{\rm AU}\times 15.5~{\rm AU}$) in order to highlight the formation of the hydrostatic core. In the first frame ($n_{\rm n,c}=3\times 10^{11}~{\rm cm}^{-3}$), the maximum radial (vertical) velocity is $0.52~{\rm km~s}^{-1}$ ($1.03~{\rm km~s}^{-1}$). This is the last frame shown in which the innermost $r\lesssim 4~{\rm AU}$ of the cloud exhibits a disclike geometry. In the second frame ($n_{\rm n,c} = 3\times 10^{12}~{\rm cm}^{-3}$), this region has begun to assume a spherical geometry due to the increasing relative importance of thermal pressure. The magnetic field has essentially decoupled from the matter and the field lines are straight inside the $\simeq 3\times 10^{11}~{\rm cm}^{-3}$ isodensity contour. The radial and vertical velocities attain their maxima ($0.64~{\rm km~s}^{-1}$ and $1.14~{\rm km~s}^{-1}$, respectively) just outside the boundary of the newly-formed hydrostatic core, where shocks occur in both radial and vertical directions. These shocks are a result of the rapid deceleration of matter as it slams into the hydrostatic core boundary. In the third frame ($n_{\rm n,c} = 3\times 10^{13}~{\rm cm}^{-3}$), the hydrostatic core is clearly visible and exhibits a spherical geometry inside $\simeq 2~{\rm AU}$. The magnetic field lines continue to remain straight within the $\simeq 3\times 10^{11}~{\rm cm}^{-3}$ isodensity contour. The maximum radial (vertical) velocity in the frame is $1.13~{\rm km~s}^{-1}$ ($1.27~{\rm km~s}^{-1}$). A closer inspection reveals a small outward motion with speed of $\sim 0.1~{\rm km~s}^{-1}$ along the vertical symmetry axis, a result of infalling gas rebounding after overshooting hydrostatic quasi-equilibrium. Throughout the evolution, the magnitude of the radial component of the magnetic field never becomes greater than the magnitude of the vertical component. Magnetic pinching forces remain small and magnetic reconnection does not occur. 

Fig.~\ref{figure:overall:final}a shows the density, magnetic field, and velocity field in the innermost 0.01\% of the computational region at the end of the simulation (when $n_{\rm n,c}\simeq 10^{14}~{\rm cm}^{-3}$). The maximum radial (vertical) velocity in the frame is $1.44~{\rm km~s}^{-1}$ ($1.24~{\rm km~s}^{-1}$). In Fig.~\ref{figure:overall:final}b, we highlight the temperature structure by overplotting isodensity contours (dashed lines) and isotherms (solid lines). The central temperature at this point is $266~{\rm K}$. Isotherms and isodensity contours coincide only within a region $\mathcalligra{r}\equiv (r^2+z^2)^{1/2}\lesssim 2~{\rm AU}$. Outside this region, the temperature structure is appreciably more spherical than the density structure. 

\subsection{Evolution of Central Quantities}\label{section:central}

\begin{figure*}
\center
\includegraphics[width=2.8in]{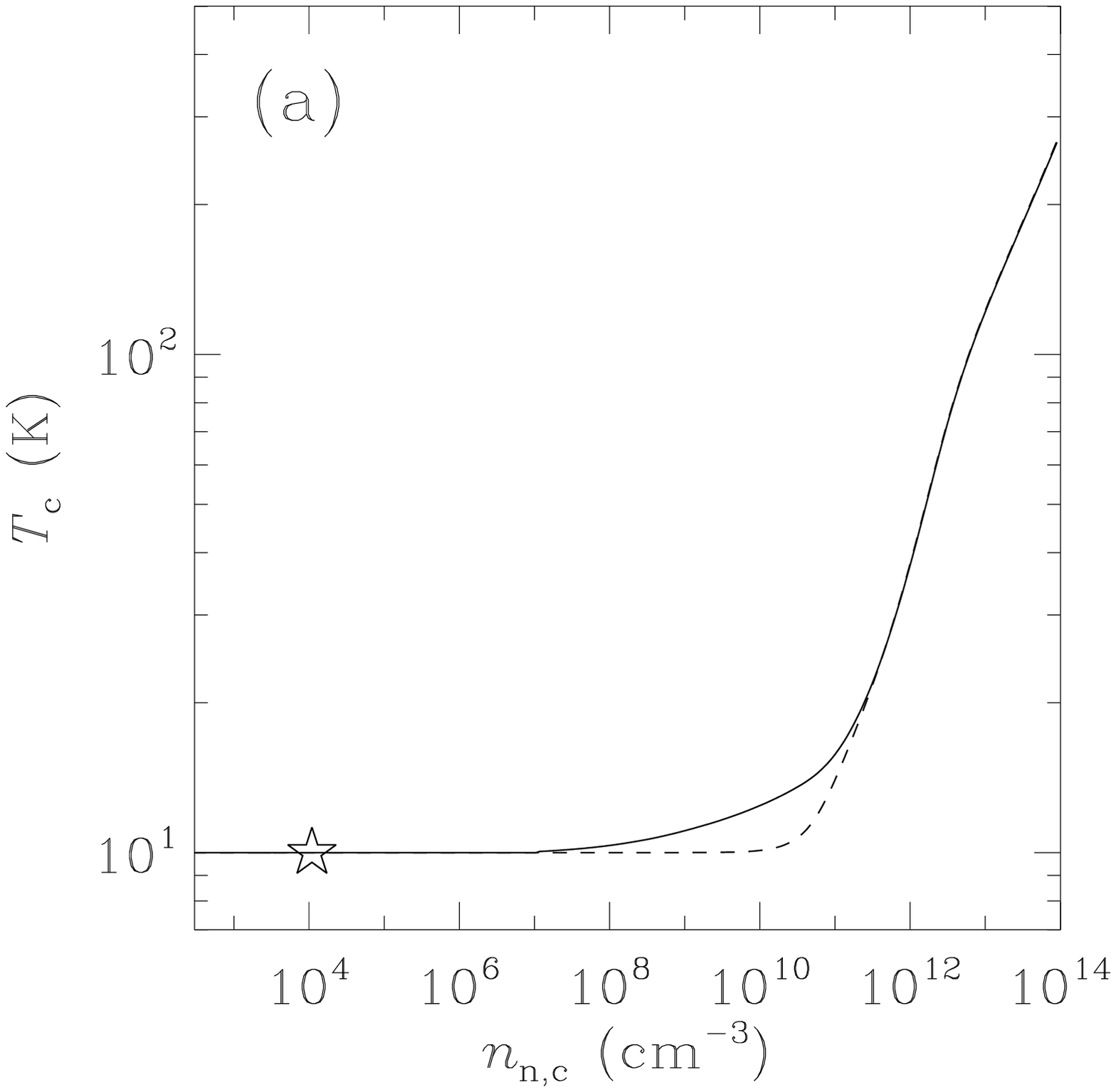}
\includegraphics[width=2.8in]{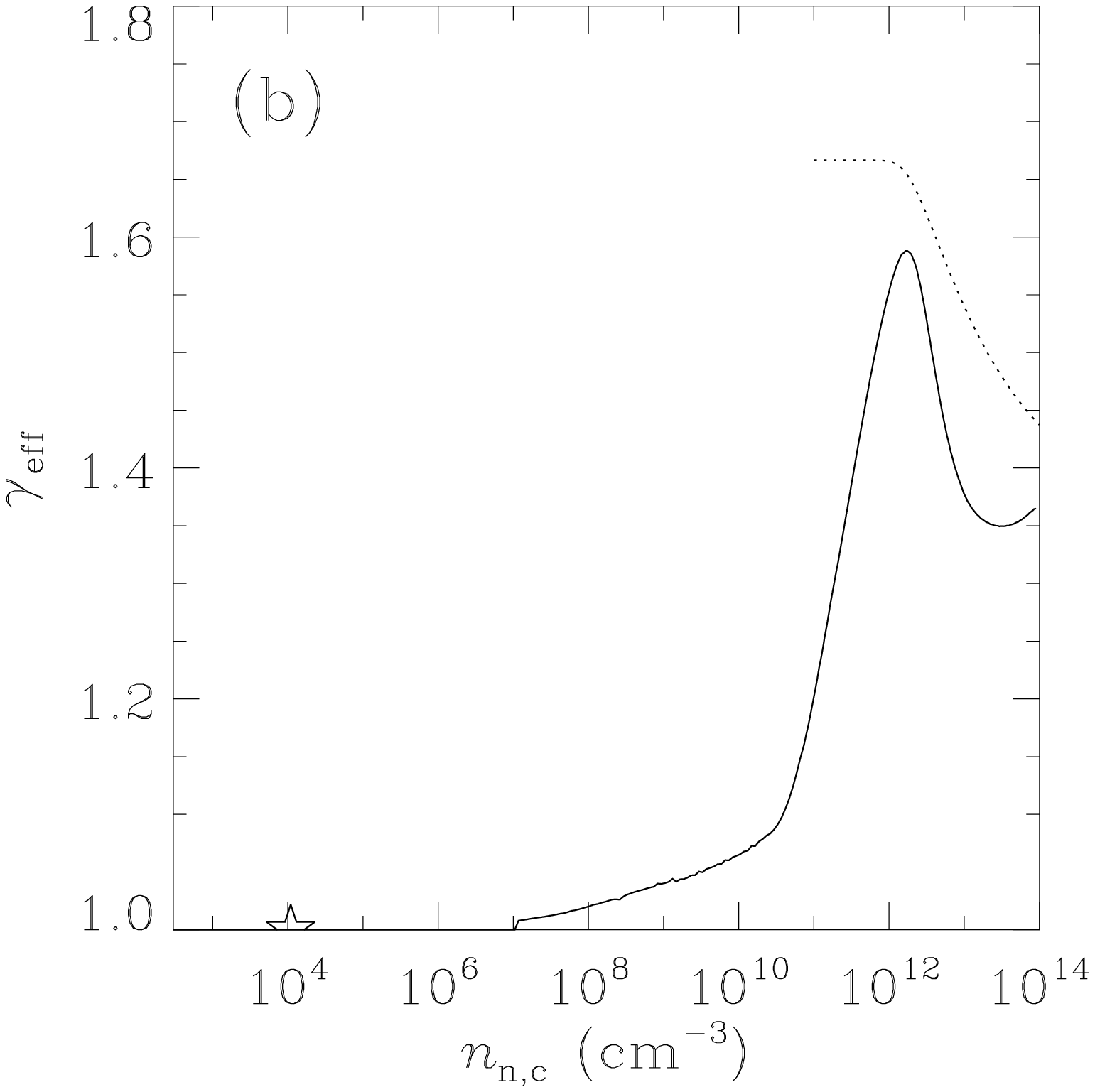}
\newline
\includegraphics[width=2.8in]{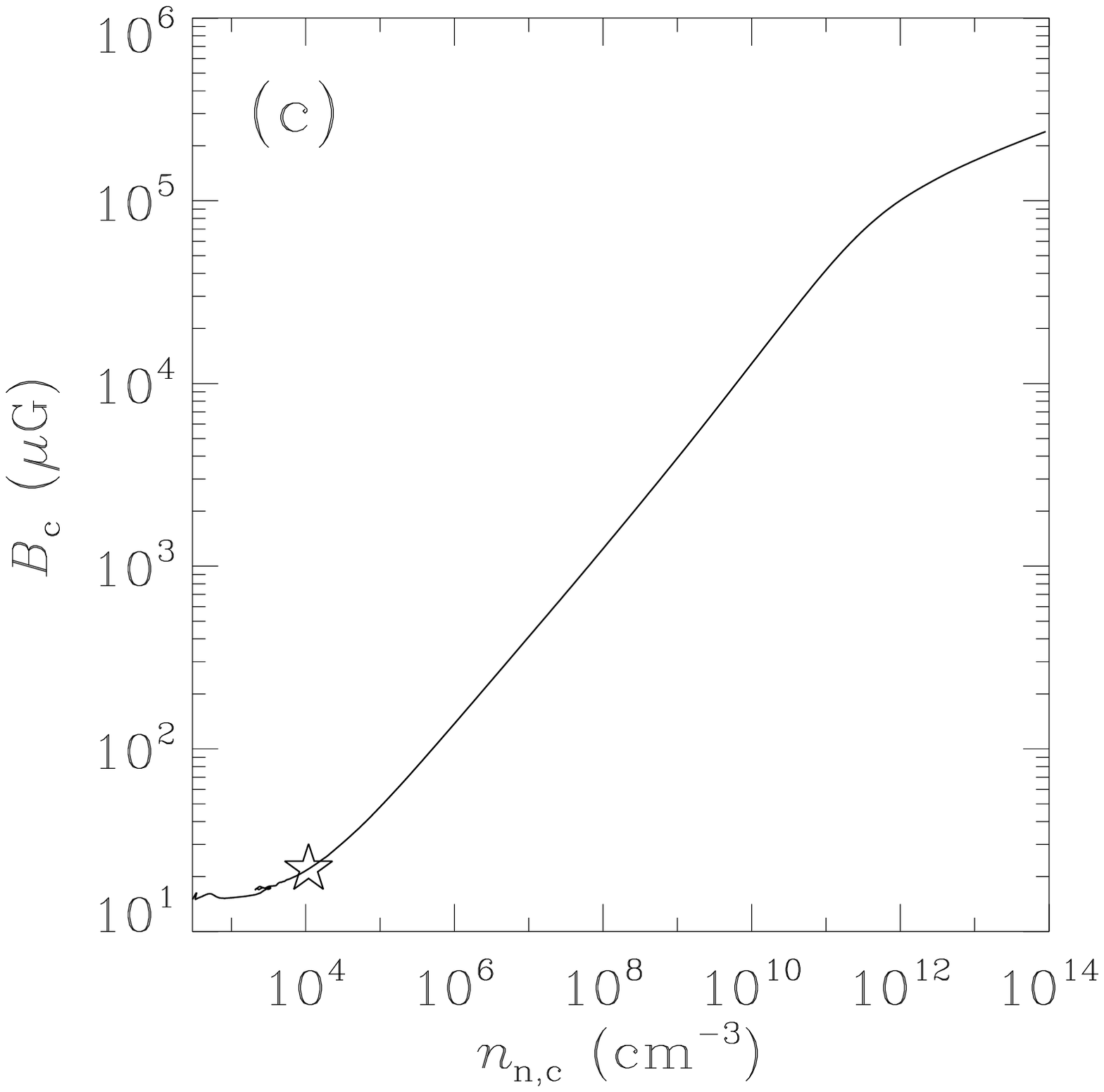}
\includegraphics[width=2.8in]{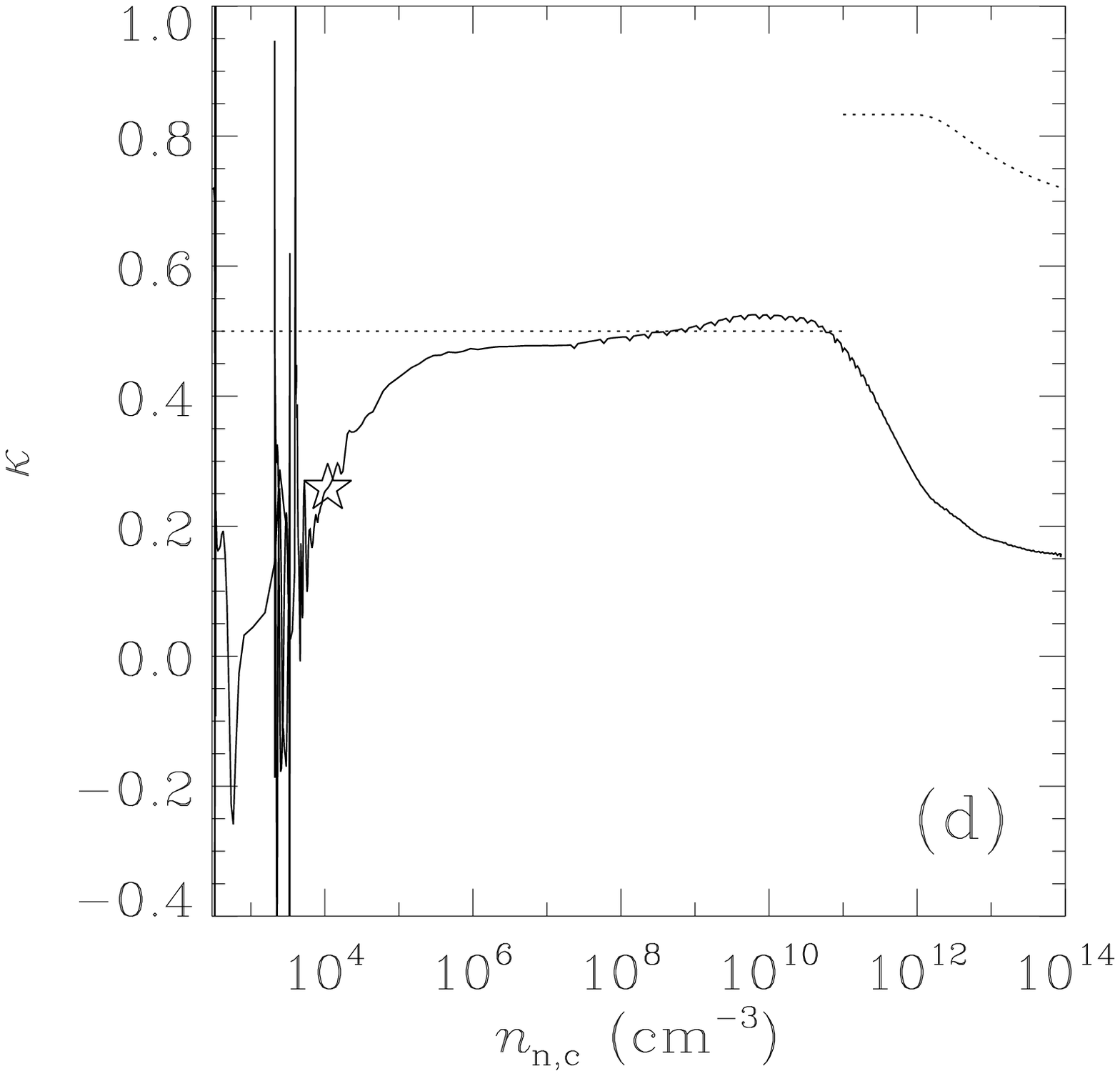}
\newline
\caption{Evolution of central quantities as functions of the central number density of neutrals, $n_{\rm n,c}$: (a) Gas (solid line) and radiation (dashed line) temperatures. (b) Exponent $\gamma_{\rm eff}$ in the relation $T_{\rm c}\propto n^{\gamma_{\rm eff}-1}_{\rm n,c}$ (solid line). The dotted line shows the exponent if the gas were to evolve adiabatically beyond a central density $10^{11}~{\rm cm}^{-3}$. (c) Magnetic field strength. (d) Exponent $\kappa$ in the relation $B_{\rm c}\propto n^\kappa_{\rm n,c}$ (solid line). The dotted line shows the exponent if the magnetic field were to evolve under strict flux-freezing and the gas were to behave adiabatically beyond a central density $10^{11}~{\rm cm}^{-3}$. The `star' marks the time at which a supercritical core forms.}
\label{figure:central:fields}
\end{figure*}

Fig.~\ref{figure:central:fields}a exhibits the central gas (solid line) and radiation (dashed line) temperatures, $T_{\rm c}$, as a function of the central number density of neutrals. The gas remains isothermal at $10~{\rm K}$ until a density $n_{\rm opq} \simeq 10^7~{\rm cm}^{-3}$, after which the heat released due gravitational contraction is unable to escape freely. The central gas temperature then rises with density at a rate $\partial\ln T_{\rm c}/\partial\ln n_{\rm n,c} \equiv \gamma_{\rm eff} - 1$, where $\gamma_{\rm eff}$ is the effective adiabatic index denoted by the solid line in Fig.~\ref{figure:central:fields}b. The dotted line in Fig.~\ref{figure:central:fields}b shows this index if the gas were to evolve adiabatically beyond a central density $10^{11}~{\rm cm}^{-3}$. {\em The gas never evolves adiabatically during the simulation}. There is some indication that adiabaticity will set in at the cloud centre by a central density $\sim 10^{15}~{\rm cm}^{-3}$, after which $\gamma_{\rm eff}\simeq 7/5$. 

A central temperature of $100~{\rm K}$ is reached when the central density is $\simeq 6\times 10^{12}~{\rm cm}^{-3}$. This is a significantly higher density than that found by, e.g., \citet{larson69}. There are two principal reasons for this difference. First, a disclike, rather than spherical, geometry allows the radiation to escape more easily, since the radiation is not isotropically confined by a spherical opacity distribution. Second, the early redistribution of mass by ambipolar diffusion is responsible for lowering the dust-to-gas ratio in the central flux tubes of the cloud from its usual interstellar value \citep[; also discussion below]{cm94}. The fewer dust grains there result in a lowered opacity, and consequently allow radiation to escape more easily. This effect is complicated somewhat, however, as our opacities are temperature-dependent (see fig.~2 of Paper I), whereas \citet{larson69} assumed a constant value $\chi_{\rm R}=15~{\rm cm}^2~{\rm g}^{-1}$ of dust (with a constant dust-to-gas ratio of $0.01$).

Fig.~\ref{figure:central:fields}c exhibits the central magnetic field strength, $B_{\rm c}$, as a function of central density. During the ambipolar-diffusion--controlled evolution (until the formation of a supercritical core, marked by the `star'), $B_{\rm c}$ increases by only $\simeq 47\%$, from $15~\mu{\rm G}$ to $22~\mu{\rm G}$. A phase of near flux-freezing follows, during which $B_{\rm c}$ resembles a power-law $B_{\rm c}\propto n_{\rm n,c}^\kappa$ with $\kappa\simeq 0.47$ (see Fig.~\ref{figure:central:fields}d), as found by \citet{fm93}. This is in excellent agreement with the value $\kappa=0.47\pm 0.08$ inferred from Zeeman observations of protostellar cores \citep{crutcher99}. For a thin disc in the flux-freezing limit, $B_{\rm c}\propto n_{\rm n,c}^{1/2} T_{\rm c}^{1/2}$, so that $\kappa=\gamma_{\rm eff}/2$ \citep{mouschovias76a,mouschovias76b,mouschovias91b}. Therefore, any nonisothermal evolution (i.e., $\gamma_{\rm eff}>1$) will result in $\kappa>1/2$ (in the flux-freezing limit). This is evident briefly in Fig.~\ref{figure:central:fields}d for the central density range $n_{\rm n,c}\sim 10^9$ -- $10^{11}~{\rm cm}^{-3}$. This trend is halted once the magnetic field begins to decouple from the matter and $\kappa$ decreases. 

Note that the rapid, large oscillations in $\kappa$ seen in fig.~1f of \citet{tm07b} after magnetic decoupling sets in are absent in our Fig.~\ref{figure:central:fields}d. Those oscillations were due to the abrupt and (relatively) rapid adiabatic temperature increase in the core, which forced the contracting core to overshoot and subsequently oscillate about hydrostatic equilibrium. In the situation presented here, the much more gradual increase in the central temperature affords the core enough time to maintain hydrostatic force balance as it contracts. A close inspection of Fig.~\ref{figure:central:fields}d does indeed reveal small physical oscillations about a hydrostatic equilibrium.

\begin{figure}
\center
\includegraphics[width=2.8in]{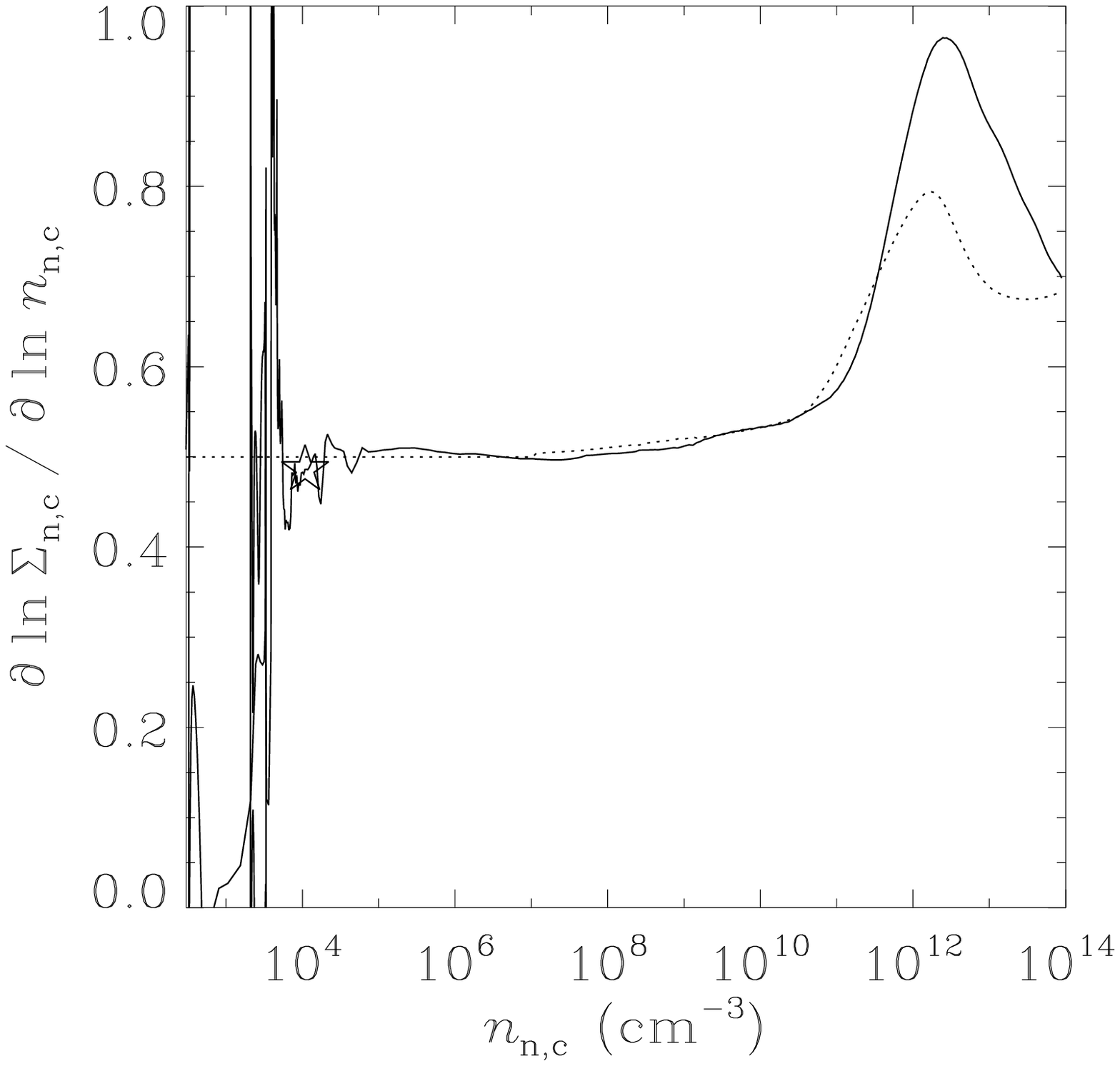}
\caption{Evolution of the exponent $\partial\ln\Sigma_{\rm n,c}/\partial\ln n_{\rm n,c}$ (solid line) as a function of the central number density of neutrals, $n_{\rm n,c}$. The dotted line shows the exponent if the geometry were to remain disclike throughout the evolution. The `star' marks the time at which a supercritical core forms.}
\label{figure:central:disc2sphere}
\end{figure}

The evolution of the central (gas) temperature assists in understanding the evolution of the quantity $\partial\ln\Sigma_{\rm n,c}/\partial\ln n_{\rm n,c}$ (shown in Fig.~\ref{figure:central:disc2sphere}), which is a quantitative indicator of the core geometry. For a disc, in which thermal-pressure forces balance gravity along magnetic field lines, or a sphere, in which thermal-pressure forces balance (spherical) gravity radially, $\partial\ln\Sigma_{\rm n,c}/\partial\ln n_{\rm n,c} = \gamma_{\rm eff}/2$ (whose value is represented by the dotted line in the figure), whereas for a sphere of constant mass, $\partial\ln\Sigma_{\rm n,c}/\partial\ln n_{\rm n,c}=2/3$. After the rapid contraction along field lines, $\partial\ln\Sigma_{\rm n,c}/\partial\ln n_{\rm n,c}$ approaches and oscillates about $1/2$; the cloud has settled into a disclike quasi-equilibrium. Even during the magnetically supercritical phase of dynamical contraction, the geometry remains disclike (as indicated by the near coincidence of the solid and dotted curves in the figure). This trend holds until central densities $\approx 3\times 10^{11}~{\rm cm}^{-3}$ are attained, after which departures from disc geometry become pronounced and, by the end of the run, the formation of an almost spherical core is evident.

\begin{figure}
\center
\includegraphics[width=2.8in]{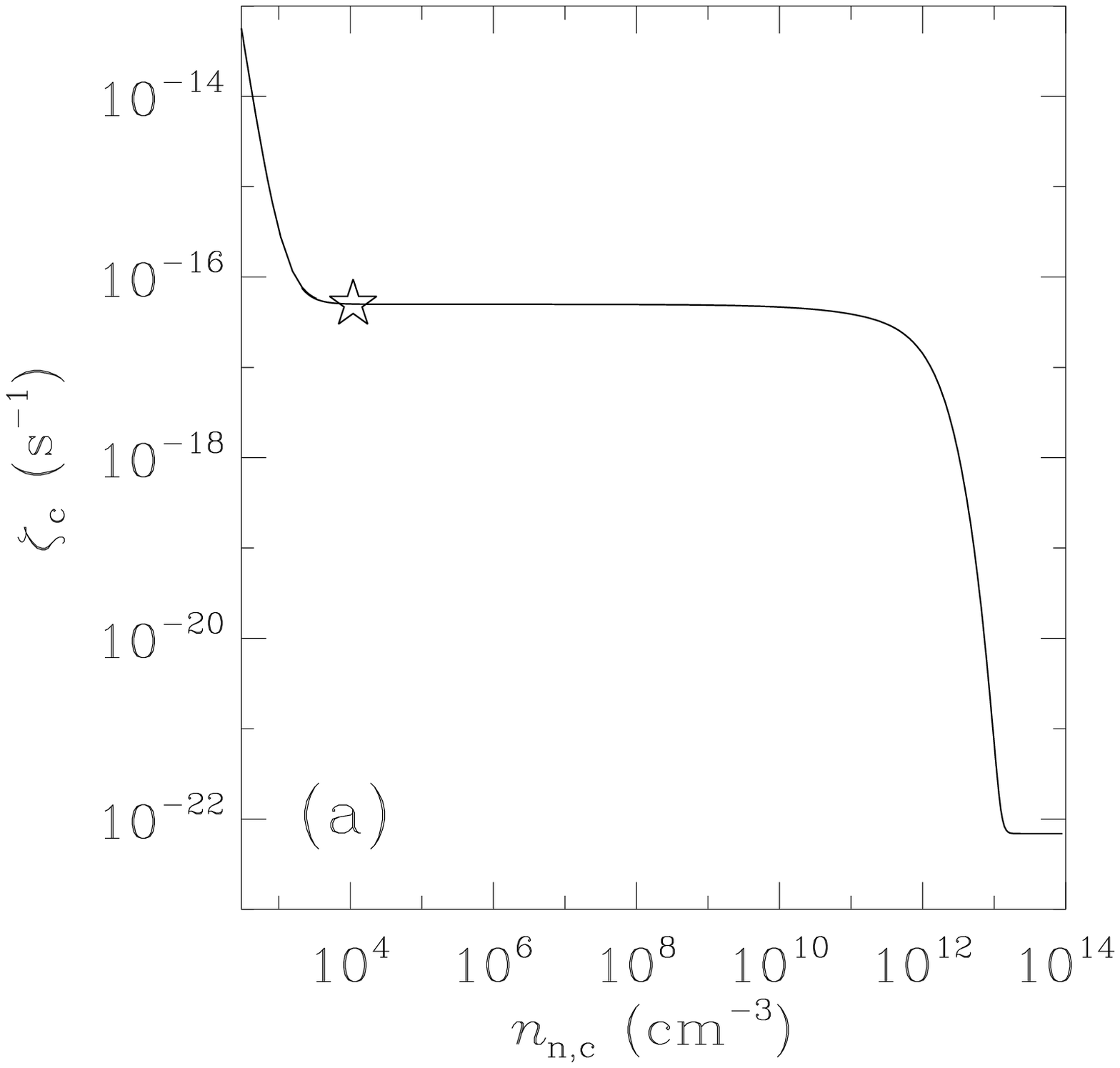}
\newline
\includegraphics[width=2.8in]{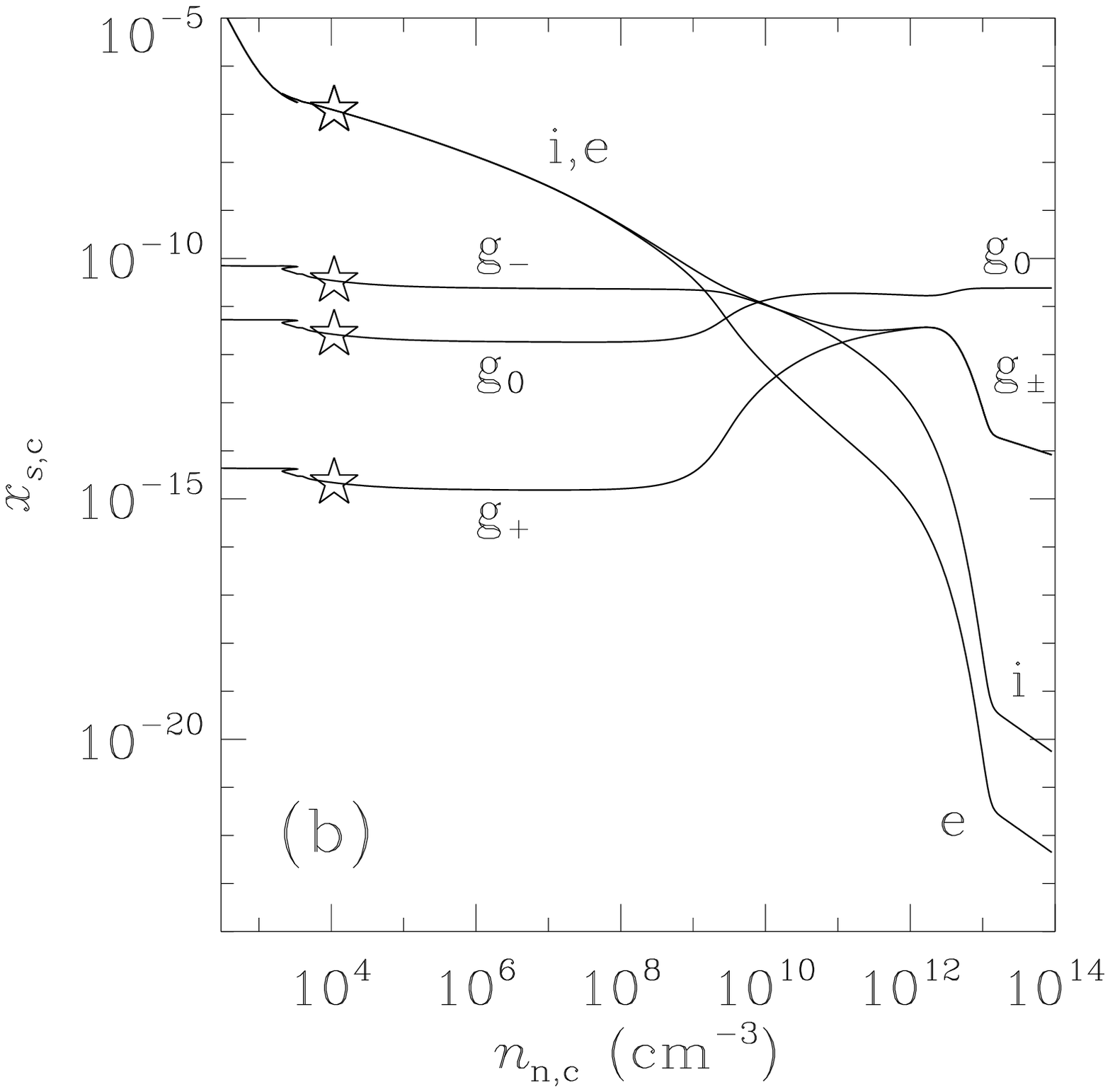}
\newline
\caption{Evolution of central quantities as functions of the central number density of neutrals, $n_{\rm n,c}$: (a) Ionisation rate due to all processes (UV radiation, cosmic rays, and radioactive decays). (b) Abundances of species. The `star' marks the time at which a supercritical core forms.}
\label{figure:central:chem}
\end{figure}

Fig.~\ref{figure:central:chem} shows the evolution of quantities related to the chemistry as functions of central density: (a) the ionisation rate at the centre of the cloud, $\zeta_{\rm c}$, due to all processes (UV radiation, cosmic rays, and radioactive decays); and (b) the central abundances of different species, $x_{s,{\rm c}}\equiv n_{s,{\rm c}}/n_{\rm n,c}$ (where $s={\rm e}$, i, g$_-$, g$_0$, g$_+$). For $n_{\rm n,c}\lesssim 1800~{\rm cm}^{-3}$, UV radiation dominates the ionisation rate (see Paper I, \S~4.1). Beyond this density, the ionisation in the core is mainly due to cosmic rays. Its value is constant, $\zeta = 5\times 10^{-17}~{\rm s}^{-1}$, for densities up to $\approx 3\times 10^{11}~{\rm cm}^{-3}$, beyond which the column density exceeds $\approx 100~{\rm g~cm}^{-2}$ and cosmic rays are appreciably attenuated. The ionisation rate then decreases monotonically, until it is dominated by radioactive decays (mainly of $^{40}$K) above a central density $\approx 10^{13}~{\rm cm}^{-3}$ and reaches a constant value $\zeta = 6.9\times 10^{-23}~{\rm s}^{-1}$. Thermal ionisation does not become important for the density range investigated here, since the temperature never reaches $\simeq 10^3~{\rm K}$. A simple extrapolation of temperature based on our results indicates that such a temperature will not be reached until a central density $\approx 3\times 10^{15}~{\rm cm}^{-3}$. It is therefore suggested that magnetic recoupling will not occur until central densities of at least several $\times 10^{15}~{\rm cm}^{-3}$ are attained.

The central species abundances are strongly affected not only by changes in the ionisation rate, but also by microscopic interactions between the different species and by the macroscopic dynamics of the cloud. [For central neutral densities $\lesssim 1800~{\rm cm}^{-3}$, the term added to the ion density \citep[as in][]{fm92,fm93}, which mimics the effect of UV ionisation at low densities in cloud envelopes, is responsible for the behaviour of the ion and electron relative abundances and is not meant to be a model or prediction for the exact behaviour of $n_{\rm i}$ and $n_{\rm e}$ in cloud envelopes. It is used mostly for computational convenience.]

At densities greater than $\approx 2 \times 10^{3}~{\rm cm^{-3}}$ ionisation is primarily due to cosmic rays, and $x_{\rm i,e}$ behave as found earlier by \citet{cm94}. It cannot be overemphasised that the `canonical' relation $x_{\rm i,e}\propto n_{\rm n}^{k-1}$ with $k={\rm const}=1/2$ is never established. In fact, during the subcritical phase of the evolution (i.e., $n_{\rm n,c}\lesssim 10^4~{\rm cm}^{-3}$), $k$ is always greater than $1/2$, an effect that is entirely due to ambipolar diffusion. As explained by \citet{cm94,cm98}, during this phase of evolution charged and neutral dust grains are well attached to magnetic field lines, which are essentially `held in place' \citep{mouschovias78,mouschovias79}, and are `left behind' by the inwardly diffusing neutrals. This results in a decrease in the central dust-to-gas ratio by a factor roughly equal to the initial (dimensionless) central mass-to-flux ratio. One consequence is a significantly reduced inelastic capture of ions and electrons onto grains and therefore a greater number of gas-phase ions and electrons than predicted by calculations that assume a constant dust-to-gas ratio.

Once a supercritical core forms, the grain relative abundances are frozen at their values until a central density of $\sim 10^9~{\rm cm}^{-3}$ is reached. Gas-phase ions and electrons are then quickly adsorbed onto grain surfaces and the grains become the main charge carriers (when $n_{\rm n,c}\sim 10^{10}~{\rm cm}^{-3}$). However, most grains remain neutral due to inadequate numbers of ions and electrons, whose abundances are determined by balancing their rates of production and their collision rates with neutral grains. Because these collision rates are inversely proportional to the square root of the ion/electron mass, the ions are more abundant than the lighter electrons by a factor $\simeq (m_{\rm i}/m_{\rm e})^{1/2}\simeq 210$ (different sticking probabilities for ions and electrons reduce this value to $\simeq 125$). Since $x_{\rm g_0}$ approaches a constant, the densities of ions and electrons approach constant values, so their abundances relative to neutrals asymptote towards a $n_{\rm n}^{-1}$ dependence. Once cosmic rays become appreciably attenuated, however, this asymptotic behaviour is interrupted and the abundances of electrons and ions decrease much more rapidly. An ionisation floor is established by the radionuclide $^{40}$K at densities $\approx 10^{13}~{\rm cm}^{-3}$, after which the electron and ion abundances once again scale roughly as $n_{\rm n}^{-1}$. The abundances of charged grains are proportional to $n_{\rm n}^{-1/2}$, for the same reasons that applied to the abundances of electrons and ions at low densities.

\begin{figure*}
\center
\includegraphics[width=2.8in]{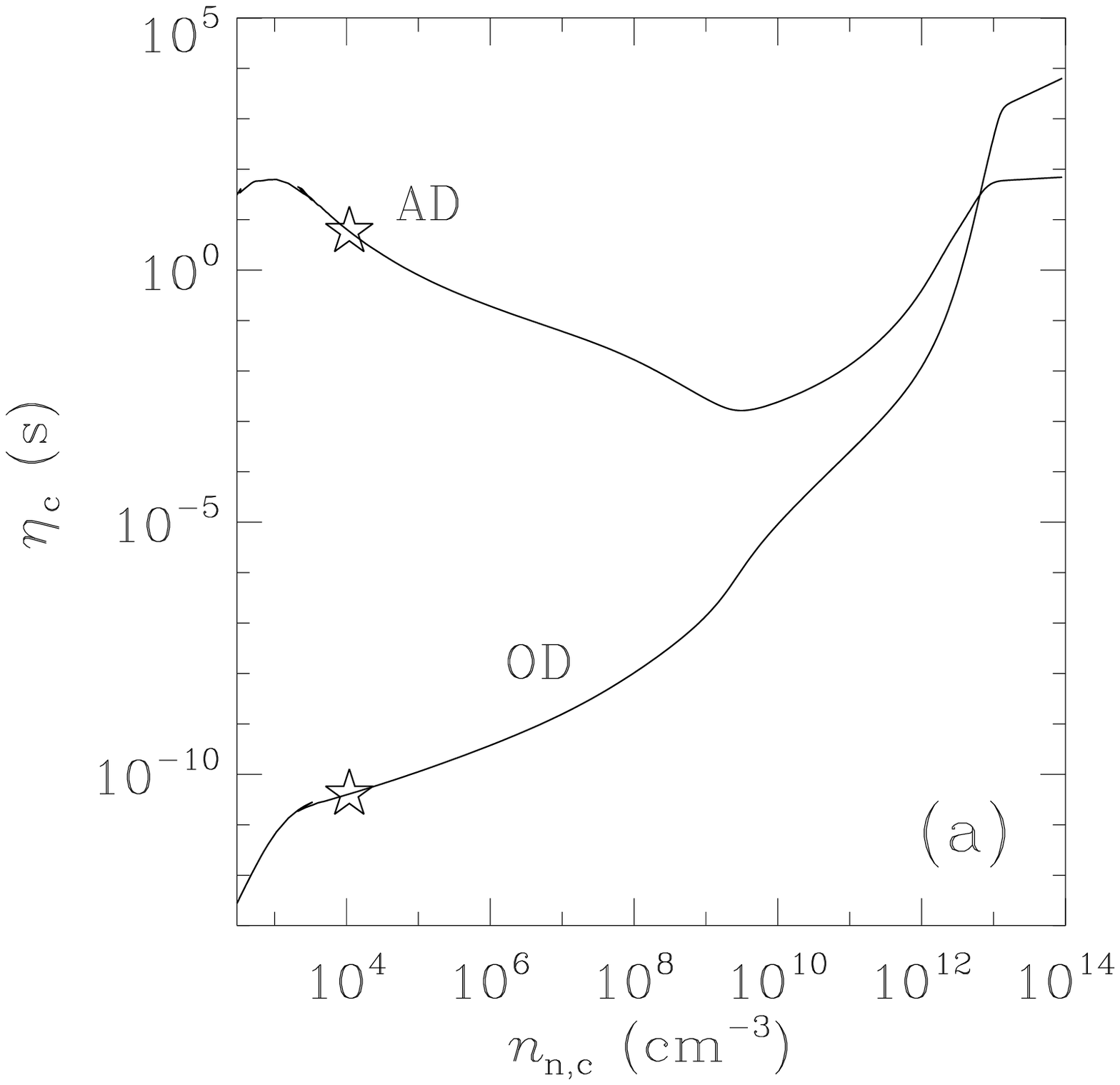}
\includegraphics[width=2.8in]{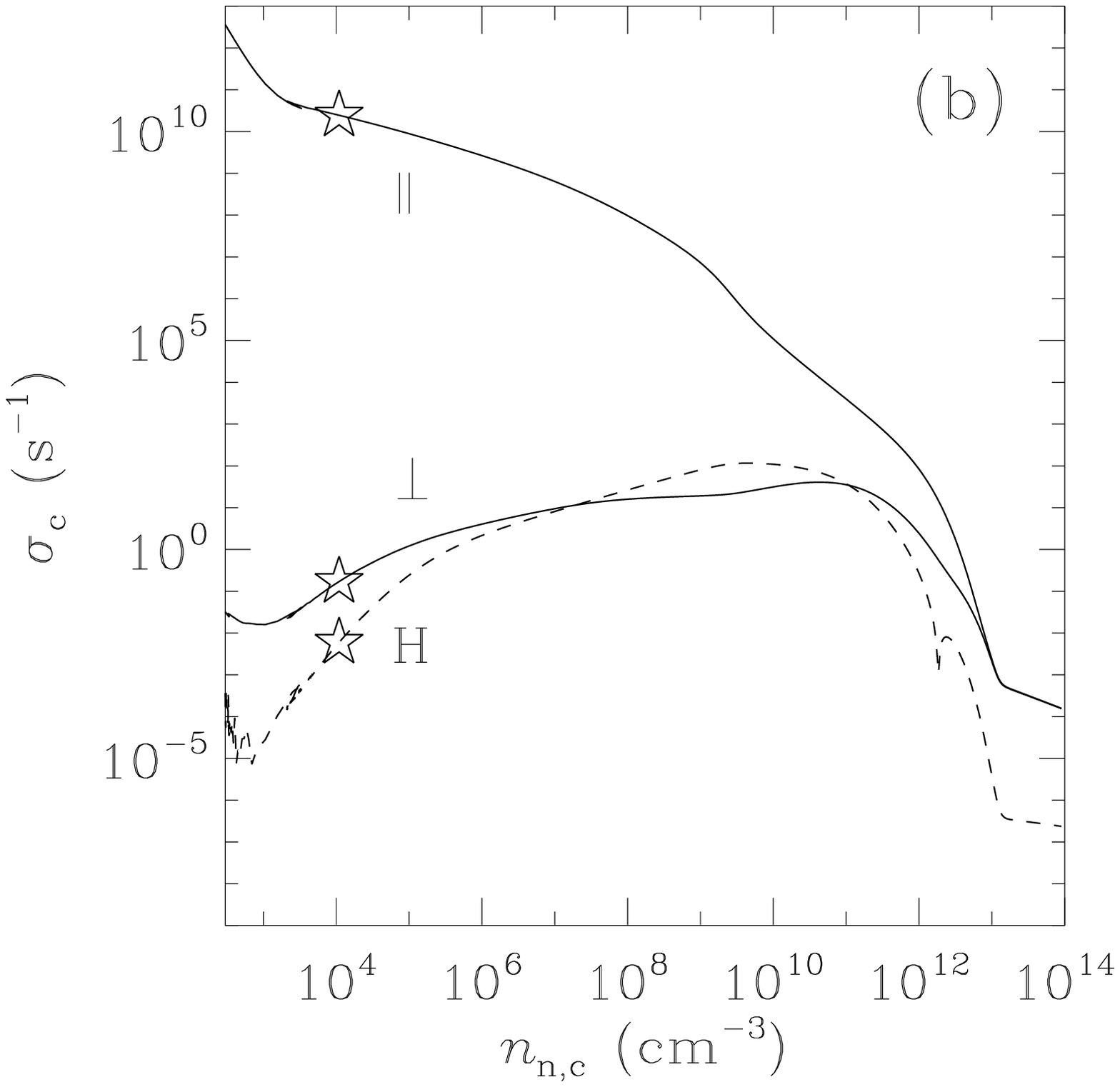}
\newline
\includegraphics[width=2.8in]{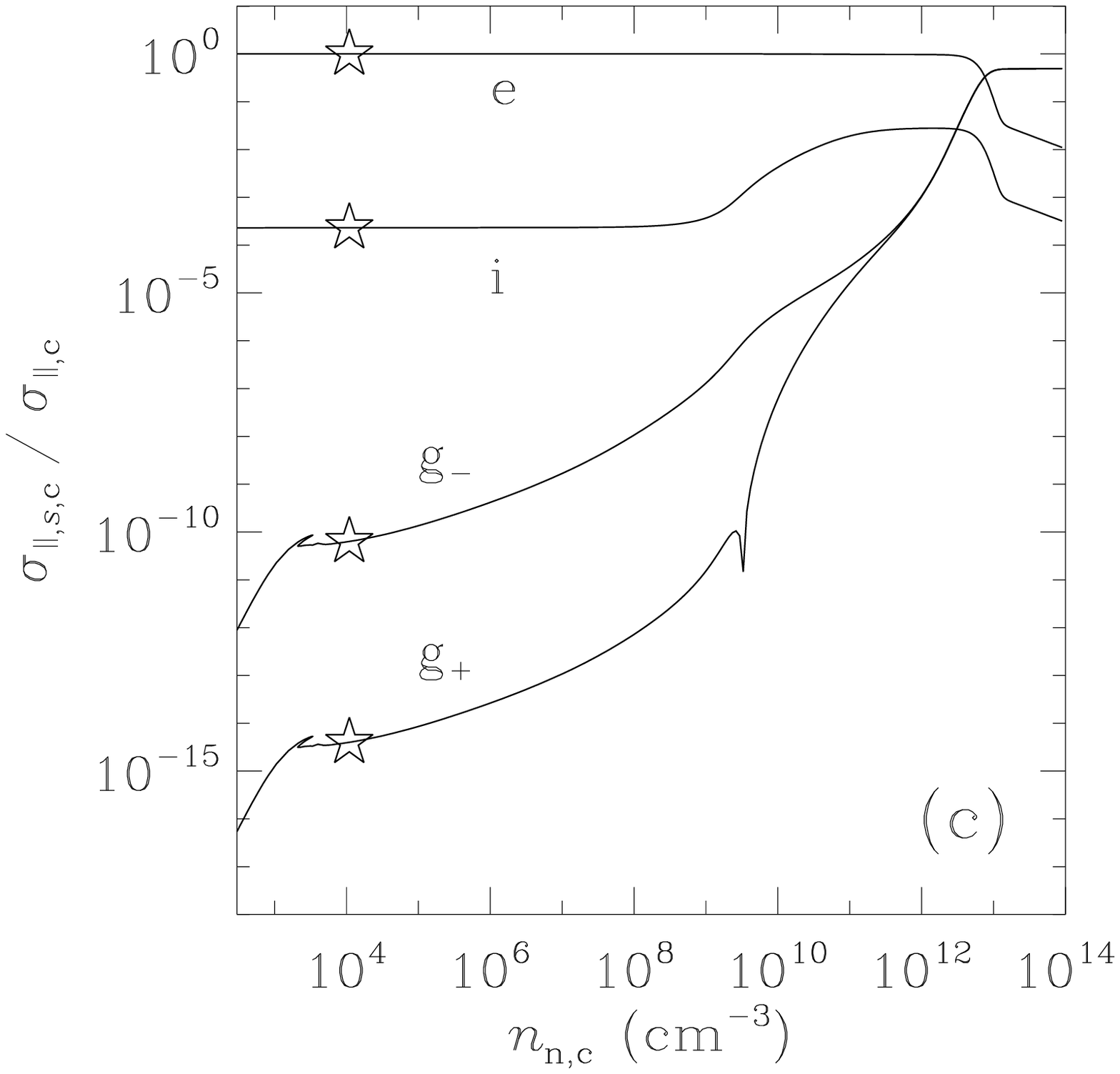}
\includegraphics[width=2.8in]{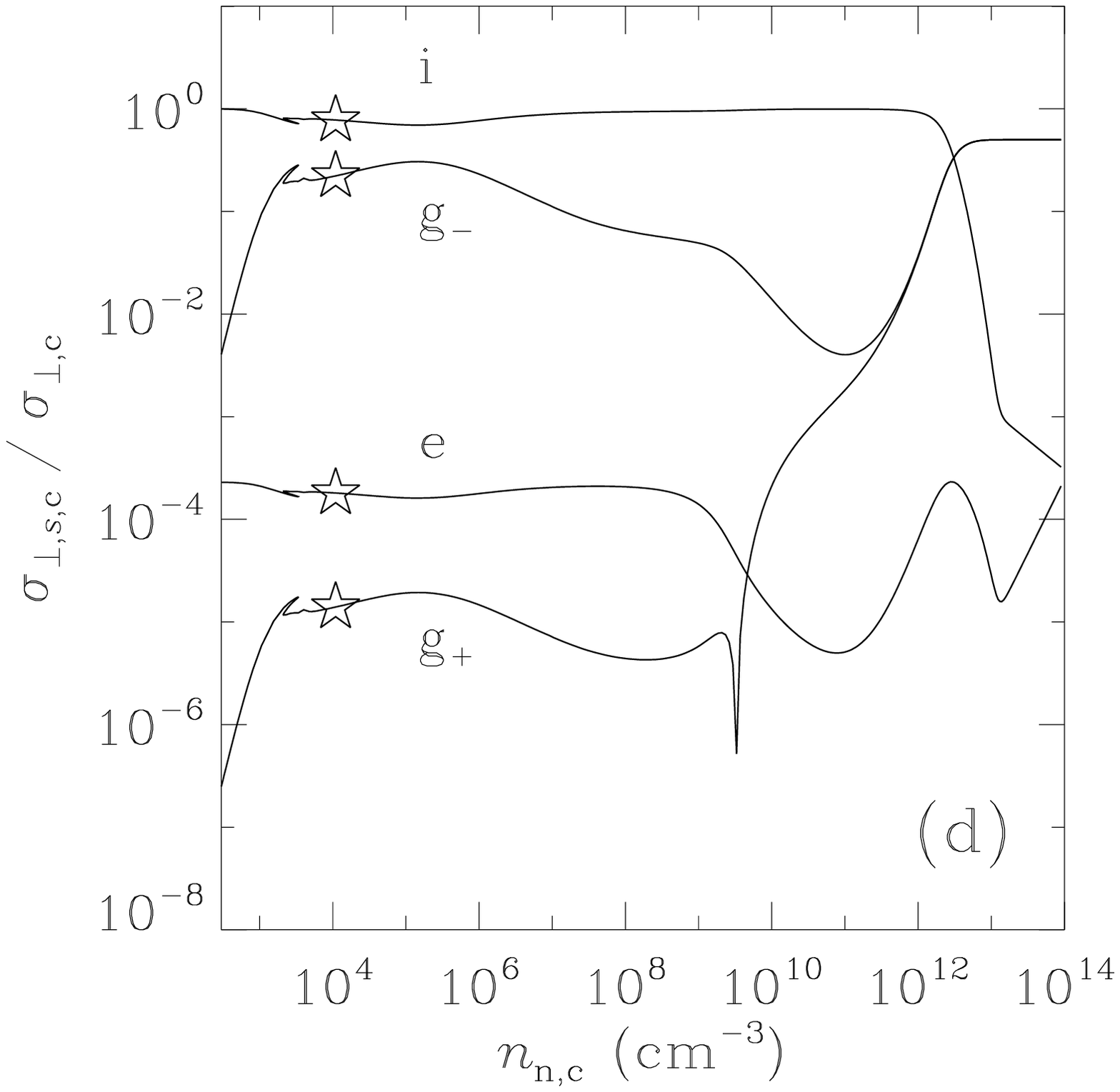}
\newline
\caption{Evolution of central quantities as functions of the central number density of neutrals, $n_{\rm n,c}$: (a) Resistivities due to ambipolar diffusion (AD) and Ohmic dissipation (OD). (b) Parallel ($||$), perpendicular ($\perp$), and Hall (H) magnetic conductivities. (c) Fraction of the parallel magnetic conductivity carried by each species. (d) Fraction of the perpendicular magnetic conductivity carried by each species. The `star' marks the time at which a supercritical core forms.}
\label{figure:central:resist}
\end{figure*}

Central quantities related to magnetic diffusion are shown in Fig.~\ref{figure:central:resist} as functions of central density. The magnetic diffusion coefficients related to ambipolar diffusion (AD) and Ohmic dissipation (OD) are shown in Fig.~\ref{figure:central:resist}a. As we are concerned here with the evolution of only the poloidal component of the magnetic field (recall that rotation has been justifiably ignored), the Hall effect plays no role in the dynamics of the cloud. Magnetic-flux redistribution is due to ambipolar diffusion until a central density $\simeq 7\times 10^{12}~{\rm cm}^{-3}$ is reached, after which Ohmic dissipation is the primary magnetic diffusion mechanism. Therefore, ambipolar diffusion still dominates Ohmic dissipation as a flux reduction mechanism when the grains become the primary charge carriers, contrary to the expectations of \citet{nu86a,nu86b}. Equality of the ambipolar-diffusion and Ohmic-dissipation diffusion rates occurs once the central mass-to-flux ratio becomes equal to $\simeq 16$ times its critical value and the grains become the dominant {\em current} carriers (see below).

In Fig.~\ref{figure:central:resist}b, we show the parallel ($||$), perpendicular ($\perp$), and Hall (H) conductivities. Figures \ref{figure:central:resist}c and \ref{figure:central:resist}d show the fraction of the parallel and perpendicular conductivities, respectively, carried by each species. (Absolute values have been taken; the cusps in the positive grain curves are due to the conductivity passing through zero as it changes sign.) The parallel conductivity is primarily due to electrons until Ohmic dissipation becomes the primary magnetic diffusion mechanism, after which the contribution of the charged grains to the parallel conductivity dominates that of the electrons. The perpendicular conductivity is mainly due to ions up to a central density $\simeq 3\times 10^{12}~{\rm cm}^{-3}$, after which the charged grains dominate the ions' contribution. In other words, the grains begin to carry most of the current at roughly the same time that Ohmic dissipation becomes important (see Section \ref{section:current}). {\em Star formation and protostellar disc calculations that study the phase when Ohmic dissipation becomes an important magnetic diffusion mechanism must include not only the chemistry but also the magnetohydrodynamics of dust grains}. 

\begin{figure*}
\center
\includegraphics[width=2.8in]{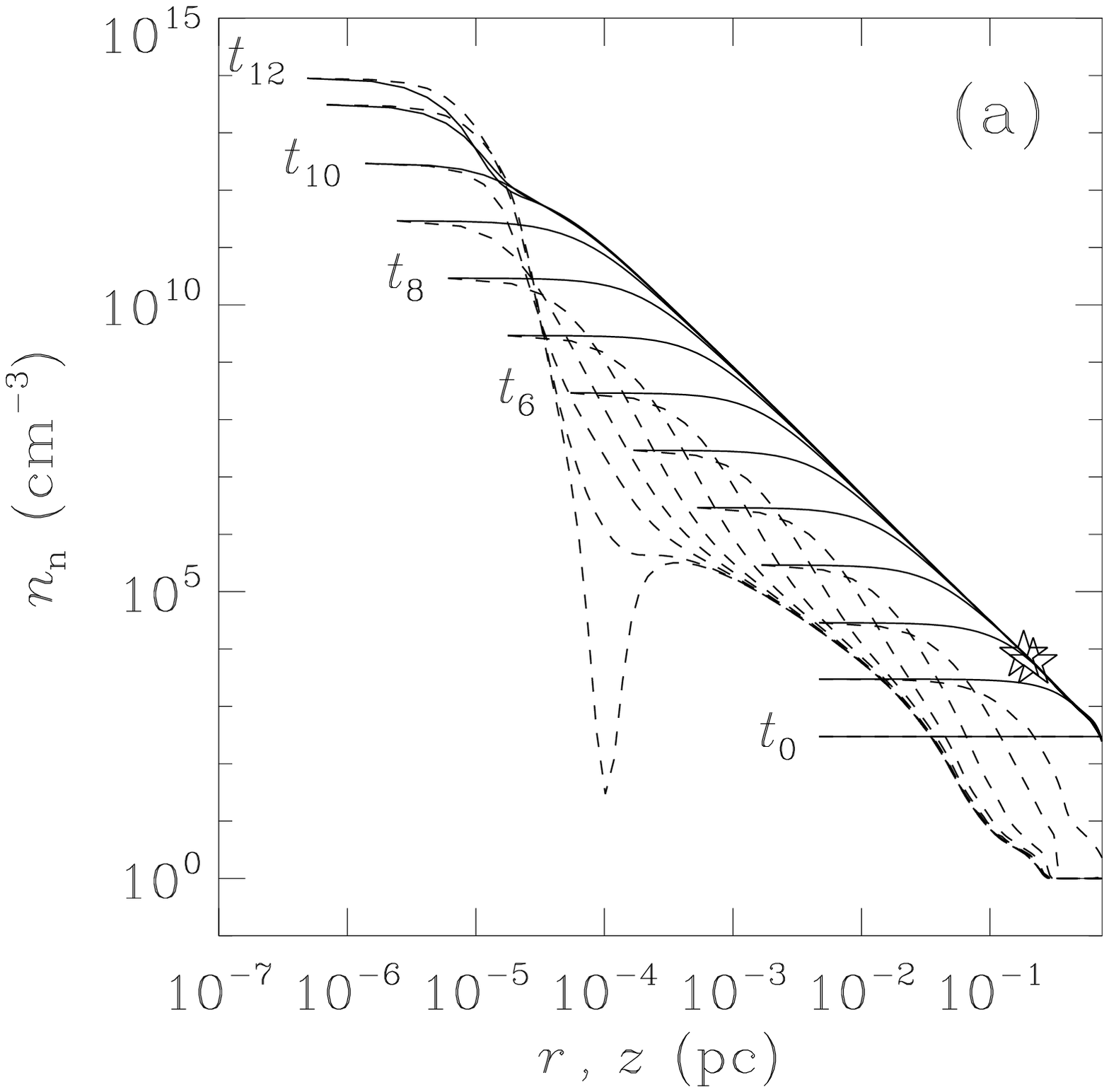}
\includegraphics[width=2.8in]{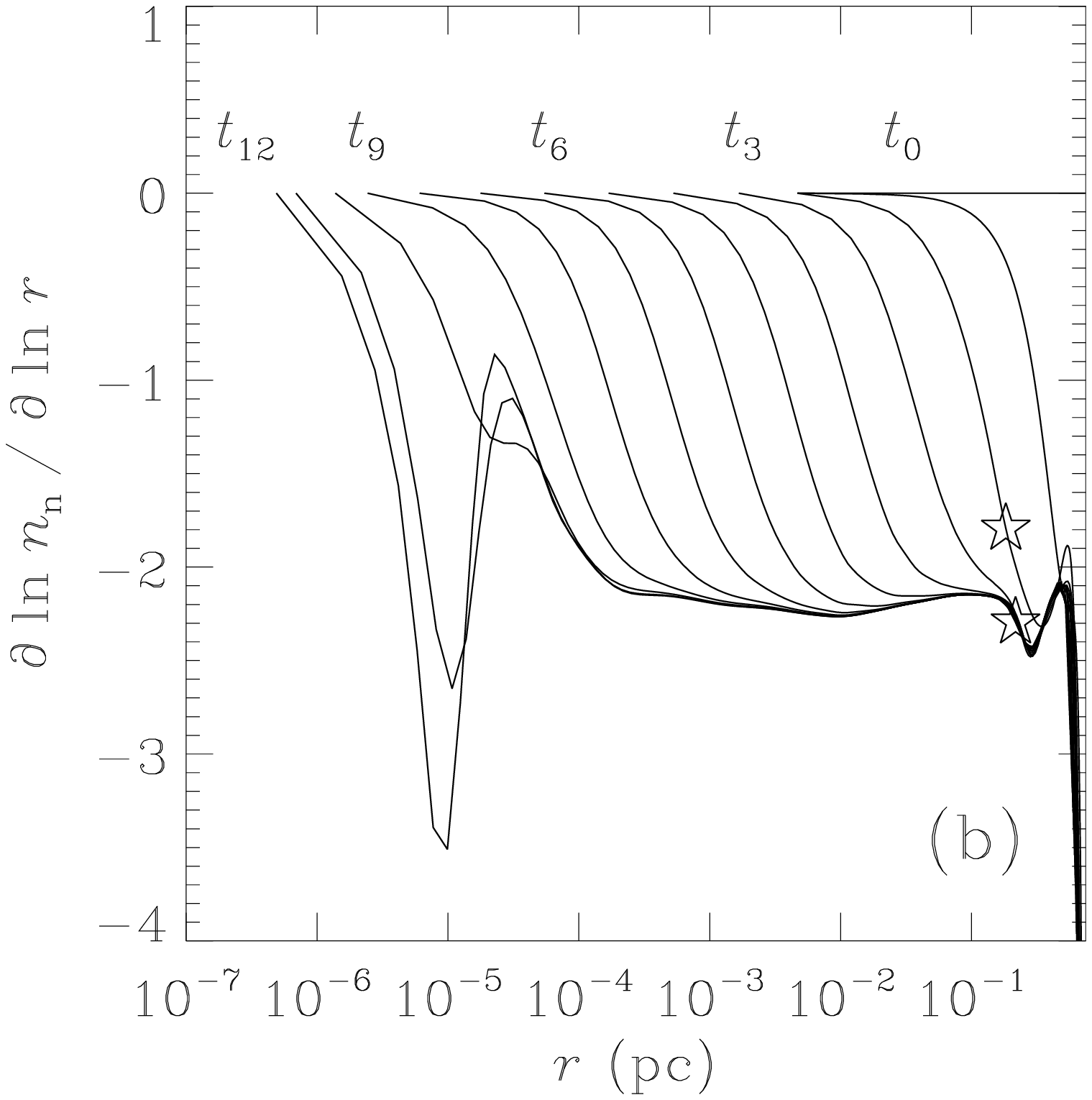}
\newline
\includegraphics[width=2.8in]{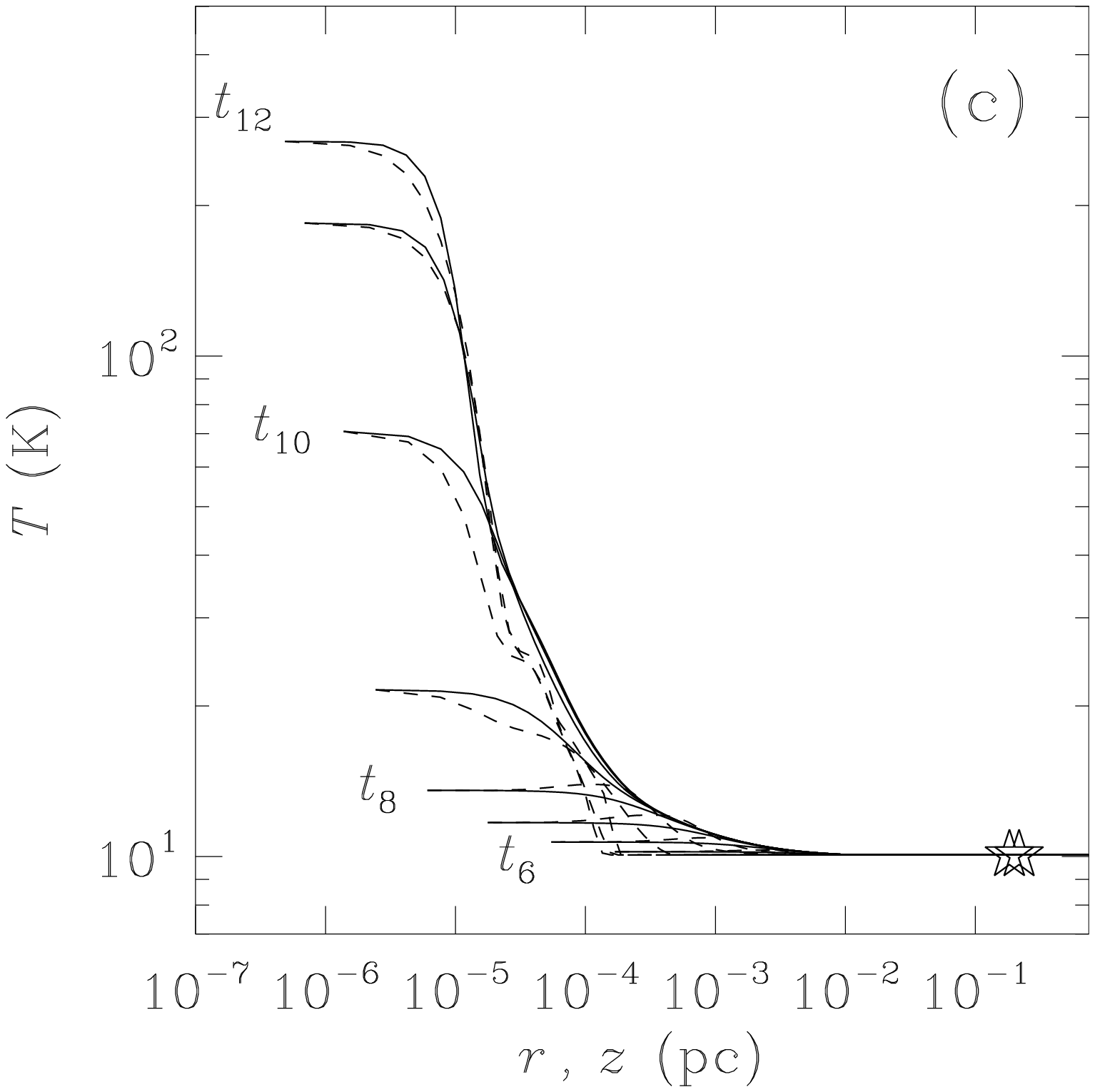}
\includegraphics[width=2.8in]{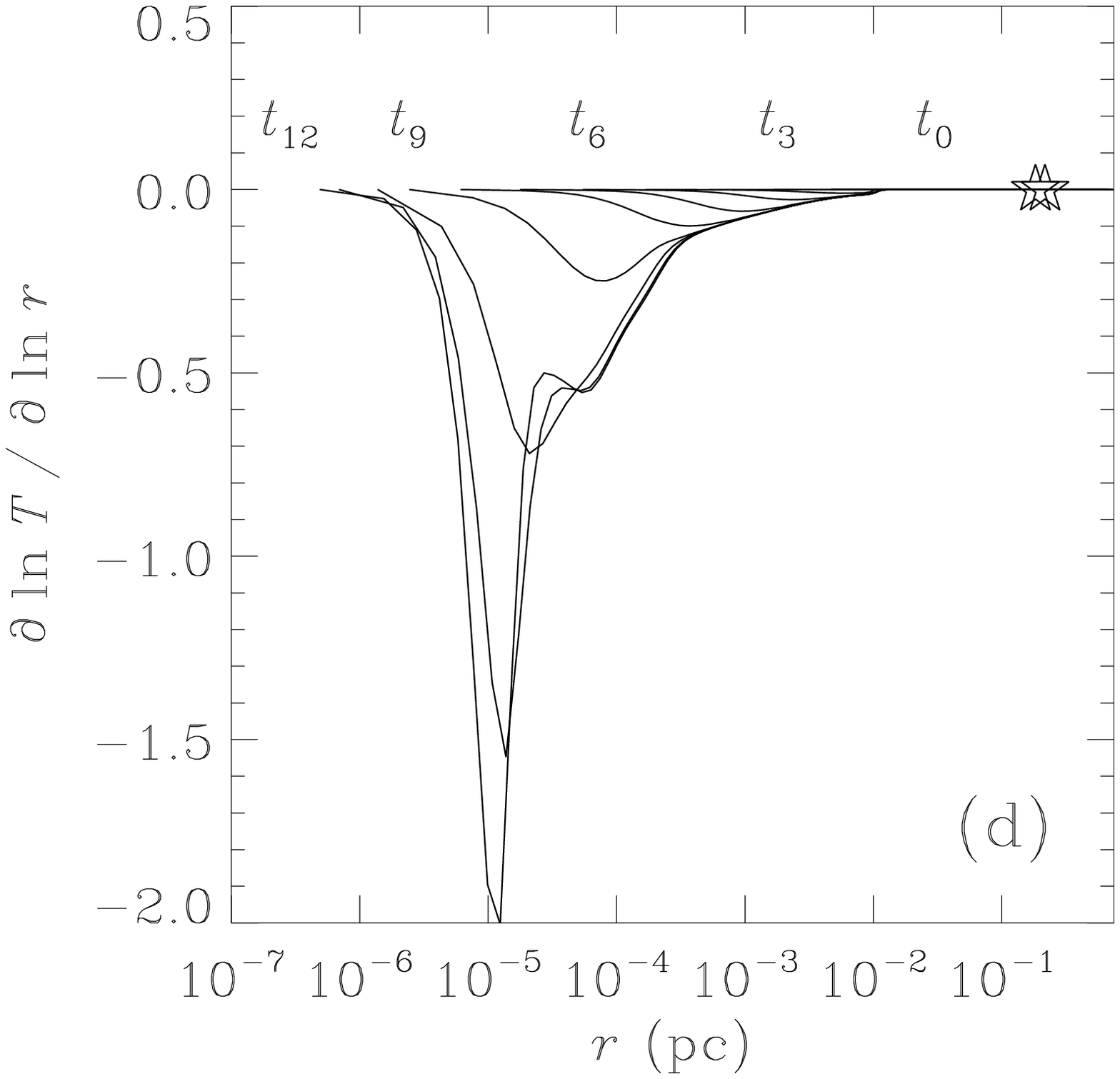}
\newline
\caption{Radial midplane (solid line) and vertical symmetry-axis (dashed line) profiles of (a) the number density of neutrals and (c) the gas temperature at thirteen different times ($t_{0} = -4.00$, $t_{1} = 3.763974$, $t_{2} = 10.022421$, $t_{3} = 10.890328$, $t_{4} = 11.021924$, $t_{5} = 11.049141$, $t_{6} = 11.056294$, $t_{7} = 11.058372$, $t_{8} = 11.059021$, $t_{9} = 11.059230$, $t_{10} = 11.059310$, $t_{11} = 11.059346$, $t_{12} = 11.059357~{\rm Myr}$). Also shown are the radial derivatives along the midplane of (b) the density profile, $\partial\ln n_{\rm n}/\partial\ln r$, and (d) the (gas) temperature profile, $\partial\ln T/\partial\ln r$. The inner (outer) `star' on a curve, present only after a supercritical core forms, marks the initial (final) radius of the supercritical core.}
\label{figure:mpax:denstemp}
\end{figure*}

\begin{figure*}
\center
\includegraphics[width=2.8in]{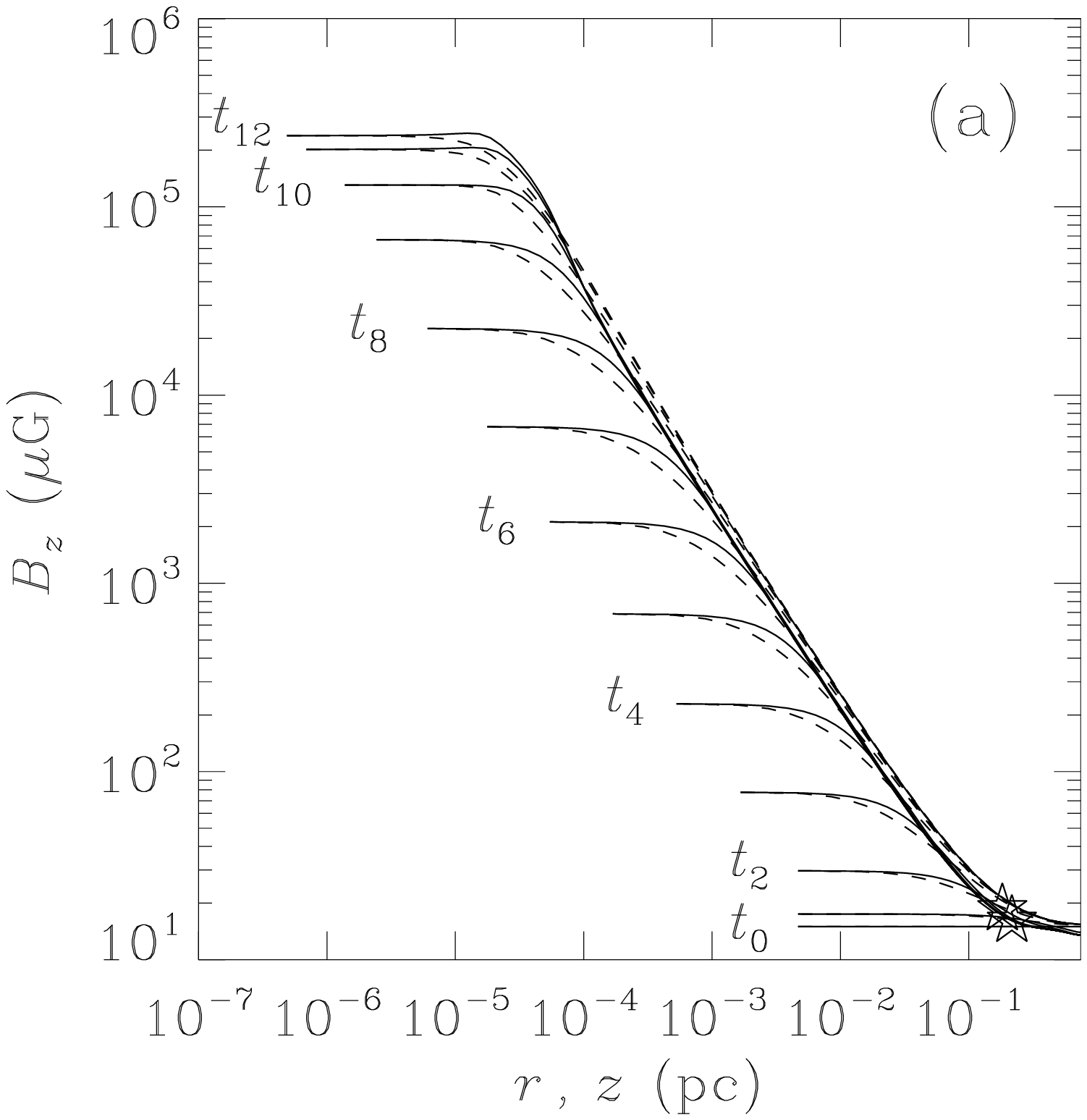}
\includegraphics[width=2.8in]{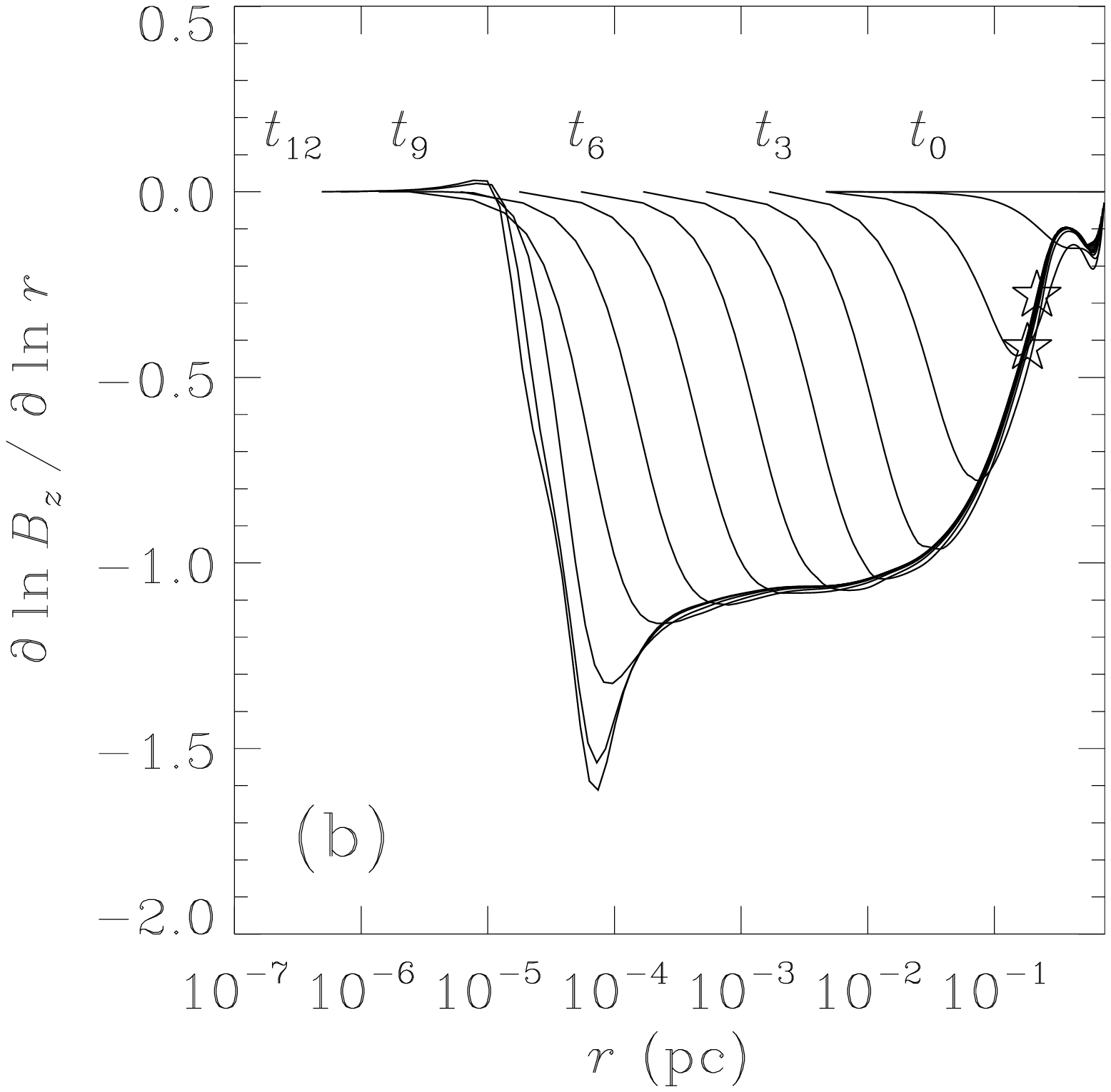}
\newline
\includegraphics[width=2.8in]{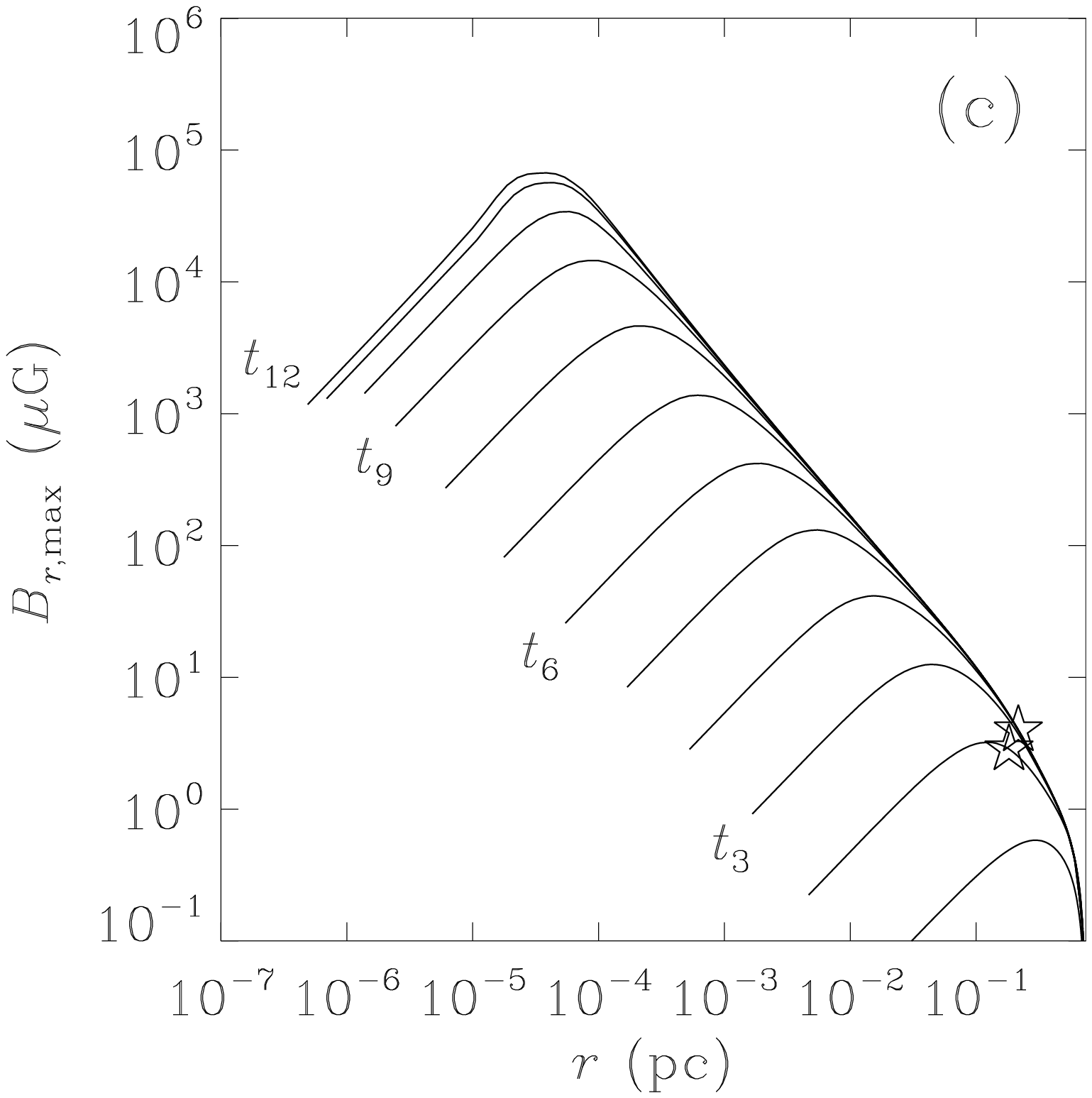}
\includegraphics[width=2.8in]{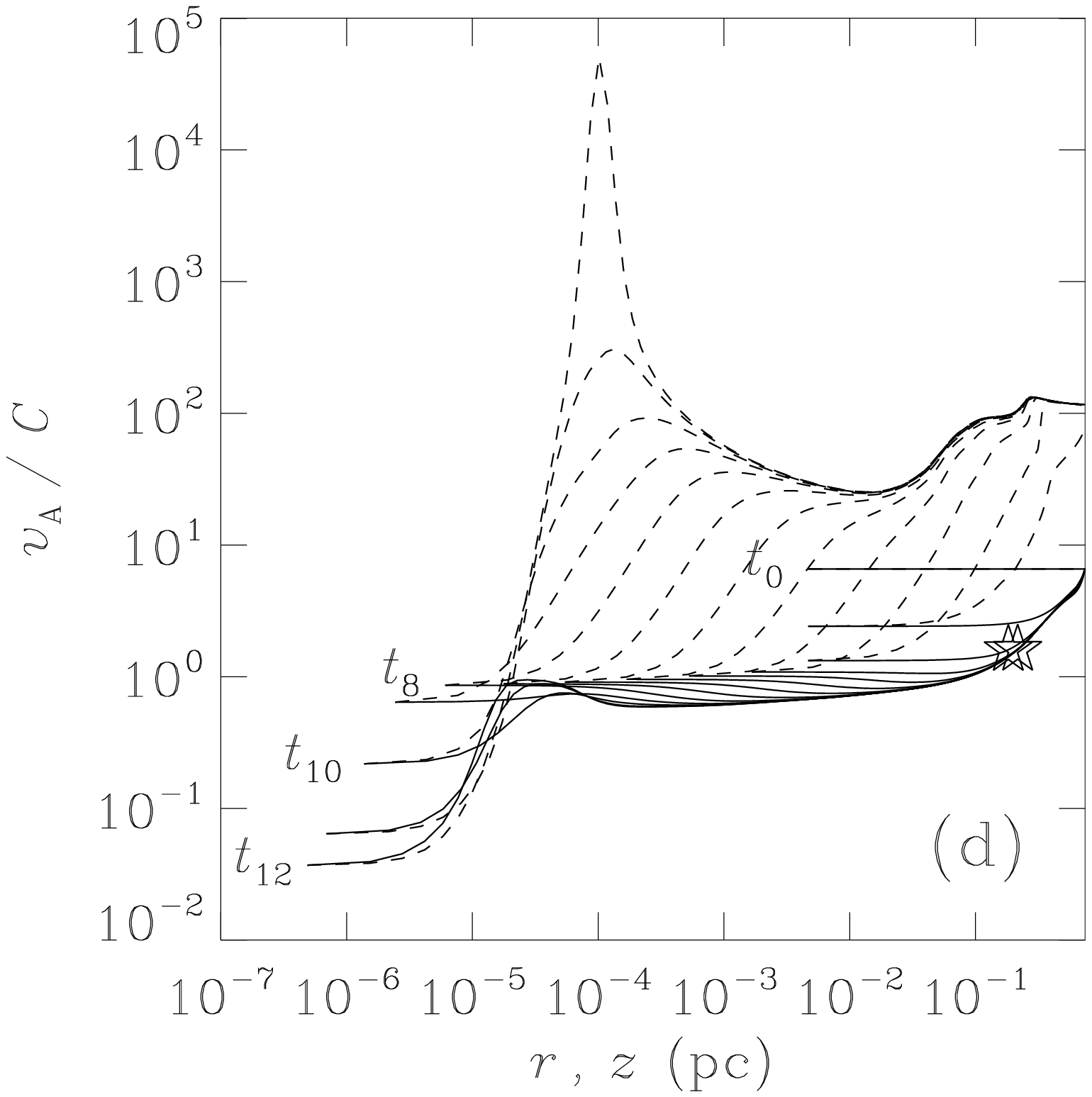}
\newline
\caption{Radial midplane (solid line) and vertical symmetry-axis (dashed line) profiles of (a) the $z$-component of the magnetic field and (d) the ratio of the Alfv\'{e}n speed and the local isothermal sound speed at different times, as in Fig.~\ref{figure:mpax:denstemp}. Also shown are (b) the radial derivative of the $z$-component of the magnetic field, $\partial\ln B_z/\partial\ln r$, along the midplane, and (c) radial profiles of the maximum (in $z$) strength of the $r$-component of the magnetic field. The inner (outer) `star' on a curve, present only after a supercritical core forms, marks the initial (final) radius of the supercritical core.}
\label{figure:mpax:bfield}
\end{figure*}

\subsection{Evolution of Spatial Structure of the Model Cloud}\label{section:spatial}

Figures \ref{figure:mpax:denstemp} -- \ref{figure:mpax:luminosity} show radial midplane (solid lines) and vertical symmetry-axis (dashed lines) profiles of physical quantities at thirteen different times. Each curve corresponds to a time $t_j$ at which the central density has increased by a factor of $10^j$ relative to the initial uniform `reference' state (i.e., the central density is $\simeq 3\times 10^{2+j}~{\rm cm}^{-3}$ at time $t_j$). The last curve (corresponding to $t_{12}$) is an exception; it refers to the time at which $n_{\rm n,c}\simeq 10^{14}~{\rm cm}^{-3}$. These times are $t_{0} = -4.00$, $t_{1} = 3.763974$, $t_{2} = 10.022421$, $t_{3} = 10.890328$, $t_{4} = 11.021924$, $t_{5} = 11.049141$, $t_{6} = 11.056294$, $t_{7} = 11.058372$, $t_{8} = 11.059021$, $t_{9} = 11.059230$, $t_{10} = 11.059310$, $t_{11} = 11.059346$, and $t_{12} = 11.059357~{\rm Myr}$.

\subsubsection{Density and Temperature}\label{section:denstemp}

Fig.~\ref{figure:mpax:denstemp}a displays the number density of neutrals as a function of radius ($r$) and height ($z$) at the 13 different times referred to above. We also show the logarithmic slope of the (midplane) density profile versus radius, $\partial\ln n_{\rm n}/\partial\ln r$, in Fig.~\ref{figure:mpax:denstemp}b. A uniform central region is maintained (in the radial direction) for $r < \lambda_{\rm T,cr}(t)$ by thermal-pressure forces. Outside this region, the density profile resembles a broken power-law. Inside the supercritical core, the slope has an average value $\simeq -2.2$ for $10^{-4}~{\rm pc} \lesssim r \lesssim 10^{-1}~{\rm pc}$. For smaller radii in the range $\approx 10^{-5}$ -- $10^{-4}~{\rm pc}$ ($\approx 2$ -- $20~{\rm AU}$), the slope increases briefly due to a local increase in magnetic support (see Section \ref{section:forces}). A transition to spherical geometry occurs at $r\simeq 4~{\rm AU}$ when the central density is $\simeq 3\times 10^{12}~{\rm cm}^{-3}$. The radial and vertical density profiles are nearly identical inside this radius, with a slope approaching $\approx -10/3$, corresponding to hydrostatic force balance (with $\gamma_{\rm eff}=7/5$). Outside the magnetically supercritical core ($\approx 0.4~{\rm pc}$), there is an abrupt break in the slope because of significant magnetic support in the envelope. Very near the radial boundary ($r=R=0.75~{\rm pc}$) of the computational region, the slope decreases again to satisfy the boundary condition that the matter has no radial velocity at this surface.

Fig.~\ref{figure:mpax:denstemp}c displays similar profiles for the (gas) temperature. The logarithmic slope of the (midplane) temperature versus radius, $\partial\ln T/\partial\ln r$, is shown in Fig.~\ref{figure:mpax:denstemp}d. The molecular cloud and supercritical core evolve isothermally for densities $n_{\rm n}\lesssim 10^{7}~{\rm cm}^{-3}$ and (cylindrical polar) radii $r\gtrsim 10^{-2}~{\rm pc}$. Just inside the hydrostatic core ($r\lesssim 2~{\rm AU}$), the slope is closely approximated by $(\gamma_{\rm eff}-1)(\partial\ln n_{\rm n}/\partial\ln r)$, whereas the temperature becomes constant very near the centre of the core. For $2~{\rm AU}\lesssim r \lesssim 2000~{\rm AU}$, no single power-law can approximate the temperature profile.

\subsubsection{Magnetic Field and Alfv\'{e}n Speed}\label{section:bfield}

The $z$-component of the magnetic field is shown in Fig.~\ref{figure:mpax:bfield}a. In common with the density profile, there is an inner region [$r \lesssim \lambda_{\rm T,cr}(t)$] in which the magnetic field is almost uniform. Outside the supercritical core ($r\approx 0.4~{\rm pc}$, marked by a `star'), where the gas is magnetically supported, the profile flattens significantly and $B_z$ is almost uniform. For radii in the range $\approx 10^{-3}-10^{-2}~{\rm pc}$ the slope $\partial\ln B_z/\partial\ln r \simeq -1.1$ (see Fig.~\ref{figure:mpax:bfield}b). As the temperature increases at smaller radii, the slope approaches $\simeq -1.4$ down to the location of the hydrostatic core boundary, inside which the slope quickly asymptotes to zero. This zero slope is due to effective Ohmic dissipation, which quickly erases any spatial variation in the magnetic field. The value of the magnetic field inside the hydrostatic core is $\simeq 0.2~{\rm G}$, in excellent agreement with the protosolar magnetic field as derived from meteoritic data \citep{slp61,hr74,levy88}.

A closer inspection of Figures \ref{figure:mpax:bfield}a and \ref{figure:mpax:bfield}b reveals that the $z$-component of the magnetic field actually has a local maximum just outside the hydrostatic core at $r\approx 10^{-5}~{\rm pc}$ ($\approx 2~{\rm AU}$), where the inwardly advected magnetic flux has piled up and formed a `magnetic wall.' This {\em local} concentration of magnetic flux outside the hydrostatic core boundary results in a magnetic shock, the consequences of which are discussed in Sections \ref{section:forces} and \ref{section:velocities}. \citet{tm05b} followed the formation and evolution of a series of these magnetic shocks and found that accretion onto the forming protostar occurs in a time-dependent, spasmodic fashion.

Fig.~\ref{figure:mpax:bfield}c shows the radial profile of the maximum strength of the $r$-component of the magnetic field, at whichever height $z$ it occurs above the midplane, at twelve different times ($t_{1}$, $t_{2}$, $\ldots$, $t_{12}$) as in Fig.~\ref{figure:mpax:denstemp}; $B_{r}$ is generated as a result of field-line deformation during core contraction. Inside the inner flat region where the magnetic field is almost uniform, $B_{r{\rm ,max}}$ declines rapidly to very small values, reaching zero at the origin. The rapid contraction outside the hydrostatic core induces significant field-line deformation, and $B_{r{\rm ,max}}$ has a sharp local maximum at the boundary of the hydrostatic core, where its value becomes comparable to, although still smaller than, that of $B_z$ ($B_r/B_z = 0.99$ at the location of $B_{r{\rm ,max}}$). It is the case, nevertheless, that  $B_z>B_r$ at all locations and for all times.

The ratio of the local Alfv\'{e}n speed and the local isothermal sound speed, $v_{\rm A}/C$, is shown in Fig.~\ref{figure:mpax:bfield}d. The behaviour of this ratio may be understood from the fact that $v_{\rm A}/C\propto n_{\rm n}^{\kappa-\gamma_{\rm eff}/2}$. Outside the magnetically supercritical core (whose radius is marked by a `star') $\kappa\approx 0$ and so $v_{\rm A}/C\propto n_{\rm n}^{-1/2}$. This accounts for the large values of $v_{\rm A}/C$ in the cloud envelope. In the magnetically supercritical core during the isothermal phase of contraction (when $\kappa\simeq 0.47$ and $\gamma_{\rm eff}=1$), the Alfv\'{e}n speed varies only slightly with the isothermal sound speed and is approximately a constant of order unity. The fact that $v_{\rm A}$ becomes comparable to (or smaller than) the sound speed inside the supercritical core provides {\em an explanation for the thermalisation of linewidths observed in molecular cloud cores} \citep{bpddc81,mb83}. In the theory of magnetic star formation, the material motions responsible for the observed linewidths are attributed to long-wavelength, standing Alfv\'{e}n waves \citep{mouschovias87,mp95}, with a remarkable quantitative agreement between theory and observations \citep{mtk06}.

A {\em local} maximum in $v_{\rm A}/C$ occurs just outside the magnetic wall at $r\approx 10~{\rm AU}$. There is a substantial decrease in $v_{\rm A}/C$ inside the hydrostatic core, where the gas is primarily thermally supported, due to the increase in $\gamma_{\rm eff}$ and decrease in $\kappa$; i.e., the gas begins to evolve almost adiabatically and the magnetic field begins to decouple from the matter. 

\subsubsection{Forces}\label{section:forces}

\begin{figure}
\center
\includegraphics[width=2.8in]{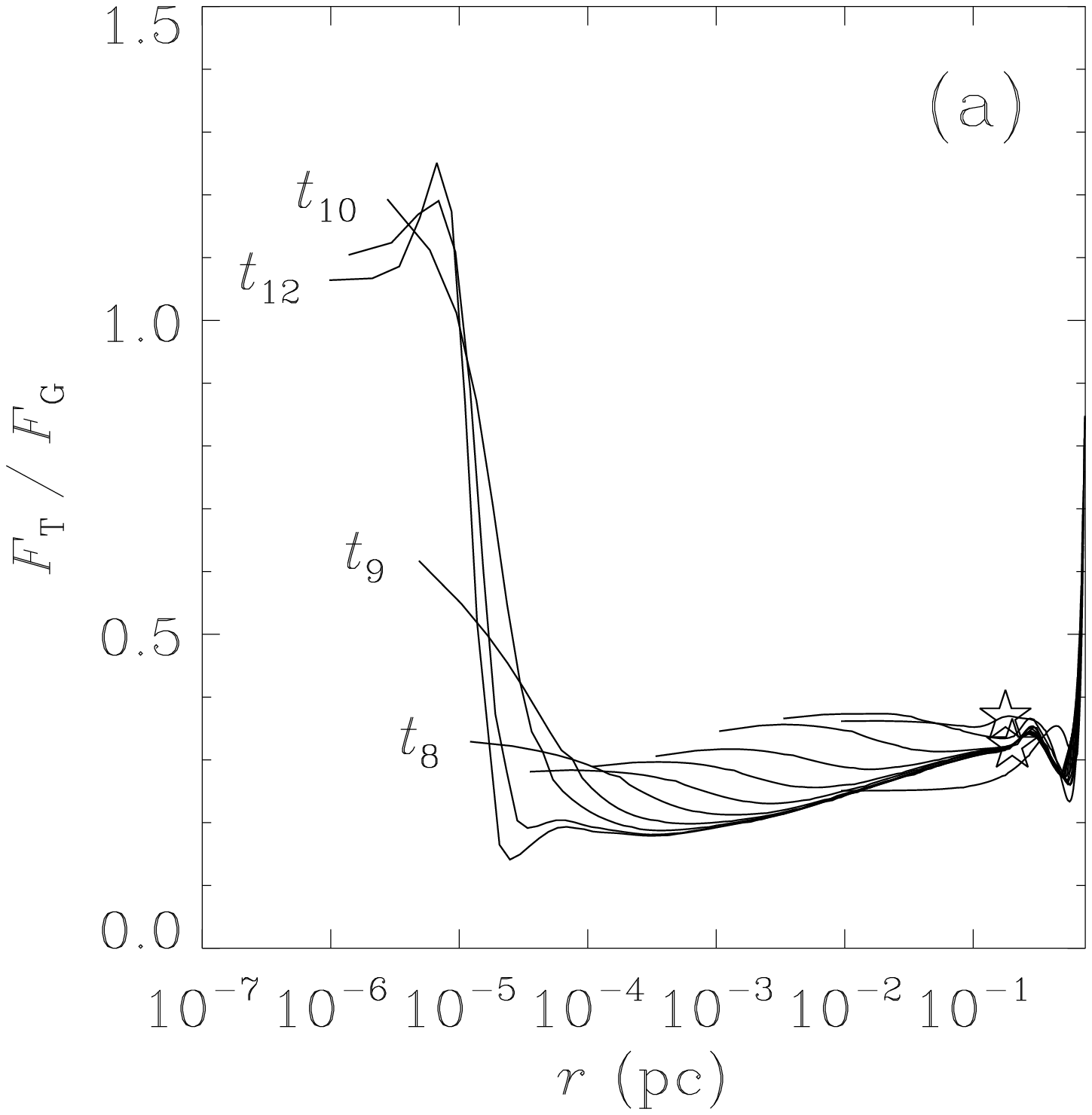}
\newline
\includegraphics[width=2.8in]{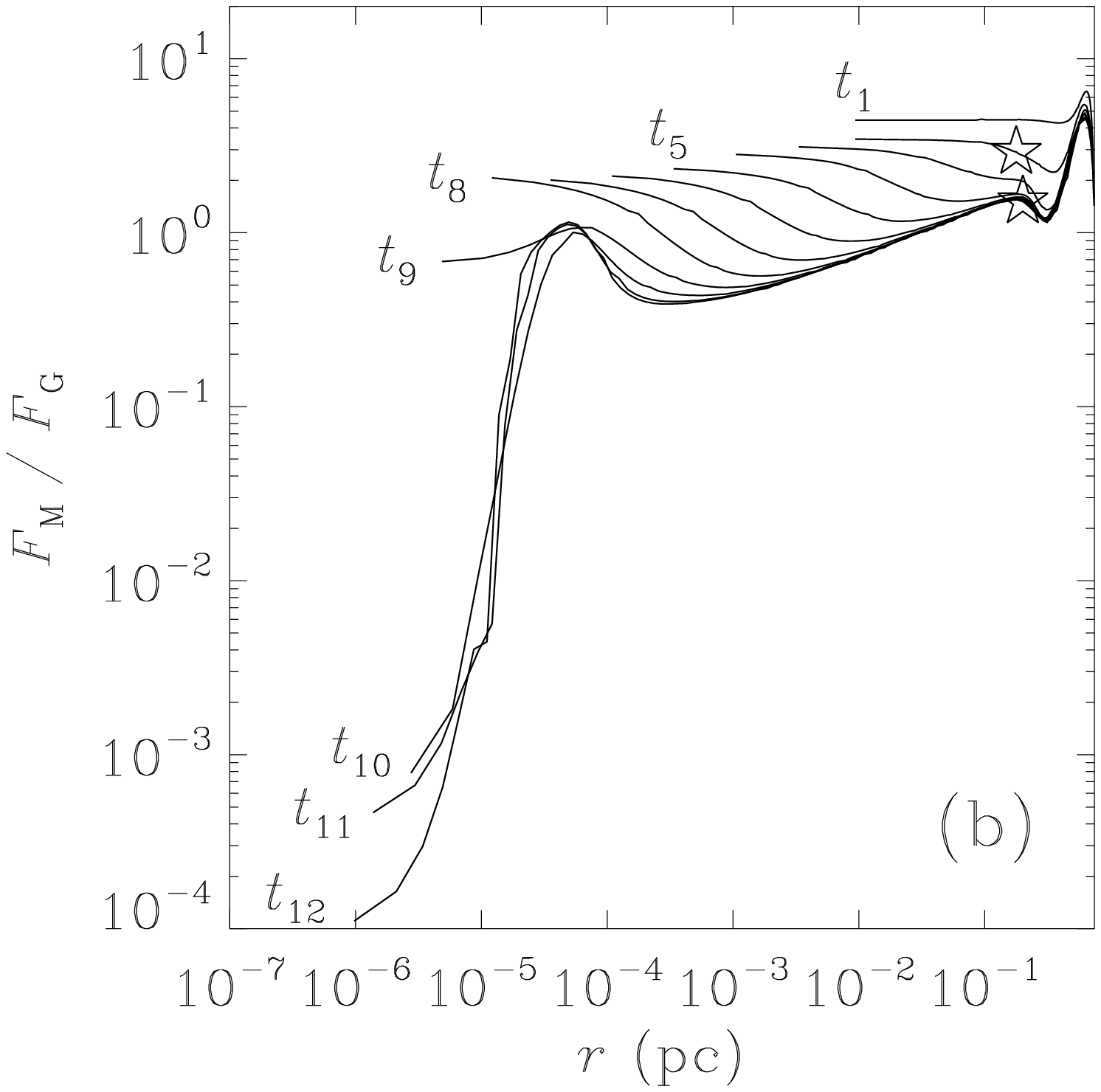}
\newline
\includegraphics[width=2.8in]{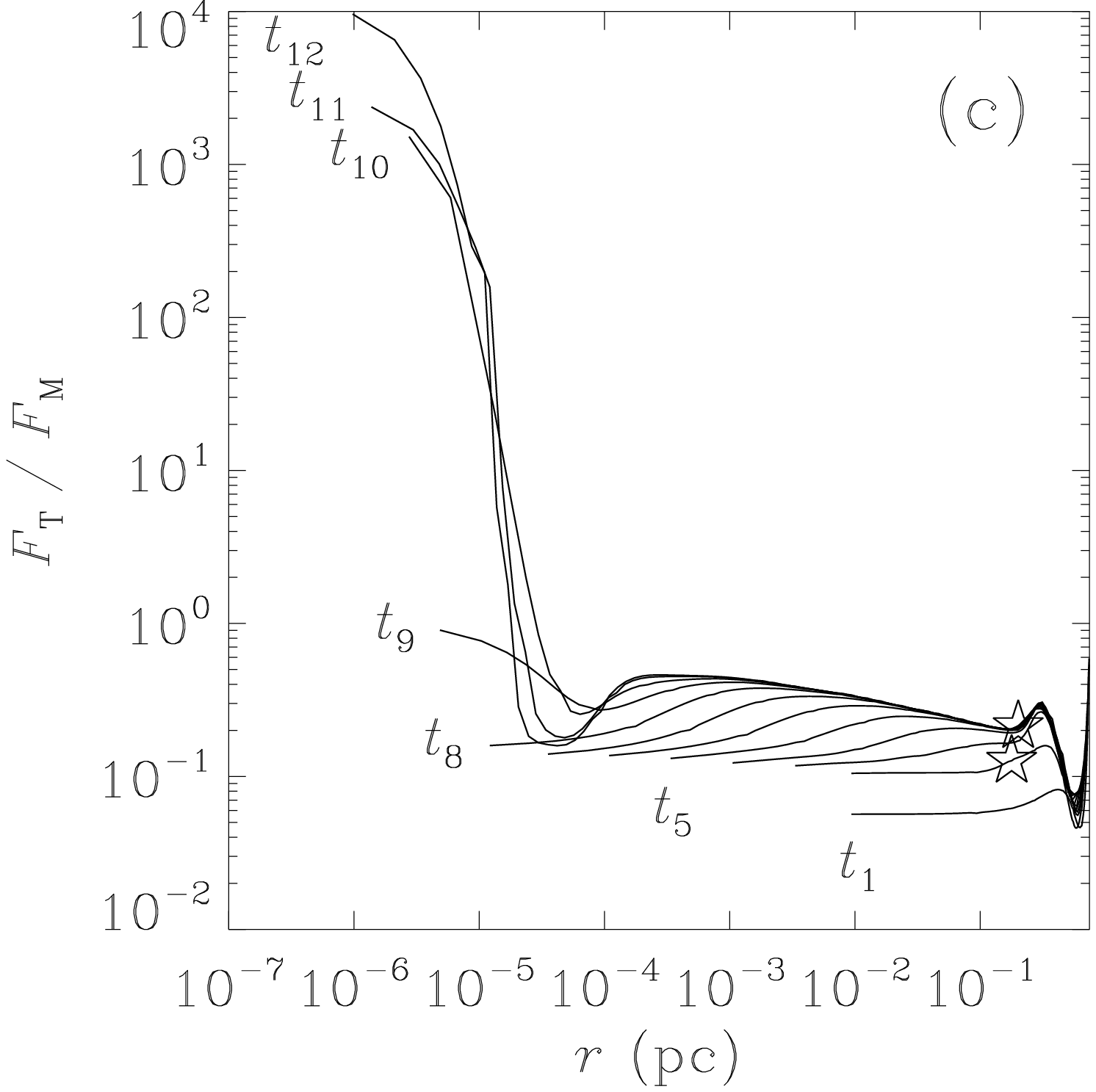}
\newline
\caption{Radial midplane profiles of (a) the ratio of thermal-pressure and gravitational forces, (b) the ratio of magnetic and gravitational forces, and (c) the ratio of thermal-pressure and magnetic forces at different times, as in Fig.~\ref{figure:mpax:denstemp}. The inner (outer) `star' on a curve, present only after a supercritical core forms, marks the initial (final) radius of the supercritical core.}
\label{figure:mpax:forces}
\end{figure}

Fig.~\ref{figure:mpax:forces} exhibits the relative magnitude of thermal-pressure forces, magnetic-pressure and tension forces, and gravitational forces during the evolution of the forming protostar. Fig.~\ref{figure:mpax:forces}a shows the ratio of thermal-pressure and gravitational forces. The gravitational force dominates the thermal-pressure force everywhere (since the cloud is thermally supercritical), except in the hydrostatic core ($r\lesssim 2~{\rm AU}$). Just outside the hydrostatic core, there is an abrupt increase in the thermal-pressure force which is ultimately responsible for the formation of a thermal shock there (see Section \ref{section:velocities}). At time $t_{10}$, the ratio of the thermal-pressure and gravitational forces inside the hydrostatic core is greater than unity. This is because the collapsing core has overshot its equilibrium, which results in radial (spherical) pulsations (see Section \ref{section:velocities}).

Fig.~\ref{figure:mpax:forces}b exhibits the ratio of magnetic and gravitational forces. The magnetic force is an appreciable fraction of the gravitational force everywhere except inside the hydrostatic core (where effective Ohmic dissipation is at work --- see Section \ref{section:adod}). This is the case even inside the dynamically contracting supercritical core. Hence, {\em the dynamical contraction of the magnetically supercritical core is significantly slower than free fall}. The magnetic force has a local maximum with respect to the gravitational force at $r \approx 10~{\rm AU}$, just outside the location of the maximum of the magnetic field seen in Fig.~\ref{figure:mpax:bfield}a. There the abrupt increase in the magnetic force by a factor $\gtrsim 2.5$ represents a magnetic wall and results in the formation of a magnetic shock (see Section \ref{section:velocities}).

In order to determine which agent is mainly responsible for diluting the effects of gravity at any given radius, we also give in Fig.~\ref{figure:mpax:forces}c the ratio of thermal-pressure and magnetic forces. The magnetic force is greater than the thermal-pressure force (and hence provides the main opposition to gravity) everywhere except in the hydrostatic core, where the thermal-pressure force dominates the magnetic force by a few orders of magnitude. There is a local minimum in the thermal-to-magnetic force ratio outside the hydrostatic core, where the magnetic field strength increases briefly.

\subsubsection{Velocities}\label{section:velocities}

\begin{figure}
\center
\includegraphics[width=2.8in]{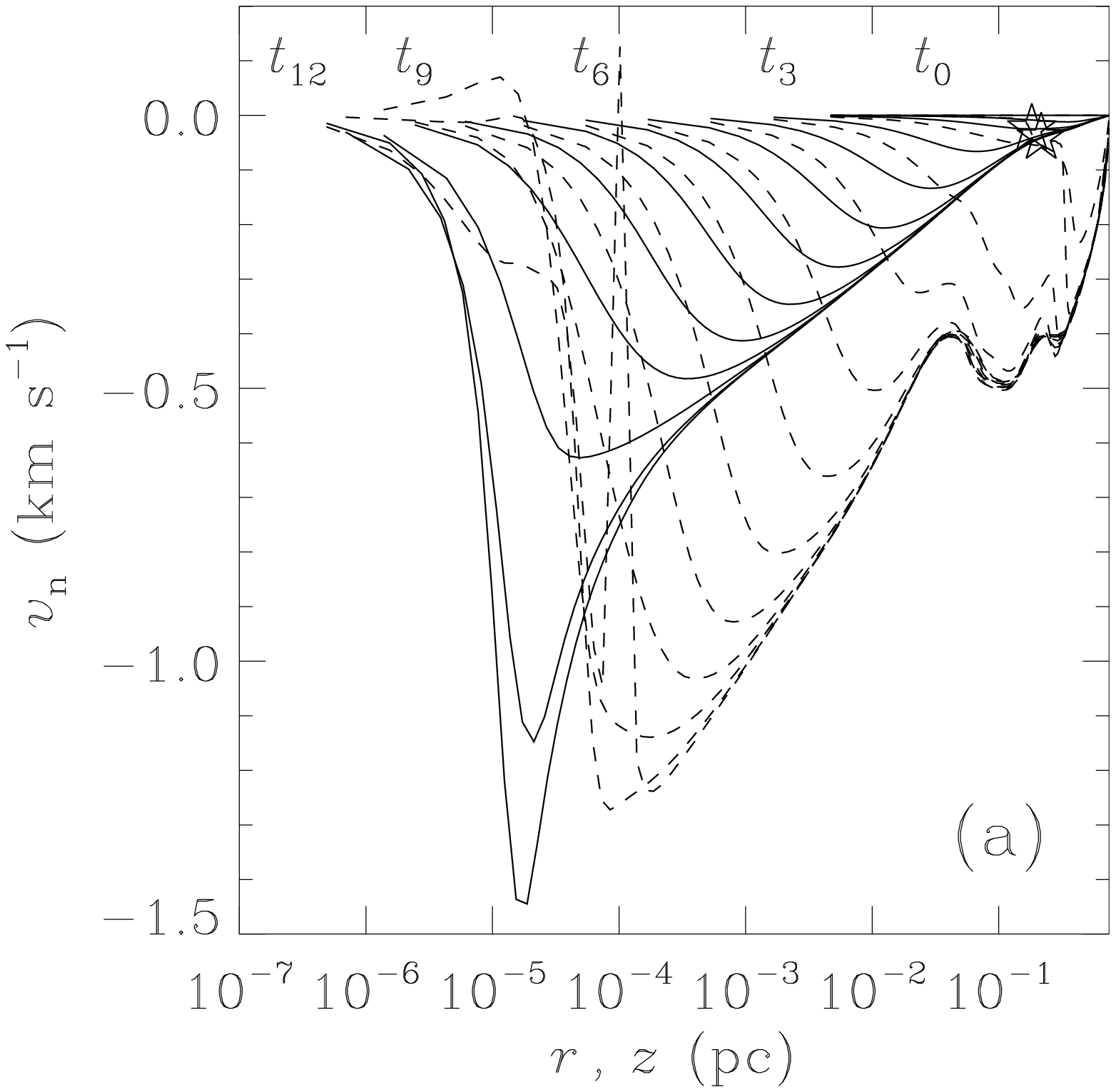}
\newline
\includegraphics[width=2.8in]{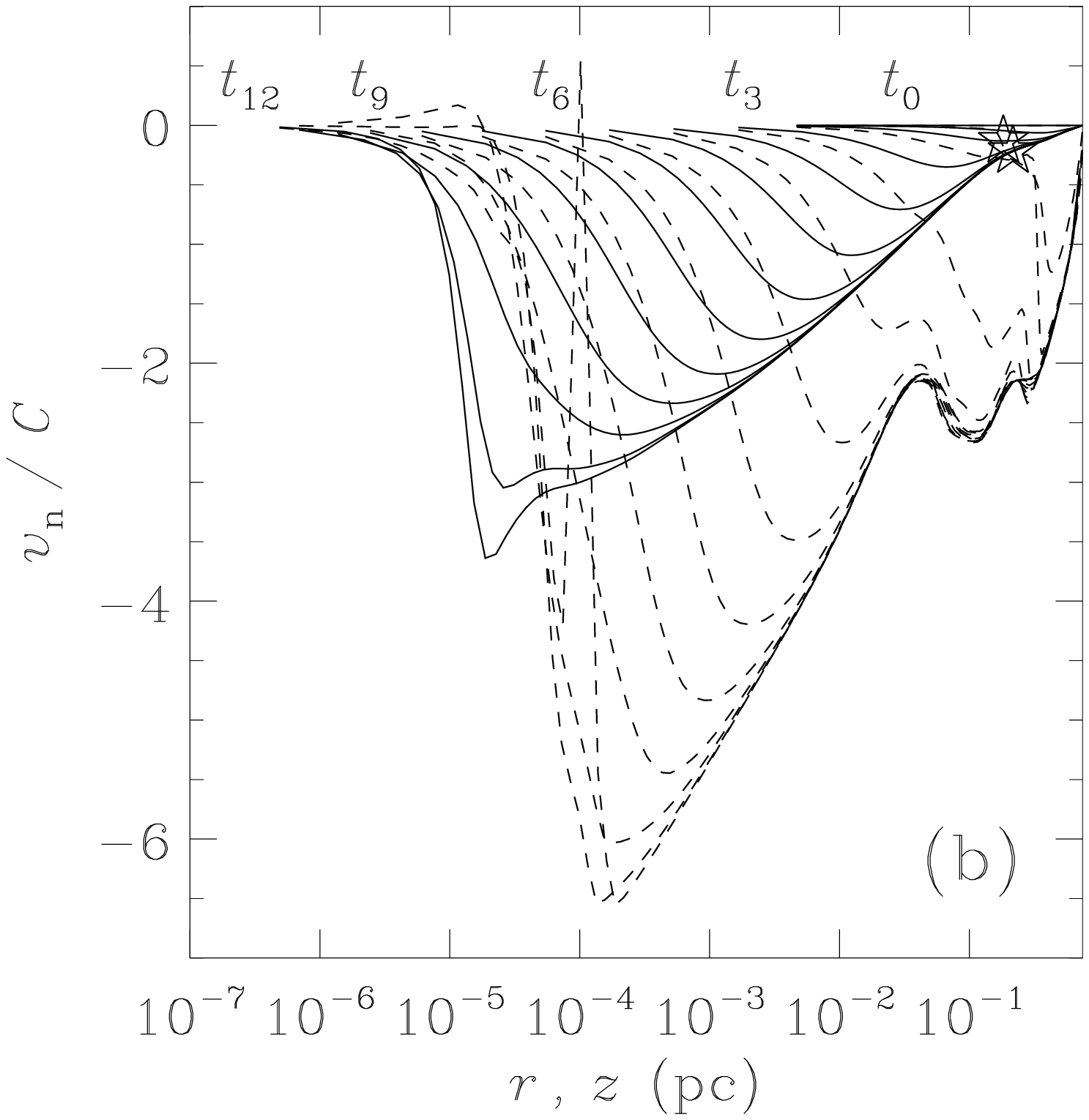}
\newline
\caption{Radial midplane (solid line) and vertical symmetry-axis (dashed line) profiles of the velocity of the neutrals (a) in ${\rm km~s}^{-1}$ and (b) in units of the instantaneous {\em local} sound speed, $C$. Different curves correspond to different times, as in Fig.~\ref{figure:mpax:denstemp}. The inner (outer) `star' on a curve, present only after a supercritical core forms, marks the initial (final) radius of the supercritical core.}
\label{figure:mpax:nzvelocity}
\end{figure}

The velocity of the neutrals is given in Fig.~\ref{figure:mpax:nzvelocity} (a) in ${\rm km~s}^{-1}$ and (b) in units of the instantaneous {\em local} isothermal sound speed, $C$. For reference, the isothermal sound speed at a temperature of $10~{\rm K}$ is $0.188~{\rm km~s}^{-1}$. The ambipolar-diffusion--controlled, quasistatic phase is marked by small radial velocities ($\simeq 0.01-0.03~{\rm km~s}^{-1}$). Once a magnetically and thermally supercritical core forms, dynamical contraction ensues inside the core and the radial velocity increases to $\simeq 0.5C$ -- $1.0C$. It is noteworthy that the radial velocity is $\approx 0.1~{\rm km~s}^{-1}$ at $r\sim 0.1~{\rm pc}$, in excellent agreement with the observed infall motions of L1544 by \citet{tmmcbb98}. The predicted radial velocities also compare well with the observed infall motions near the starless molecular cloud core MC27 in Taurus, where speeds of $0.2-0.3~{\rm km~s}^{-1}$ have been detected at $r\simeq 2000-3000~{\rm AU}$ ($\sim 10^{-2}~{\rm pc}$) \citep{omf99}. The radial velocities outside the supercritical core, where the gas is magnetically supported, remain subsonic by a factor of a few.

At approximately the same time that isothermality breaks down in the supercritical core, the radial velocities become supersonic. The mass accumulating in the hydrostatic core causes an accelerated infall and the radial velocity approaches and exceeds $\sim 1~{\rm km~s}^{-1}$. A shock forms near the boundary of the hydrostatic core due to rapid increases in the local magnetic and thermal-pressure forces, and the velocity rapidly falls to zero toward the centre of the hydrostatic core. The large vertical velocities throughout the evolution result from the cloud collapsing along field lines and constantly responding to changes in the thermal-pressure gradient along magnetic field lines. At the boundary of the hydrostatic core, shocks appear along the (vertical) symmetry axis. Vertical velocities become positive after the core overshoots hydrostatic equilibrium, rebounds, and oscillates. Radial velocities remain negative, but form a thermal and magnetic shock outside the hydrostatic core.

The parameter study by \citet{tm07c} found that, if the onset of adiabaticity is delayed beyond its oft-assumed critical density $10^{11}~{\rm cm}^{-3}$, as is naturally the case in the RMHD simulation presented here, the maximum infall velocity of both the neutrals and magnetic field lines increases somewhat as the density at which adiabaticity sets in increases. This is because the delay of the onset of adiabaticity implies accelerated infall for longer times. Using the results from a simulation in which isothermality was assumed to always hold, \citet{tm07c} found that the longer the gas remains isothermal, the weaker the magnetic shock is. In other words, the enhanced `pile-up' of matter and magnetic flux outside the hydrostatic core becomes increasingly absent. Given that the results of our RMHD simulation must necessarily be bracketed by those in which isothermal or adiabatic equations of state are assumed, the fact that our magnetic wall is more pronounced than found in their isothermal control run, yet less pronounced than found in their adiabatic run, is not surprising.

\begin{figure*}
\center
\includegraphics[width=6in]{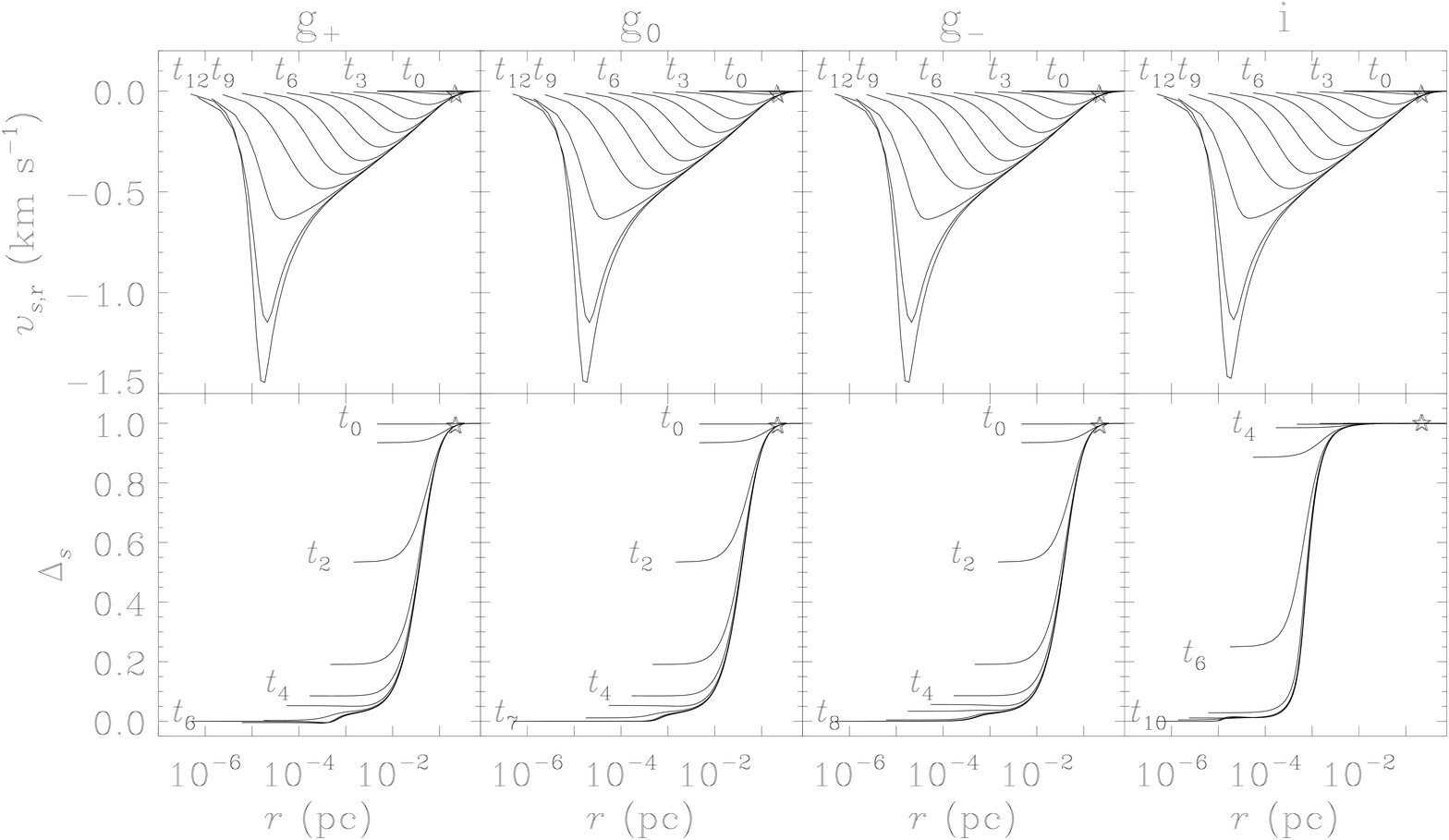}
\caption{Radial midplane profiles of (top row) velocities and (bottom row) attachment parameters for (left to right) positively-charged grains, neutral grains, negatively-charged grains, and ions. Different curves correspond to different times, as in Fig.~\ref{figure:mpax:denstemp}. The inner (outer) `star' on a curve, present only after a supercritical core forms, marks the initial (final) radius of the supercritical core.}
\label{figure:mpax:spvelocity}
\end{figure*}

Fig.~\ref{figure:mpax:spvelocity} shows radial profiles of the velocities and attachment parameters for the grains and the ions. (The electrons remain well-coupled to the magnetic field lines throughout the evolution.) The attachment parameter for species $s$ is defined by
\begin{equation}
\Delta_s \equiv \frac{v_{s,r} - v_{{\rm n},r}}{v_{{\rm f},r} - v_{{\rm n},r}}
\end{equation}
and quantifies the degree to which species $s$ is coupled to the magnetic field both directly (by gyrations of individual particles about field lines) and indirectly (through electrostatic attraction by oppositely-charged particles that are at least partially directly attached to the field lines). If $\Delta_s\approx 1$, then species $s$ is attached to the field lines, whereas if $\Delta_s\ll 1$, then species $s$ is detached from the field lines and its motion follows that of the neutrals. 

Outside the magnetically supercritical core, all charged species and the neutral grains remain well attached to the magnetic field lines. It is only inside the supercritical core that one after another the charged species begin to detach from the field lines. The negatively-charged grains are detached from the field lines by the time $t_6$, the neutral grains by $t_7$, and the positively-charged grains by $t_8$. The ions do not begin to detach from the field lines until $t_4$, well into the dynamical contraction phase, and do not fully detach until $t_{10}$. Indeed, at late times, the radial velocity profiles of these species shown in Fig. \ref{figure:mpax:spvelocity} appear identical to those of the neutrals in Fig. \ref{figure:mpax:nzvelocity}a.

A closer inspection of the attachment parameter profiles reveals more detailed and interesting behaviour than one would expect based solely on magnetic and collisional forces. These parameters do not simply asymptote directly to zero, but rather asymptote first to a small but nonzero value before finally settling at zero. For example, $\Delta_{\rm g_-}$ asymptotes to $\approx 0.05$ before falling to zero. Thus, negatively-charged grains remain mildly attached to the field lines even when simpler estimates (e.g., based on the criterion $\omega_{\rm g}\tau_{\rm gn}\ll 1$) would suggest otherwise. This is a result of electrostatic attraction between negatively-charged grains and electron-shielded ions, which have yet to fully decouple from the magnetic field. Neutral grains also couple to the magnetic field by inelastic charge-capture processes. Slightly more complex behaviour can be seen for the positively-charged grains, which are repelled from the field lines by electron-shielded ions. Even electrostatic attraction between ions and (fully-attached) electrons keeps the ions very mildly attached for $t_8$ -- $t_{10}$, though this effect is much subtler than for the grains because of the extremely low electron abundance at these high densities.

\begin{figure*}
\center
\includegraphics[width=2.8in]{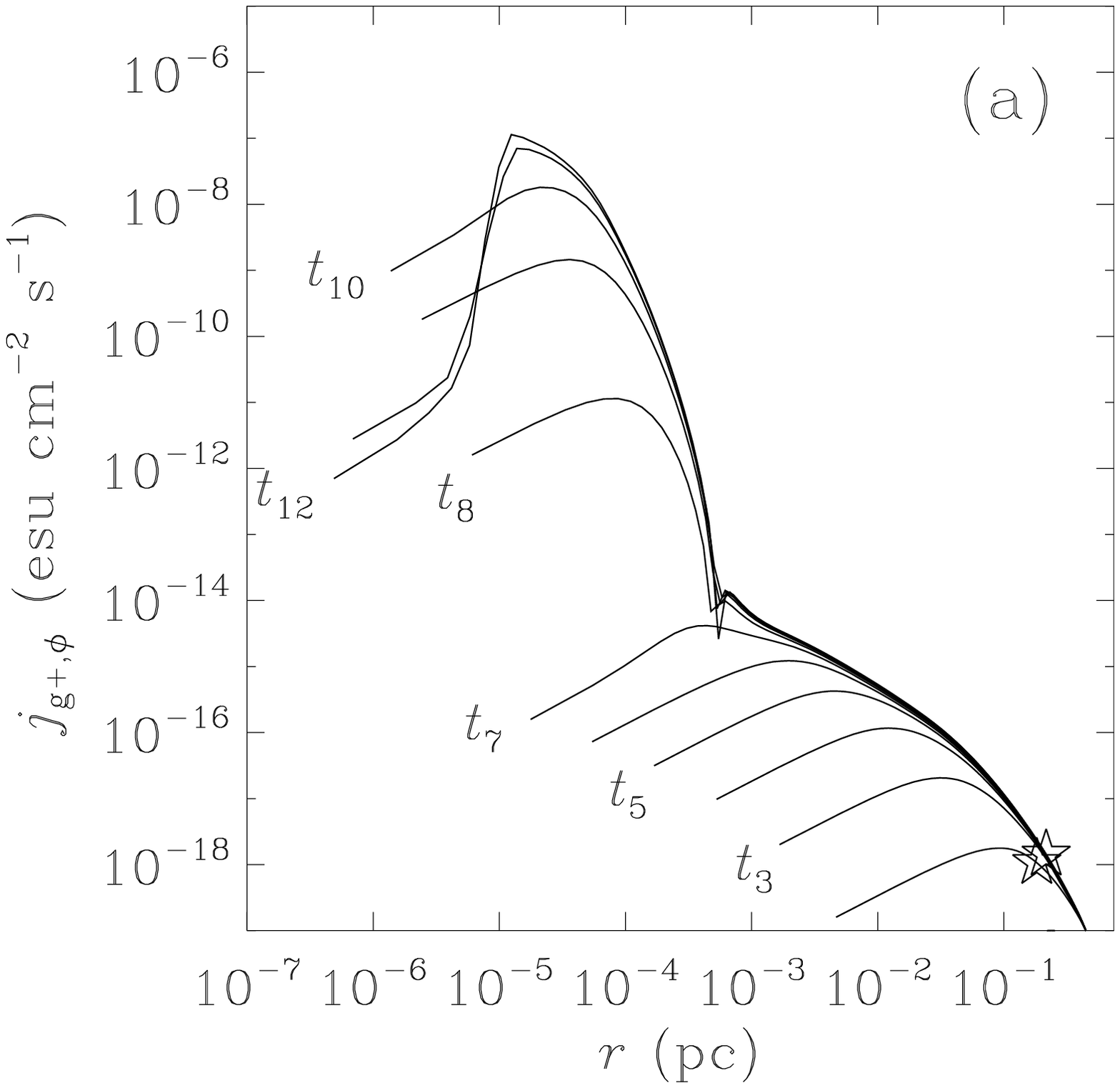}
\includegraphics[width=2.8in]{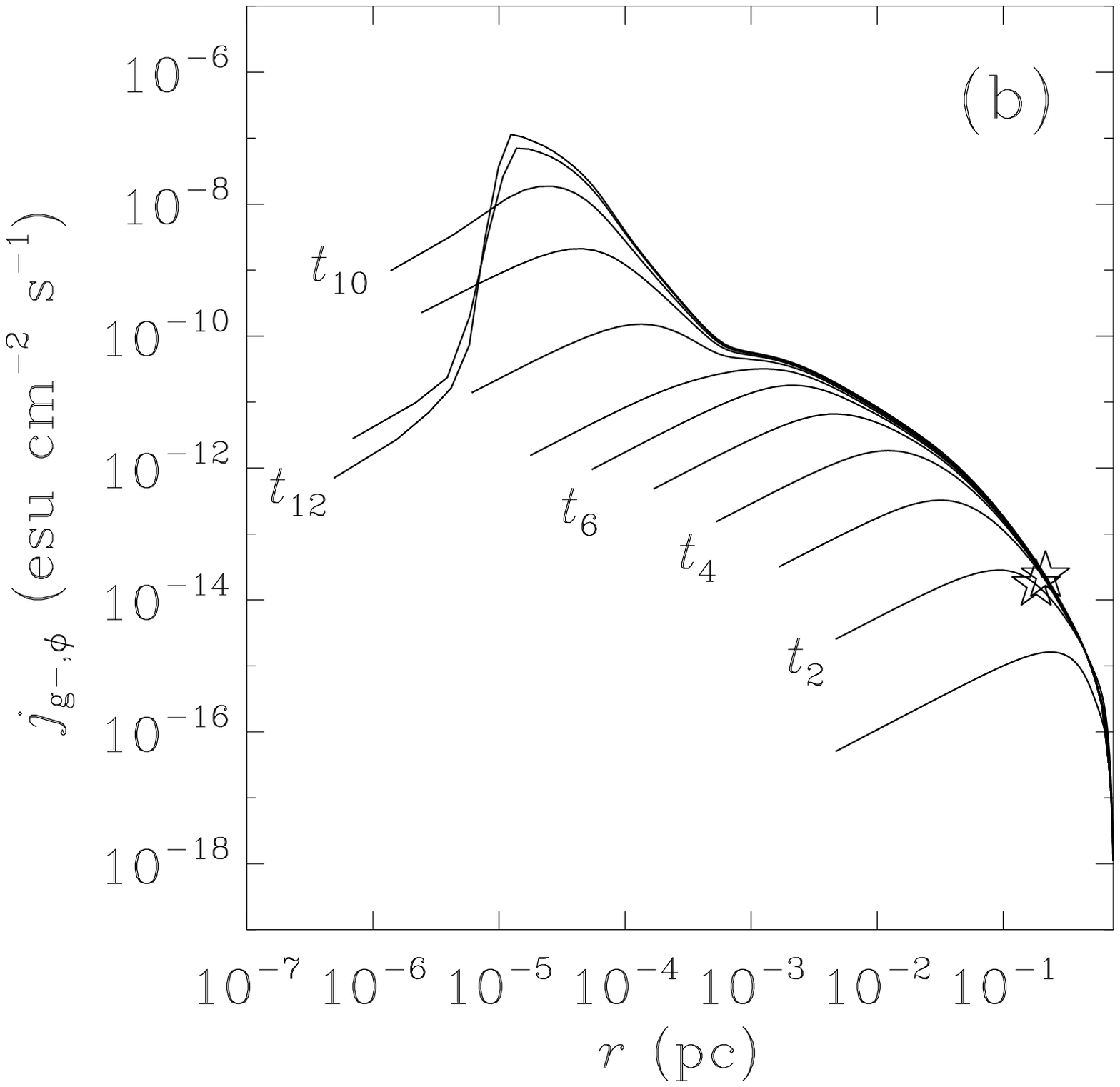}
\newline
\includegraphics[width=2.8in]{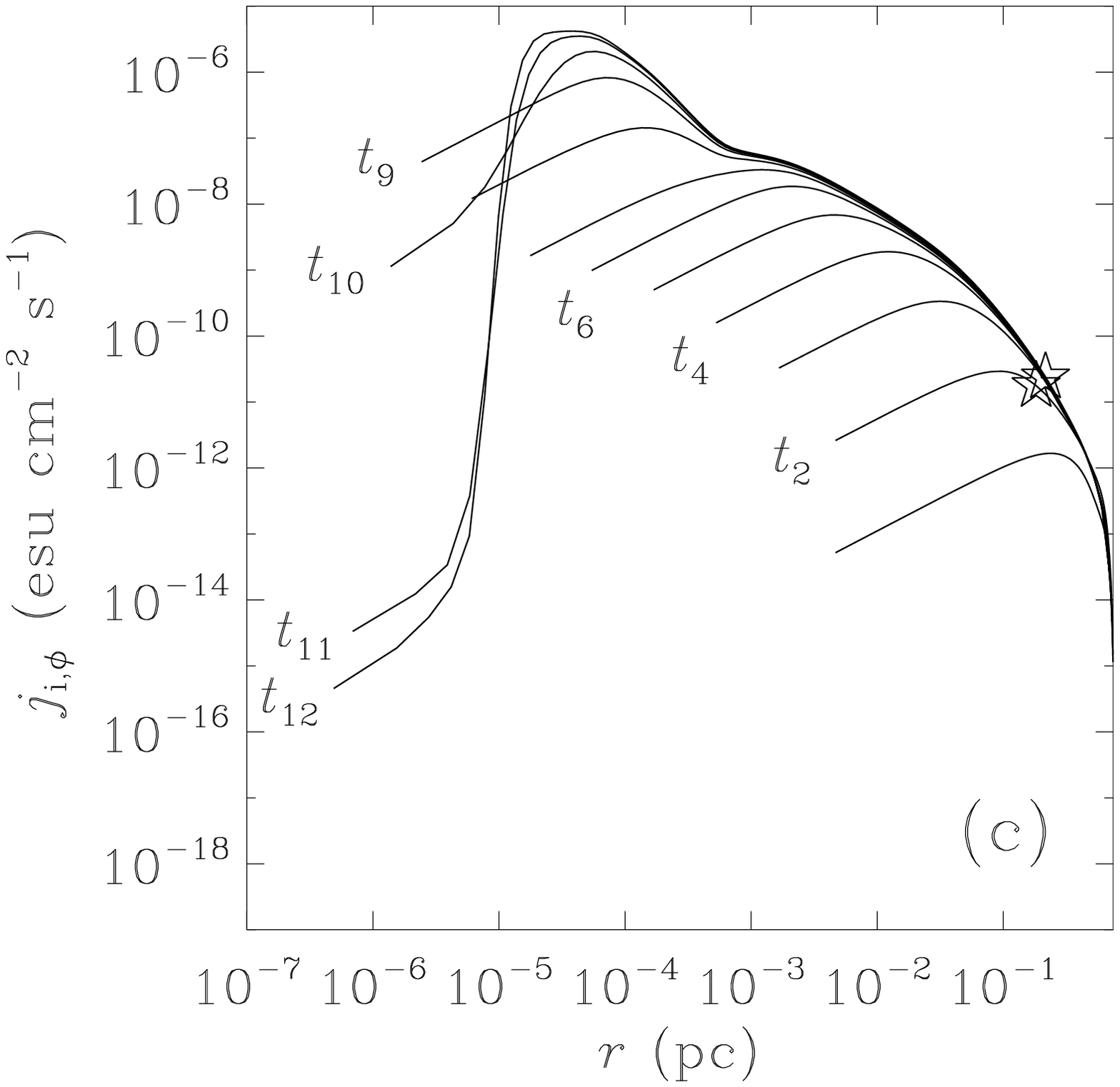}
\includegraphics[width=2.8in]{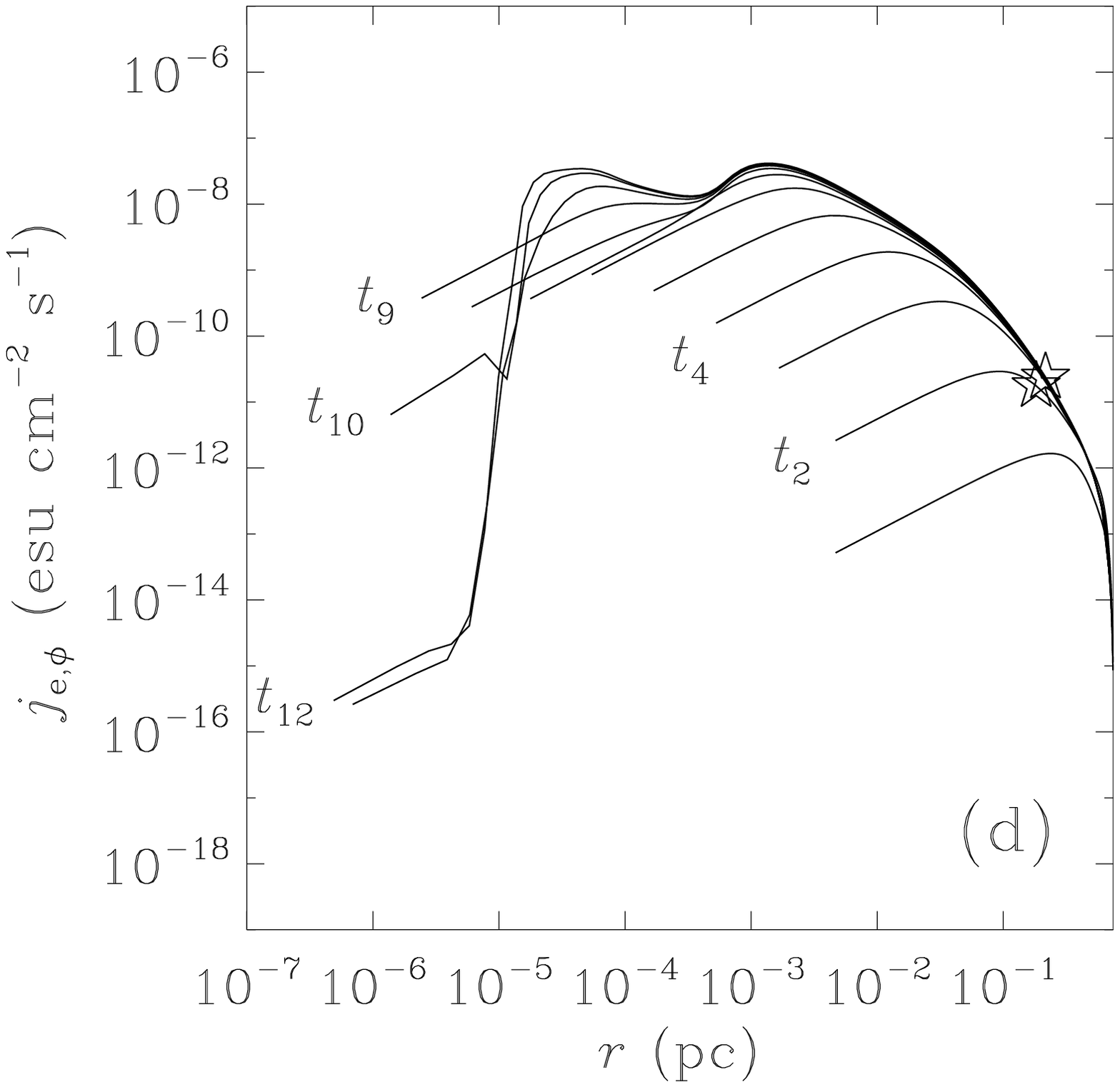}
\newline
\caption{Radial profiles of the {\em magnitude} of the electric current density in the $\phi$-direction carried by (a) the positively-charged grains, (b) the negatively-charged grains, (c) the ions, and (d) the electrons. Different curves correspond to different times, as in Fig.~\ref{figure:mpax:denstemp}. The inner (outer) `star' on a curve, present only after a supercritical core forms, marks the initial (final) radius of the supercritical core.}
\label{figure:mpax:spcurrent}
\end{figure*}

\subsubsection{Contribution to Electric Current Density by Different Species}\label{section:current}

\begin{figure}
\center
\includegraphics[width=2.8in]{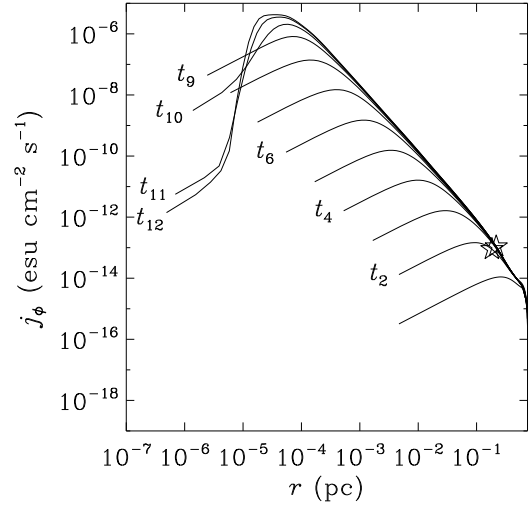}
\caption{Radial profiles of the magnitude of the total electric current density at different times, as in Fig.~\ref{figure:mpax:denstemp}. The inner (outer) `star' on a curve, present only after a supercritical core forms, marks the initial (final) radius of the supercritical core.}
\label{figure:mpax:current}
\end{figure}

Fig.~\ref{figure:mpax:spcurrent} exhibits the contribution of different charged species to the total electric current density (in the midplane) as a function of radius at different times. The current density $j_{s,\phi}$ (in the midplane) carried by species $s$ may be obtained from equation (C1) of Paper I and written in terms of the total current density $j_\phi$ (shown in Fig.~\ref{figure:mpax:current}) and various components of the resistivity tensor:
\begin{equation}
j_{s,\phi} \equiv n_s q_s v_{s,\phi} = \bigl(\sigma_{\perp,s}\eta_\perp + \sigma_{{\rm H},s}\eta_{\rm H}\bigr) j_\phi .
\end{equation}
The current density (in the midplane) due to species $s$ is given by $j_{s,\phi} = n_s q_s v_{s,\phi}$. Absolute values have been taken; the cusps in the positive-grain curves are due to the current passing through zero as it changes sign. This occurs at the same time that the positive-grain conductivities change sign (see Figs.~\ref{figure:central:resist}c and \ref{figure:central:resist}d) and at the same spatial location that the positive-grain attachment parameter passes through zero and becomes slightly negative (see Fig.~\ref{figure:mpax:spvelocity}).

At early times the contributions of electrons and ions to the electric current density are comparable and are approximately three orders of magnitude greater than the contribution from negatively-charged grains. By $t_9$, the increasing disparity between electron and ion abundances (due to electron depletion onto grains) causes the ion current to become greater than the electron current by a factor $\approx 100$ (recall that, at this stage, $x_i/x_e\simeq 125$ --- see Fig.~\ref{figure:central:chem}b). Finally, in the innermost region, the current density decays sharply due to the increasing effect of Ohmic dissipation (see Fig.~\ref{figure:mpax:adod}). As discussed in Section \ref{section:central}, grains become the main current carriers at roughly the same time as Ohmic dissipation becomes important. Their contribution to the electric current density is more than three orders of magnitude greater than that of the ions and electrons.

\subsubsection{Ambipolar Diffusion and Ohmic Dissipation}\label{section:adod}

\begin{figure}
\center
\includegraphics[width=2.8in]{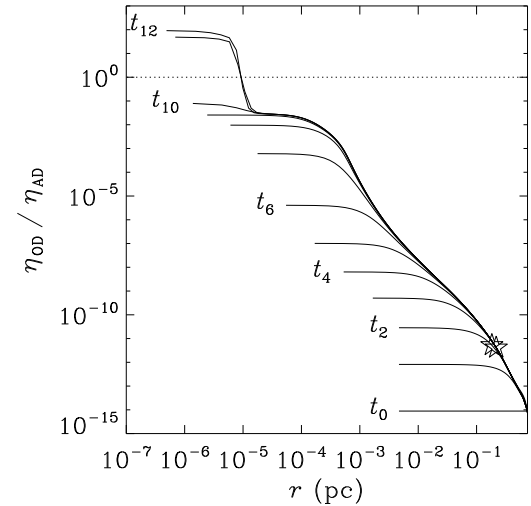}
\caption{Radial profiles of the ratio of Ohmic and ambipolar resistivities. The dotted line denotes the boundary when ambipolar diffusion and Ohmic dissipation become equally important magnetic diffusion mechanisms. Above (below) this line, Ohmic dissipation (ambipolar diffusion) dominates. Different curves correspond to different times, as in Fig.~\ref{figure:mpax:denstemp}.}
\label{figure:mpax:adod}
\end{figure}

The relative importance of ambipolar diffusion and Ohmic dissipation as agents of magnetic diffusion is demonstrated in Fig.~\ref{figure:mpax:adod}, which shows radial midplane profiles of the ratio of the ambipolar-diffusion and Ohmic-dissipation resistivities. At low densities and large radii (and height), ambipolar diffusion dominates Ohmic dissipation. The two processes become equally important only at a density $\simeq 7\times 10^{12}~{\rm cm}^{-3}$, while at higher densities Ohmic dissipation operates on a timescale shorter than that of ambipolar diffusion (see also Fig.~\ref{figure:central:resist}a.) The region in which this occurs coincides with the region in which the electric currents diminish ($r\lesssim 2~{\rm AU}$). 

Note that, by the time Ohmic dissipation becomes important, all species except the electrons have detached from the field lines. Hence, magnetic decoupling is essentially complete. Calculations that arrived at the conclusion that Ohmic dissipation is the cause of magnetic decoupling \citep[e.g.,][]{nu86a, nu86b} were based on a value of the e-H$_2$ cross section which is too large by a factor of about 100 \citep[see review by][\S~2.1]{mouschovias96}.

\subsubsection{Mass, Magnetic Flux, and Mass Infall Rate}\label{section:massflux}

\begin{figure*}
\center
\includegraphics[width=2.8in]{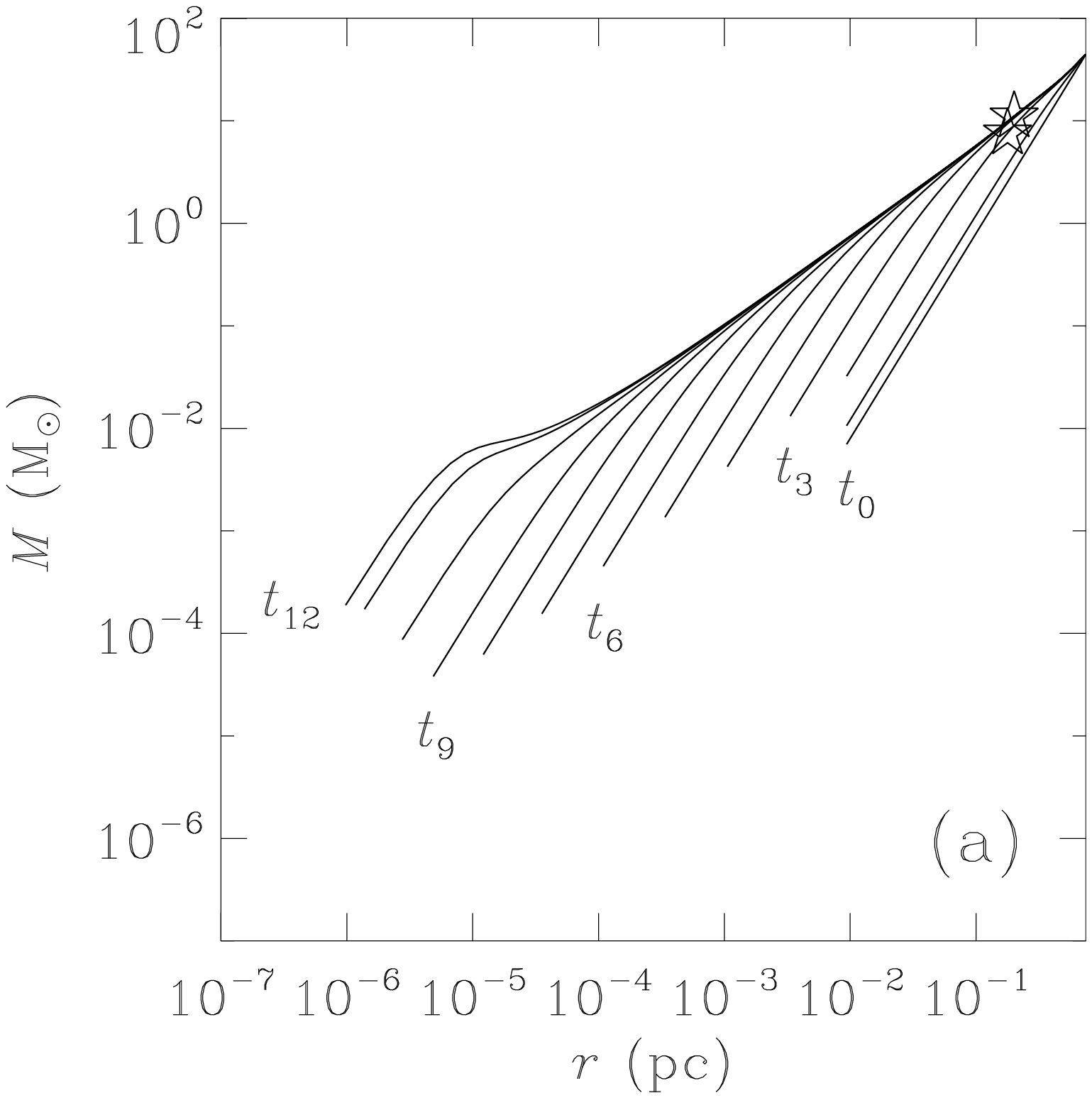}
\includegraphics[width=2.8in]{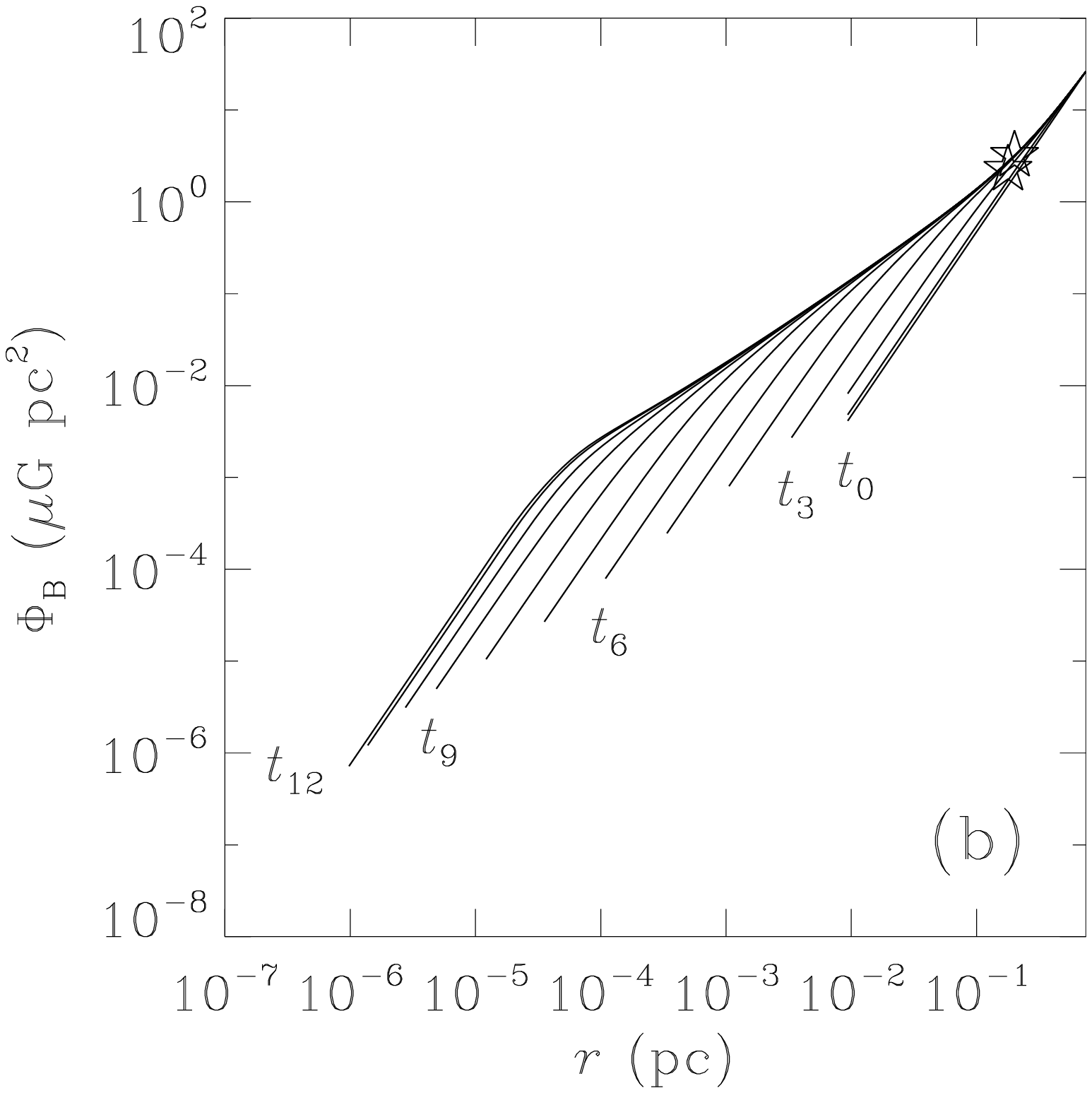}
\newline
\includegraphics[width=2.8in]{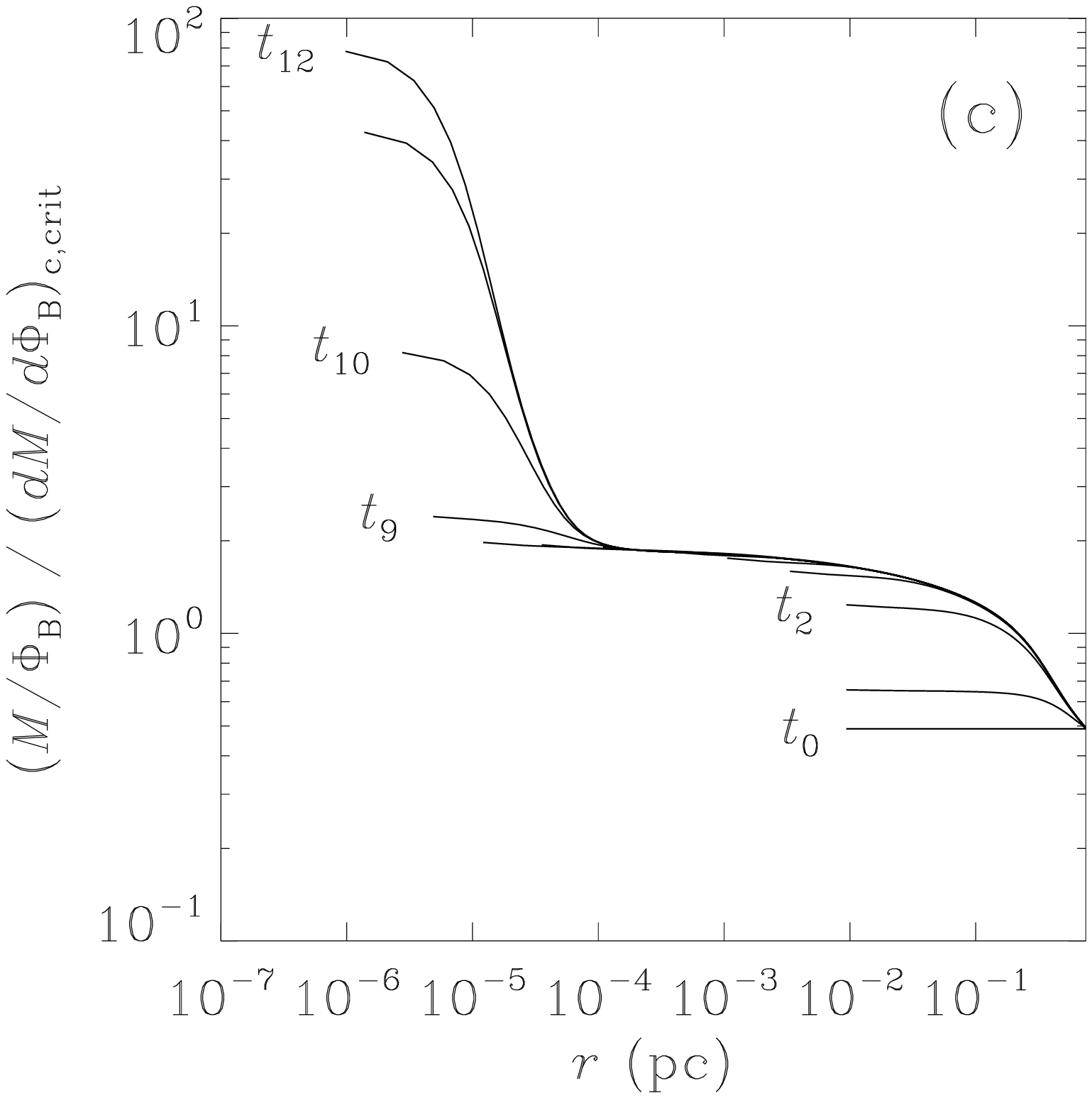}
\includegraphics[width=2.8in]{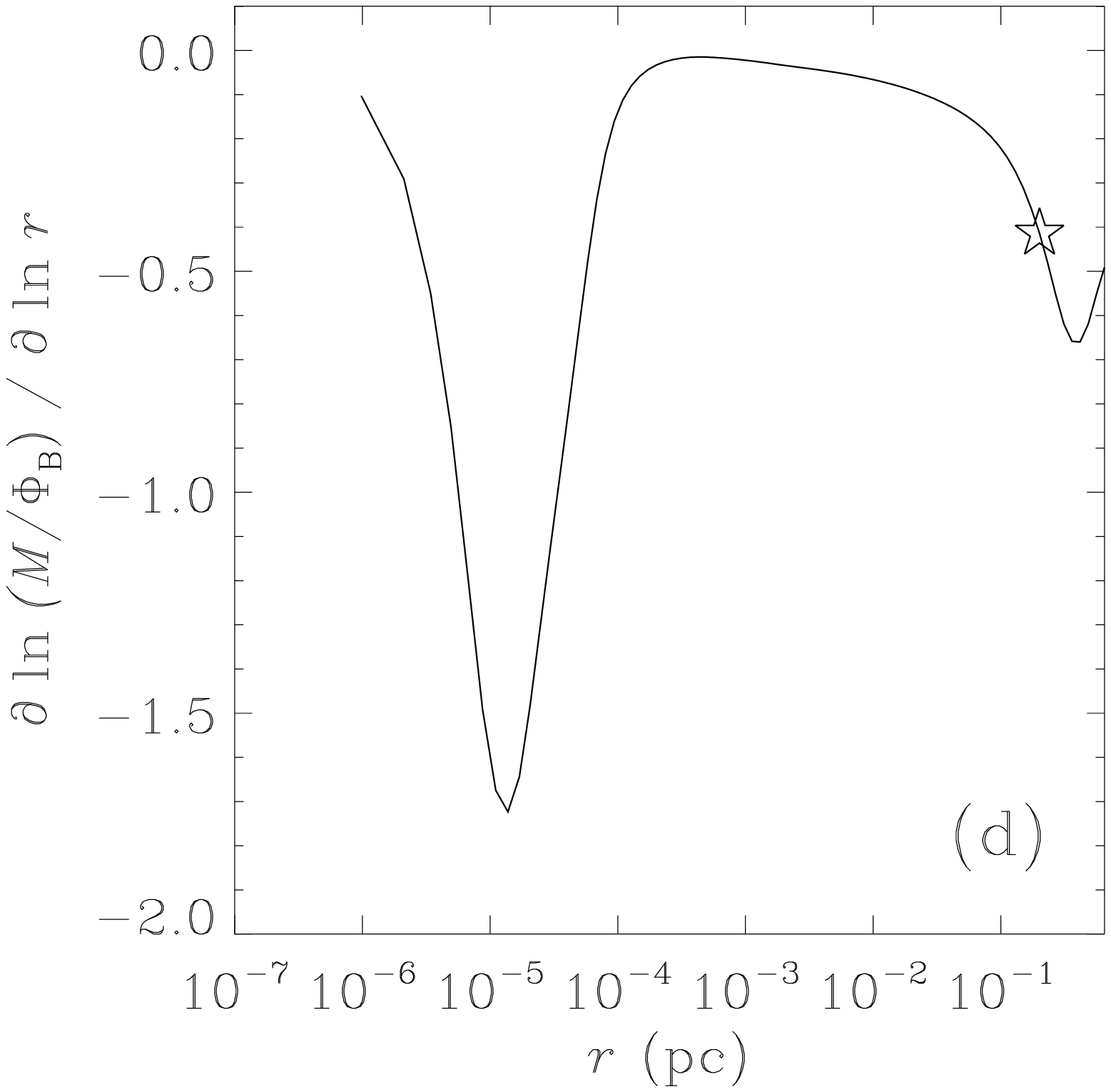}
\newline
\caption{Radial profiles of (a) the cumulative mass within a radius $r$, (b) the cumulative magnetic flux within a radius $r$, and (c) the mass-to-flux ratio (normalised to the central critical value for collapse) at different times, as in Fig.~\ref{figure:mpax:denstemp}. (d) Radial derivative of the mass-to-flux ratio, $\partial\ln (M/\Phi_{\rm B})/\partial\ln r$, at $t_{12}$. The inner (outer) `star' on a curve, present only after a supercritical core forms, marks the initial (final) radius of the supercritical core.}
\label{figure:mpax:massflux}
\end{figure*}

In Fig.~\ref{figure:mpax:massflux}a we show radial profiles of the cumulative mass within a radius $r$. Within the thermally and magnetically supercritical core, there are three different regions that exhibit characteristically different (logarithmic) slopes of the mass spatial profile: (1) Inside the hydrostatic core, both the density and the temperature are essentially uniform (see Figs.~\ref{figure:mpax:denstemp}a and \ref{figure:mpax:denstemp}c), because the size of this region is smaller than the instantaneous value of the thermal critical lengthscale $\lambda_{\rm T,cr}(t)$ and therefore thermal-pressure forces rapidly smooth out any nonuniformity in these quantities. Consequently, $\partial\ln n_{\rm n}/\partial\ln r\simeq 0$, from which it follows that $\partial\ln M/\partial\ln r \simeq 2$ (see Fig.~\ref{figure:mpax:massflux}a, region $r \lesssim 10^{-5}~{\rm pc}$). (2) Just outside the boundary of the hydrostatic core, the infall velocity has its maximum value (see Fig.~\ref{figure:mpax:nzvelocity}), thereby tending to empty this region of mass and to cause a steepening of the density profile. Since relatively little mass exists in this region, the mass profile flattens (see Fig.~\ref{figure:mpax:massflux}a, region $10^{-5}~{\rm pc} \lesssim r \lesssim 10^{-4}~{\rm pc}$). (3) Farther out in the supercritical core, where the geometry is disclike, we have that $M = \pi r^2 \Sigma \propto r^2 \rho^{\gamma_{\rm eff}/2}$, from which it follows that $\partial\ln M/\partial\ln r \simeq 2 + (\gamma_{\rm eff}/2)(\partial\ln n_{\rm n}/\partial\ln r)\approx 0.9$. By the end of the run, $\simeq 12~{\rm M}_\odot$ has accumulated within the magnetically supercritical core and $\simeq 0.006~{\rm M}_\odot$ within the hydrostatic core. {\em Hence, the often made statement that ambipolar diffusion can only form low-mass cores (and low-mass stars) is not correct.}

Similar behaviour is evident in Fig.~\ref{figure:mpax:massflux}b, which shows radial profiles of the cumulative magnetic flux within a radius $r$. Note, however, that the clear break in the mass-radius profile at the boundary of the hydrostatic core is hardly evident in the magnetic-flux--radius profile. The magnetic flux does not make the full transition to spherical geometry like the mass does, since the magnetic field is decoupled from every species except the tenuous electron fluid (see Section \ref{section:velocities}). The magnetic flux threading the hydrostatic core is $\approx 5\times 10^{-5}~\mu{\rm G~pc}^2\approx 4.8\times 10^{18}~{\rm Wb}$. It is intriguing that {\em this is close to the magnetic flux of a typical Ap star} such as $\theta$ Aurigae, which has a radius $4.5~{\rm R}_\odot$ and magnetic field strength $B\sim 1~{\rm kG}$, giving a magnetic flux $\sim 3\times 10^{18}~{\rm Wb}$ \citep{vrha84}. Ap stars are believed to avoid the convective stage as protostars and so retain the magnetic flux leftover from their formation. Coupled with $\theta$ Aurigae's young age ($\sim 200~{\rm Myr}$), this suggests that its $\sim 1~{\rm kG}$ magnetic field is a fossil field.

The radial profile of the mass-to-flux ratio (normalised to the central critical value for collapse) is given in Fig.~\ref{figure:mpax:massflux}c. A visual examination of the profile at the last time output suggests that the curve exhibits a three-slope power law with a flat inner region, which corresponds to the flat-density uniform-magnetic-field core. However, a plot of the profile of the slope itself shows that there is no region in which it is constant (see Fig.~\ref{figure:mpax:massflux}d). Conclusions based on visual inspections, especially of log-log plots, can be very misleading. Relatively abrupt changes in the slope separate regions governed by different physics. The first steepening (at small $r$) occurs just outside the hydrostatic core and marks the transition from the magnetically-decoupled hydrostatic core to the magnetically supercritical core, in which the magnetic flux is almost frozen in the matter. From radii $\sim 10^{-4}~{\rm pc}$ to a few $\times 10^{-2}~{\rm pc}$, the slope varies relatively slowly. In this region, the ions are still attached to the field lines and the mass-to-flux ratio hardly changes in time. Beyond the supercritical core, where magnetic forces continue to support the cloud envelope, the slope steepens again.

\begin{figure}
\center
\includegraphics[width=2.8in]{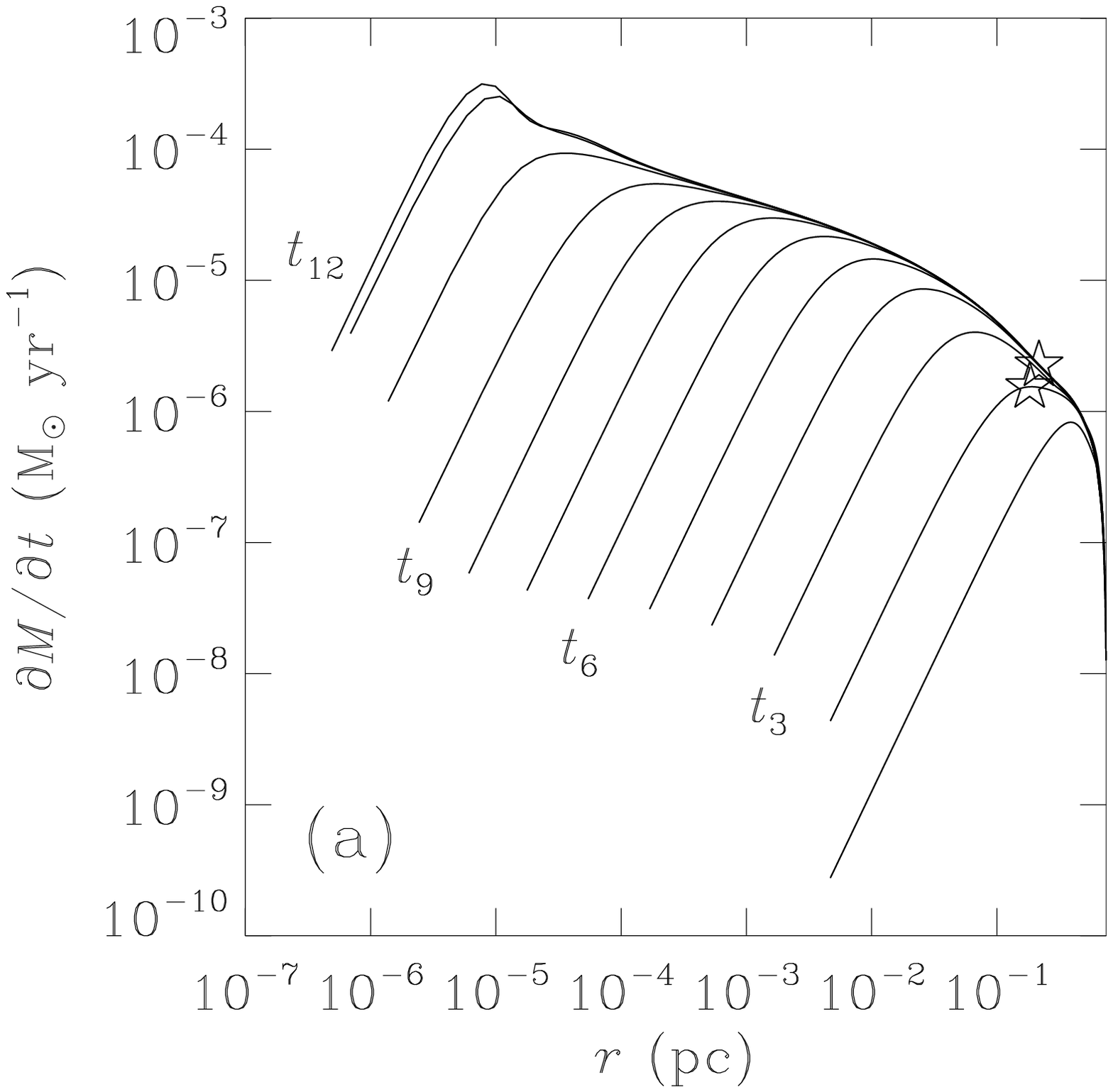}
\newline
\includegraphics[width=2.8in]{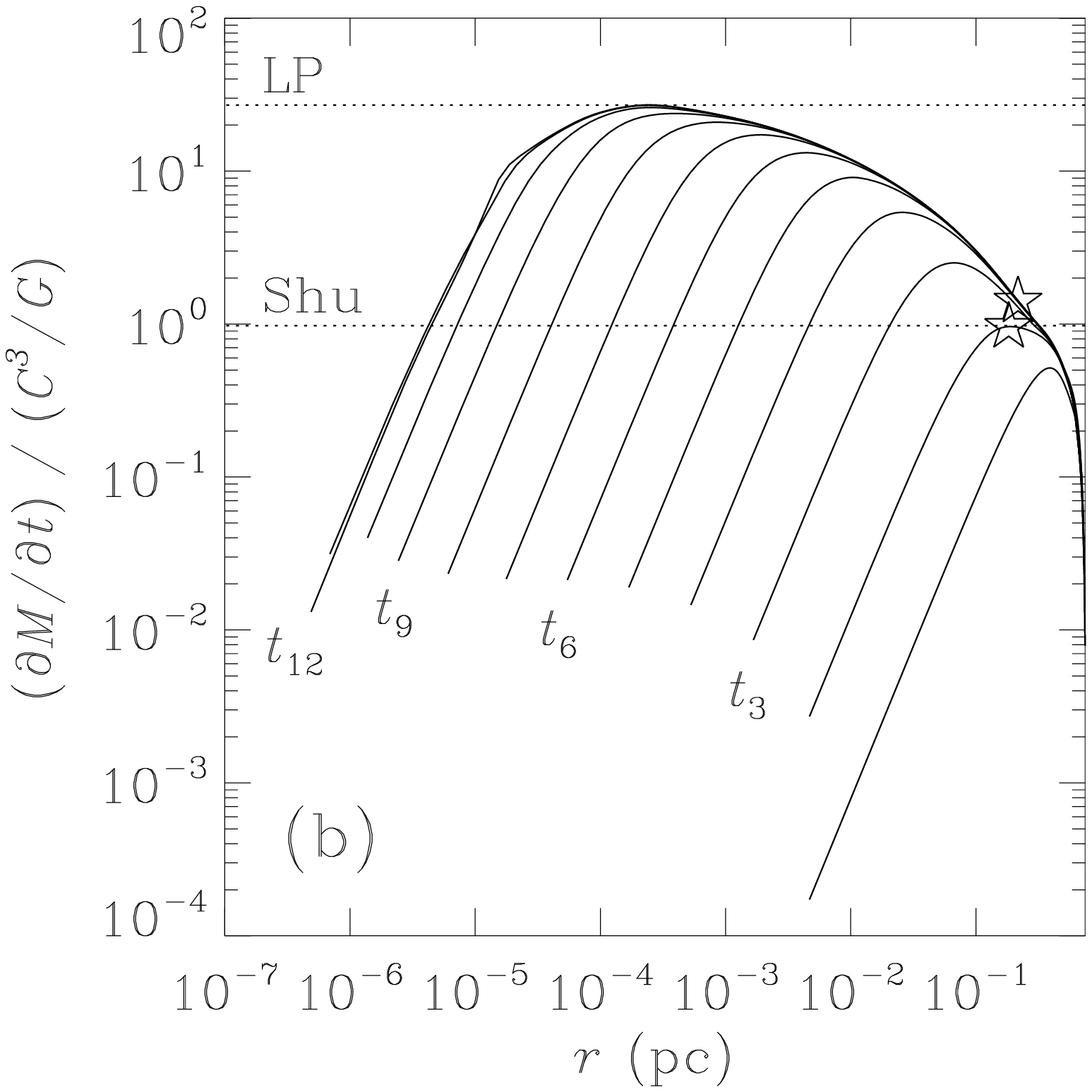}
\newline
\caption{Radial profiles of the mass infall rate through any (cylindrical) radius $r$ in units of (a) ${\rm M}_\odot~{\rm yr}^{-1}$ and (b) the instantaneous {\em local} value of $C^{3}/G$. Different curves correspond to different times, as in Fig.~\ref{figure:mpax:denstemp}. The inner (outer) `star' on a radial profile curve, present only after a supercritical core forms, marks the initial (final) radius of the supercritical core. The horizontal dotted lines denote the constant mass infall rates predicted by the \citet{shu77} and Larson-Penston (1969) solutions.}
\label{figure:mpax:mdot}
\end{figure}

The neutral density (Fig.~\ref{figure:mpax:denstemp}a) and velocity (Fig.~\ref{figure:mpax:nzvelocity}) profiles are used to calculate the mass infall rate, $\partial M/\partial t$, through any (cylindrical) radius $r$. The infall rate is given in Fig.~\ref{figure:mpax:mdot}a in units of ${\rm M}_\odot~{\rm yr}^{-1}$. It varies considerably both spatially and temporally, a behaviour also noted in magnetic calculations by, e.g., \citet{mm92a,mm92b}, \citet{cm94}, \citet{basu97}, and \citet{dm01}, and in nonmagnetic calculations by \citet{fc93}. This behaviour contrasts sharply with that of the singular isothermal sphere model, which would predict a {\em constant} and {\em uniform} mass infall rate equal to $0.975~C^3/G$ \citep{shu77}. (The inclusion of magnetic fields to this model, which results in isothermal toroids, gives a greater but still time-independent and spatially uniform infall rate --- see \citealt{ls96}.)

In order to compare the mass infall rate more quantitatively with the predictions by the various self-similar solutions mentioned earlier, we also plot in Fig.~\ref{figure:mpax:mdot}b the mass infall rate normalised to $C^{3}/G$, where $C$ is the instantaneous {\em local} isothermal sound speed. The horizontal dotted lines denote the constant mass infall rates predicted by the \citet{shu77} and Larson-Penston \citep{larson69,penston69} solutions. At the boundary of the magnetically supercritical core the mass infall rate is initially $1.54\times 10^{-6}~{\rm M}_\odot~{\rm yr}^{-1}$ ($=0.96~C^3/G$), increasing to $2.26\times 10^{-6}~{\rm M}_\odot~{\rm yr}^{-1}$ ($=1.41~C^3/G$) by the end of the run. The {\em maximum} infall rate is $3.15\times 10^{-4}~{\rm M}_\odot~{\rm yr}^{-1}$ (at $r = 1.60~{\rm AU}$), whereas the maximum value of $(\partial M/\partial t)/(C^3/G) = 27.08$ (at $r = 52.2~{\rm AU}$). That the early ambipolar-diffusion--controlled phase is characterised by a maximum mass infall rate comparable to the constant mass infall rate predicted by the \citet{shu77} solution intuitively makes sense; the \citet{shu77} solution was obtained by assuming quasi-static (i.e., negligible velocity) initial conditions, which is similar to the quasistatic (i.e., negligible acceleration) evolution of the early, ambipolar-diffusion--controlled phase that exists before the creation of a magnetically supercritical core. Likewise, the mass infall rate in the Larson-Penston (1969) solution was obtained by assuming highly dynamical initial conditions, and so it is not surprising that the maximum mass infall rate found in our simulation, which occurs once a `point mass' (the hydrostatic core) is formed in the central region, is bounded from above by the (spatially uniform) mass infall rate found in the Larson-Penston (1969) solution.

\subsection{Radiation Flux and Luminosity}\label{section:luminosity}

\begin{figure}
\center
\includegraphics[width=2.8in]{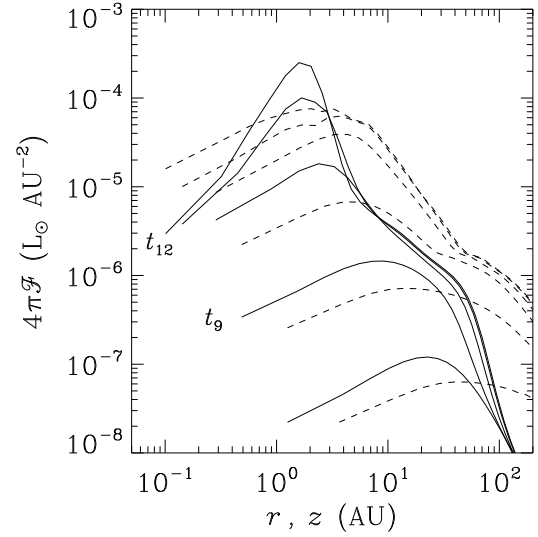}
\caption{Radial midplane (solid line) and vertical symmetry-axis (dashed line) profiles of the radiation flux at different times, as in Fig.~\ref{figure:mpax:denstemp}. To ease conversion to luminosity, we have multiplied the flux by $4\pi$ and have given it in units of ${\rm L}_\odot~{\rm AU}^{-2}$.}
\label{figure:mpax:luminosity}
\end{figure}

Under the FLD approximation, the radiation flux is given by
\begin{equation}
\bb{\mathcal{F}} = -\left(\frac{c\lambda_{\rm FLD}}{\chi_{\rm R}}\right)\del\mathcal{E} ,
\end{equation}
where $\lambda_{\rm FLD}$ is the flux limiter, $\chi_{\rm R}$ is the Rosseland mean extinction coefficient, and $\mathcal{E}$ is the (frequency-integrated) radiation energy density (see Paper I). In Fig.~\ref{figure:mpax:luminosity}, we give the $r$-component of the radiation flux along the midplane (solid line) and the $z$-component of the radiation flux along the symmetry axis (dashed line). To ease conversion to luminosity,
\begin{equation}
L(\mathcalligra{r}) = \mathcalligra{r}^2 \oint \bb{\mathcal{F}}(\mathcalligra{r},\theta)\bcdot\hat{\bb{\mathcalligra{r}}}~~d\Omega ,
\end{equation}
in the figure we have multiplied the flux by $4\pi$ and given it in units of ${\rm L}_\odot~{\rm AU}^{-2}$. (To obtain the luminosity at the hydrostatic core boundary in ${\rm L}_\odot$, one simply multiplies the value of the plotted quantity evaluated at $r\simeq 2~{\rm AU}$ by $\simeq 2^2$.) The radiation flux is substantially greater along the symmetry axis than along the midplane, because there is much less absorbing material in the former direction than in the latter. Near the boundary of the hydrostatic core the radial flux attains its maximum value $\simeq 2.5\times 10^{-4}~{\rm L}_\odot~{\rm AU}^{-2}$, corresponding to a luminosity $\sim10^{-3}~{\rm L}_\odot$.

\section{Summary and Discussion}\label{section:summary}

As shown by earlier calculations of our group, ambipolar diffusion in an initially magnetically subcritical cloud allows the neutrals to contract quasistatically due to their self-gravity through almost stationary magnetic field lines, redistributing mass in the central flux tubes of the cloud. Once the central mass-to-flux ratio exceeds the critical value for collapse (at a central neutral number density $\simeq 10^4~{\rm cm}^{-3}$ under the conditions investigated here), a core begins to contract dynamically and is referred to as a supercritical core. Its contraction is dynamic, but slower than free fall, with its magnetic flux almost frozen in the matter. Resurrection of ambipolar diffusion (at a density $\sim 10^{11}~{\rm cm^{-3}}$) marks the onset of magnetic decoupling. All species except the electrons gradually, over several orders of magnitude density enhancement, detach from the field lines. Dust grains play a crucial role in the evolution of both the cloud and the core, and by the time these high densities are reached, the grains carry most of the electric charge, despite the fact that most grains are electrically neutral. At densities $\gtrsim 3\times 10^{12}~{\rm cm}^{-3}$, the grains are also the dominant electric {\em current} carriers. These densities are comparable to those at which Ohmic dissipation dominates ambipolar diffusion ($\gtrsim 7\times 10^{12}~{\rm cm}^{-3}$), which is found to occur in a central region of size $\approx 2~{\rm AU}$.

Once magnetic decoupling occurs in the central region, the magnetic flux in this region does not increase in time. Instead, infalling matter entering this region leaves its magnetic flux behind, thus creating a magnetic wall (at a radius $R_{\rm wall} \approx 10~{\rm AU}$), the boundary of the decoupling region. Inside, the magnetic field becomes and remains spatially uniform, with magnitude $B_{\rm dec} \approx 0.2~{\rm G}$. It therefore follows that the protostellar/protoplanetary disc which will eventually form will be exposed to this $B_{\rm dec} \approx 0.2~{\rm G}$. Hence, it may be no accident that measurements of the remanent magnetisation in meteorites imply magnetic fields in the early solar nebula of strength $\approx 0.1-0.2~{\rm G}$.

Magnetic fields are also responsible for quickly establishing and maintaining a disclike geometry all the way down to radial lengthscales on the order of $10~{\rm AU}$, inside which thermal-pressure effects become important. This provides a natural explanation of observations of star-forming cores, such as L1551-IRS5, that reveal a disclike geometry over a broad range of lengthscales.

Throughout the entire collapse phase, the released gravitational potential energy escapes only as rapidly or slowly as allowed by the dust grains, which are mainly responsible for the opacity at temperatures $\lesssim 1500~{\rm K}$. Although the gas remains strictly isothermal only for central densities $\lesssim 10^7~{\rm cm}^{-3}$, the temperature does not exhibit an appreciably rapid increase until densities $\approx 10^{11}~{\rm cm}^{-3}$ are attained in the core. A central temperature of $100~{\rm K}$ is reached when the central density is $\simeq 6\times 10^{12}~{\rm cm}^{-3}$. For the densities considered here ($\lesssim 10^{14}~{\rm cm}^{-3}$), the gas never evolves adiabatically.

The dramatic temperature increase inside $\approx 4~{\rm AU}$ leads to the formation of a hydrostatic core, in which approximate balance is achieved between thermal-pressure and gravitational forces. At the boundary of the hydrostatic core ($\simeq 2~{\rm AU}$), there are prominent shocks, particularly along the vertical symmetry axis where velocities rapidly decrease from $\approx -1.2~{\rm km}~{\rm s}^{-1}$ to zero in only $\approx 8~{\rm AU}$. The infalling matter overshoots the hydrostatic equilibrium state and subsequently oscillates about it. A simple extrapolation of temperature based on our results indicates that a temperature of $1000~{\rm K}$ will not be reached until a central density $\approx 3\times 10^{15}~{\rm cm}^{-3}$. Therefore, the recoupling of the gas to the magnetic field, via thermal ionisation and grain sublimation, will most likely not occur until central densities of at least several $\times 10^{15}~{\rm cm}^{-3}$ are attained.

At the end of the calculation, the mass and magnetic flux in the hydrostatic core are $0.006~{\rm M}_\odot$ and $5\times 10^{-5}~\mu{\rm G}~{\rm pc}^2$, respectively. The mass-to-flux ratio in the central flux tube is $\simeq 80$ times the critical central value for collapse (or $\simeq 169$ times the uniform thin-disc critical mass-to-flux ratio). The luminosity at the hydrostatic core boundary is $\simeq 10^{-3}~{\rm L}_\odot$. The mass infall rate is highly nonhomologous and time-dependent, rising from $\sim C^{3}/G\simeq 1.5\times 10^{-6}~{\rm M}_\odot~{\rm yr}^{-1}$ at the initial boundary of the magnetically supercritical core to a maximum value $\approx 3\times 10^{-4}~{\rm M}_\odot~{\rm yr}^{-1}$ at a radius of $1.6~{\rm AU}$, near the boundary of the hydrostatic core. The maximum value of the mass infall rate normalised to $C^{3}/G$ is $\simeq 27$. The quantitative similarity between this value and the (spatially constant) mass infall rate associated with the Larson-Penston self-similar hydrodynamic collapse solution is no coincidence. By the time a spherical hydrostatic core is formed in our simulations, the infall velocities are dynamic and the magnetic field in the core has effectively decoupled from the matter.

Despite a concerted effort to treat the nonisothermal stage of magnetic star formation as rigorously as possible, there still remains a great deal of work to be done, including both improvements to and extensions of what has already been accomplished here. 

First, dust coagulation ought to be computed in a time-dependent fashion. Unfortunately, the exact details of how and even when grain growth occurs in protostellar cores and discs are poorly constrained both experimentally and observationally. Due to recent improvements in instrumentation, however, observational prospects are becoming increasingly auspicious. A recent comprehensive study of the circumstellar disc surrounding the T Tauri star IM Lupi using photometry, spectroscopy, millimetre interferometry, and multi-wavelength imaging has obtained {\em quantitative} evidence of dust evolution in the disc \citep{pinte08}. Similar results have also been found for a sample of protoplanetary discs around 21 Class II T Tauri stars in the Taurus-Auriga star forming region \citep{ricci09}. The details, nevertheless, may not be very important, since the effect of grain size on the temperature evolution is expected to be minimal, at least for the range of densities studied here; different grain sizes share roughly the same opacity for temperatures less than $\approx 400~{\rm K}$. On the other hand, the strength of the magnetic field at decoupling is a sensitive function of grain size \citep{dm01}. One possible way to circumvent the uncertainties of grain growth, at least at the present time, would be to adopt a nonuniform grain size distribution (such as the MRN distribution) and rerun the numerical simulation presented here. As explained in Paper I, we have already modified the Zeus-MP code to handle such a distribution. Results will be presented in a separate paper, as part of a parameter study.

Second, although rotation was justifiably ignored for the densities investigated here (see Paper I), it cannot be ignored in a study of the more advanced stages of evolution. It is essential for understanding the formation and evolution of protoplanetary discs. It is unknown at present whether magnetic braking plays a role in the formation of such discs and in their angular velocity structure or whether turbulent transfer of angular momentum may be important.

A natural extension as well as an important test of our present work and predictions is the production of synthetic observations to be compared with ongoing actual 3 mm dust and N$_2$H$^+$ observations of the very young Class 0 source HH211-mm by colleagues using the Combined Array for Research in Millimeter Astronomy (CARMA). Such close interplay between theory and observations is now under way.

\section*{Acknowledgments}

We thank Duncan Christie, Leslie Looney, Vasiliki Pavlidou, Paul Ricker, Alex Schekochihin and Konstantinos Tassis for useful discussions. All computations were performed on the {\it Turing} cluster (a 1536-processor Apple G5 X-serve cluster devoted to high-performance computing in engineering and science) maintained and operated by the Computational Science and Engineering Program at the University of Illinois at Urbana-Champaign. We acknowledge, in particular, Michael T. Campbell for his assistance in providing liberal use of the {\it Turing} cluster. TChM acknowledges partial support from the National Science Foundation under grant NSF AST-07-09206.

\bsp
\label{lastpage}

\end{document}